\documentclass[twocolumn,amsmath,amssymb,showpacs]{revtex4}
\usepackage{graphicx}
\usepackage{dcolumn}
\usepackage{bm}

\begin{document}

\title{Families of Vortex Solitons in Periodic Media}
\author{Jiandong Wang, Jianke Yang}
\affiliation{%
Department of Mathematics and Statistics, University of Vermont,
Burlington, VT 05401, USA
}%

\begin{abstract}
Various families of charge-one vortex solitons in two-dimensional
periodic media are reported. These vortices reside either in the
semi-infinite gap or higher band gaps of the media. For both Kerr
and saturable nonlinearities (either focusing or defocusing),
infinite vortex families are found. All these families do not
bifurcate from Bloch bands; rather, they turn around before reaching
edges of Bloch bands. It is further revealed that vortices with
drastically different topological shapes can belong to the same
vortex family, which is quite surprising.
\end{abstract}


\maketitle

\section{Introduction}
Nonlinear wave propagation in periodic media is an important
phenomena, and it arises in a wide array of physical settings such
as photonic-crystal fibers \cite{Jonapolous,Russell},
photorefractive crystals imprinted with a photonic lattice
\cite{Christodoulides,Kivshar_review}, and Bose-Einstein condensates
loaded in an optical lattice \cite{BEC1,BEC2,BEC3}. The interplay
between nonlinearity and periodicity results in a lot of new
physical effects of the wave propagation. For instance, in a
periodic media, solitons can exist not only under focusing
nonlinearity, but also under defocusing nonlinearity
\cite{Kivshar93,Segev_Nature,ChenYanggap_train}. A distinctive
feature of the periodic media is that it can support a wide variety
of solitons residing in different band gaps. Examples include
fundamental solitons \cite{Christodoulides88,
Silberberg,Segev_Nature,Peli_PRE04,YangMuss,Christodoulides03,ChenPRL04,Stegeman},
dipole solitons \cite{Neshev,YangStudies}, vortex solitons
\cite{Malomed,YangMuss,Chen_vortex,Segev_vortex,asymmetric_vortex},
reduced-symmetry solitons \cite{Kivshar_symmetry}, higher-band
vortex solitons \cite{EAOstrovskaya2,Segev_higher_vortex,Kaiser},
embedded-soliton trains \cite{Wang_embedded}, and so on
--- many of which have been experimentally observed. Solitons in
Bessel-ring lattices have been reported too \cite{Kartashov,Chen}.
In an effort to classify solitons in two-dimensional (2D) periodic
media, Shi and Yang \cite{ZShi} studied low-amplitude solitons near
Bloch-band edges. These solitons are Bloch waves modulated by
slowly-varying envelope functions. Using asymptotic techniques, they
derived envelope equations, based on which they successfully
classified all soliton families bifurcating from Bloch-band edges.
Examples include fundamental solitons, reduced-symmetry solitons,
dipole-array solitons, and many others. However, a peculiar question
arose on vortex solitons. According to the envelope equations, those
vortex solitons should also bifurcate from Bloch-band edges.
However, numerical results suggest that they disappear before
reaching band edges \cite{MussYang,EAOstrovskaya}. This paradox has
not been resolved yet, and it clouds over the vortex-soliton
research in periodic media.

In this paper, we comprehensively investigate charge-one vortex
solitons residing in various band gaps of a 2D periodic media, and
resolve the above paradox on this subject. For both the Kerr (cubic)
and saturable nonlinearities, we find that there exist infinite
families of on-site and off-site vortex solitons in the
semi-infinite gap (for focusing nonlinearity) and in the first gap
(for defocusing nonlinearity). We further show that all these vortex
families do not bifurcate from edges of Bloch bands. Rather, before
reaching band edges, they turn back and move into band gaps again.
Within each vortex family, shapes of vortex solitons undergo drastic
changes as the propagation constant varies. As a result, vortices
with very different topological structures can actually belong to
the same family --- a phenomenon which has never been reported
before. These findings significantly deepen our understanding on
vortex solitons in general periodic media.

\section{Vortex Solitons Under Kerr Nonlinearity}
We first consider vortex solitons in 2D periodic media under Kerr
nonlinearity. The mathematical model for this situation is
\begin{equation}
iU_z+U_{xx}+U_{yy}+V(x,y)U+\sigma|U|^2U=0, \label{eq:one}
\end{equation}
where $U$ is a complex function, $\sigma$ takes the value of 1 for
focusing nonlinearity and $-1$ for defocusing nonlinearity, and
$V(x, y)$ is a 2D periodic potential. This equation arises naturally
in Bose-Einstein condensates loaded in an optical lattice
\cite{BEC1,BEC2,BEC3} as well as nonlinear light propagation in
laser-written waveguides \cite{laser_guide} and photonic crystal
fibers with weak transverse index variation. Without loss of
generality, we take the periodic potential $V(x, y)$ as
\begin{equation}
V(x,y)=-V_0(\sin^2x+\sin^2y), \label{eq:two}
\end{equation}
and take $V_0=6$ in all our calculatioins.

Vortex solitons in Eq. (\ref{eq:one}) are sought in the form
\begin{equation}
U(x,y;z)=u(x,y)\exp(-i\mu z),
\end{equation}
where $\mu$ is the propagation constant, and $u(x,y)$ is a
complex-valued localized function which satisfies the equation
\begin{equation}
u_{xx}+u_{yy}+[\mu+V(x,y)]u+\sigma|u|^2u=0. \label{eq:three}
\end{equation}
We will determine these solutions numerically using the modified
squared-operator iteration method and the power-conserving
squared-operator method developed in \cite{JYang}.

\subsection{The Case of Focusing Kerr Nonlinearity}
\begin{figure*}[t]
\centering
\includegraphics[width=0.45\textwidth]{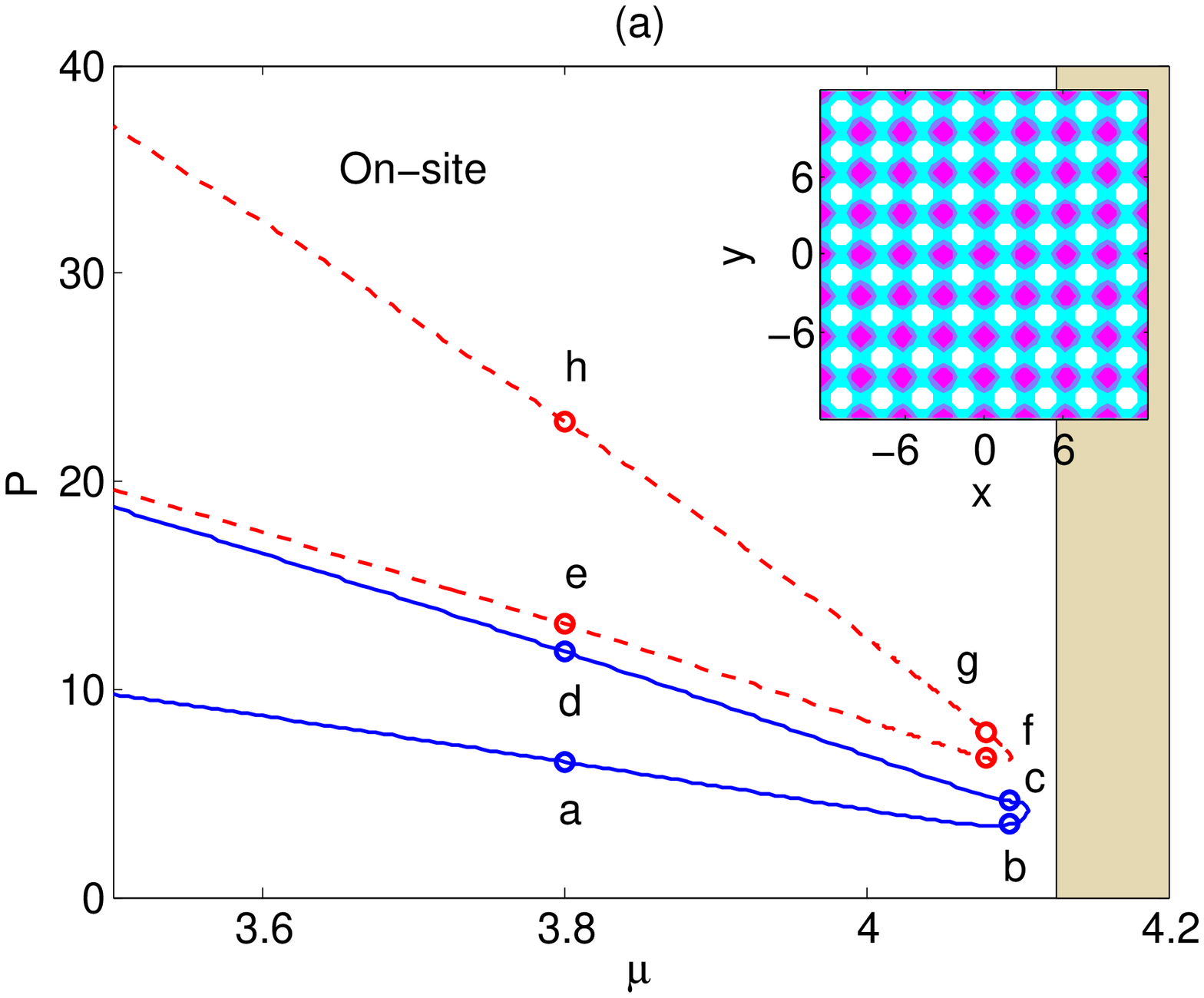}
\includegraphics[width=0.45\textwidth]{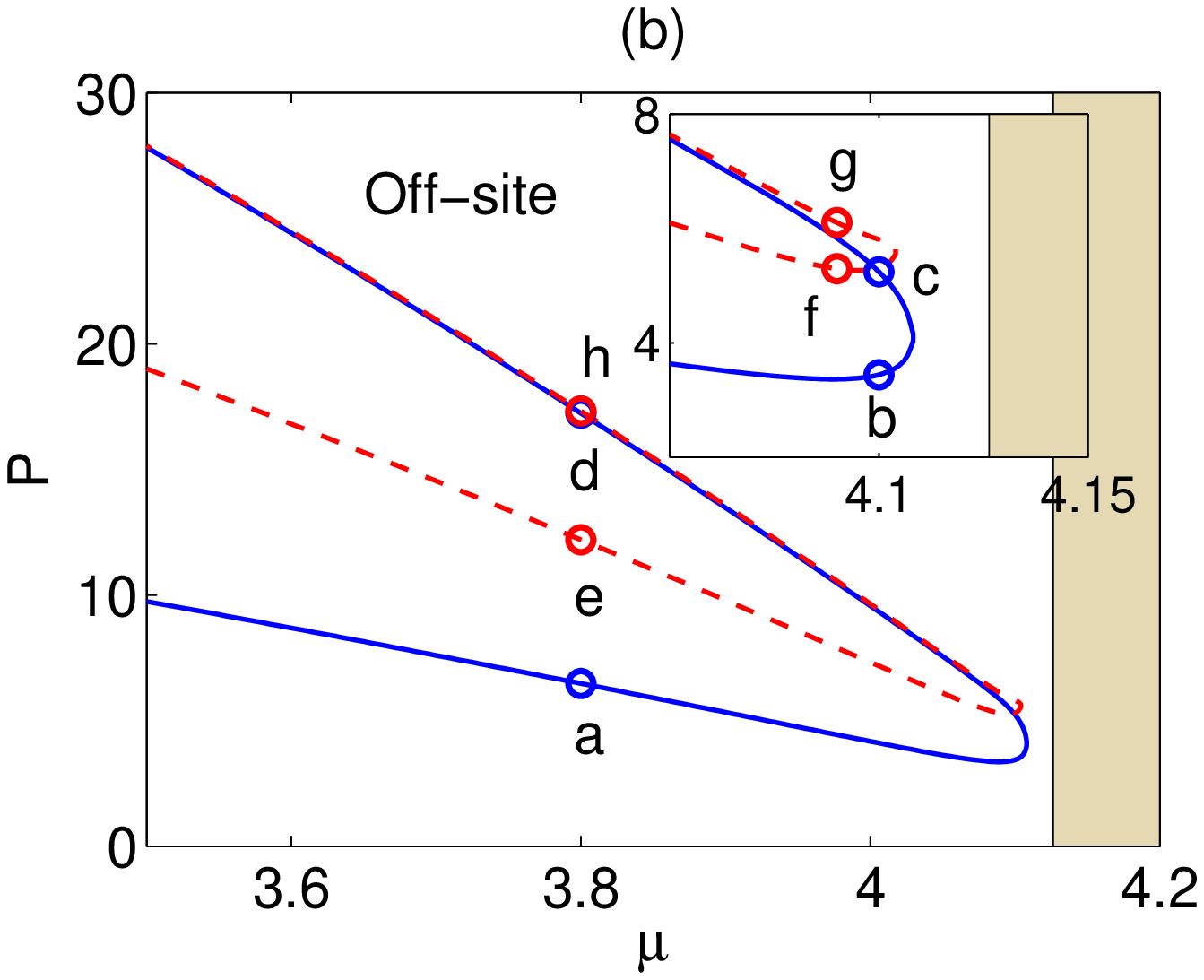}
\caption{(Color online) Power diagrams of the first two families of
on-site (a) and off-site (b) vortex solitons in the semi-infinite
gap under focusing Kerr nonlinearity. The inset in (a) shows the 2D
square lattice of Eq.~(\ref{eq:two}); the inset in (b) zooms in on
the graph near the band edge. Vortex profiles at the marked points
are shown in Fig. \ref{figure2} (on-site) and Fig. \ref{figure3}
(off-site) respectively. Shaded: the first Bloch
band.}\label{figure1}
\end{figure*}

\begin{figure*}[t]
\centering
\includegraphics[width=0.2\textwidth]{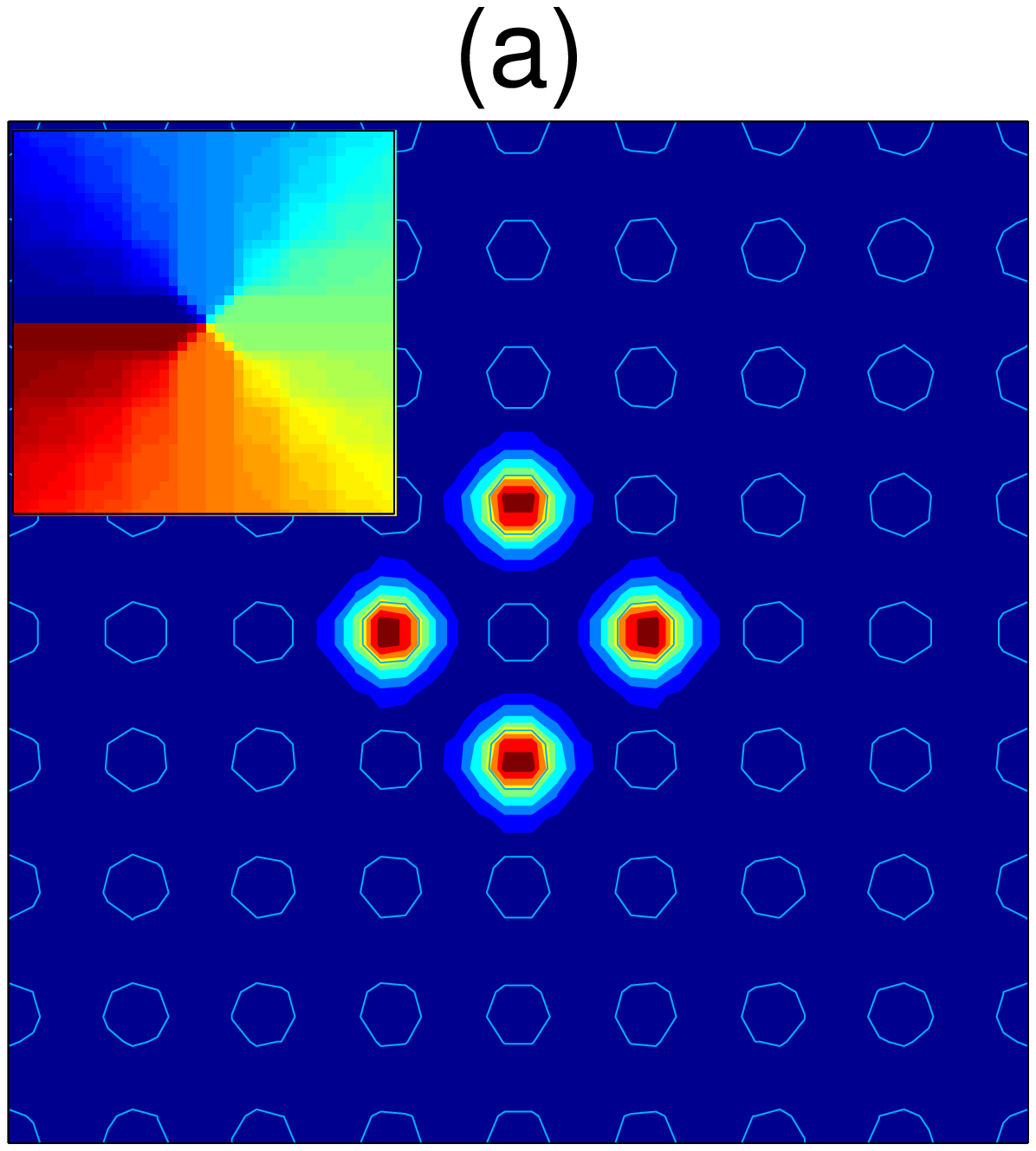}
\includegraphics[width=0.2\textwidth]{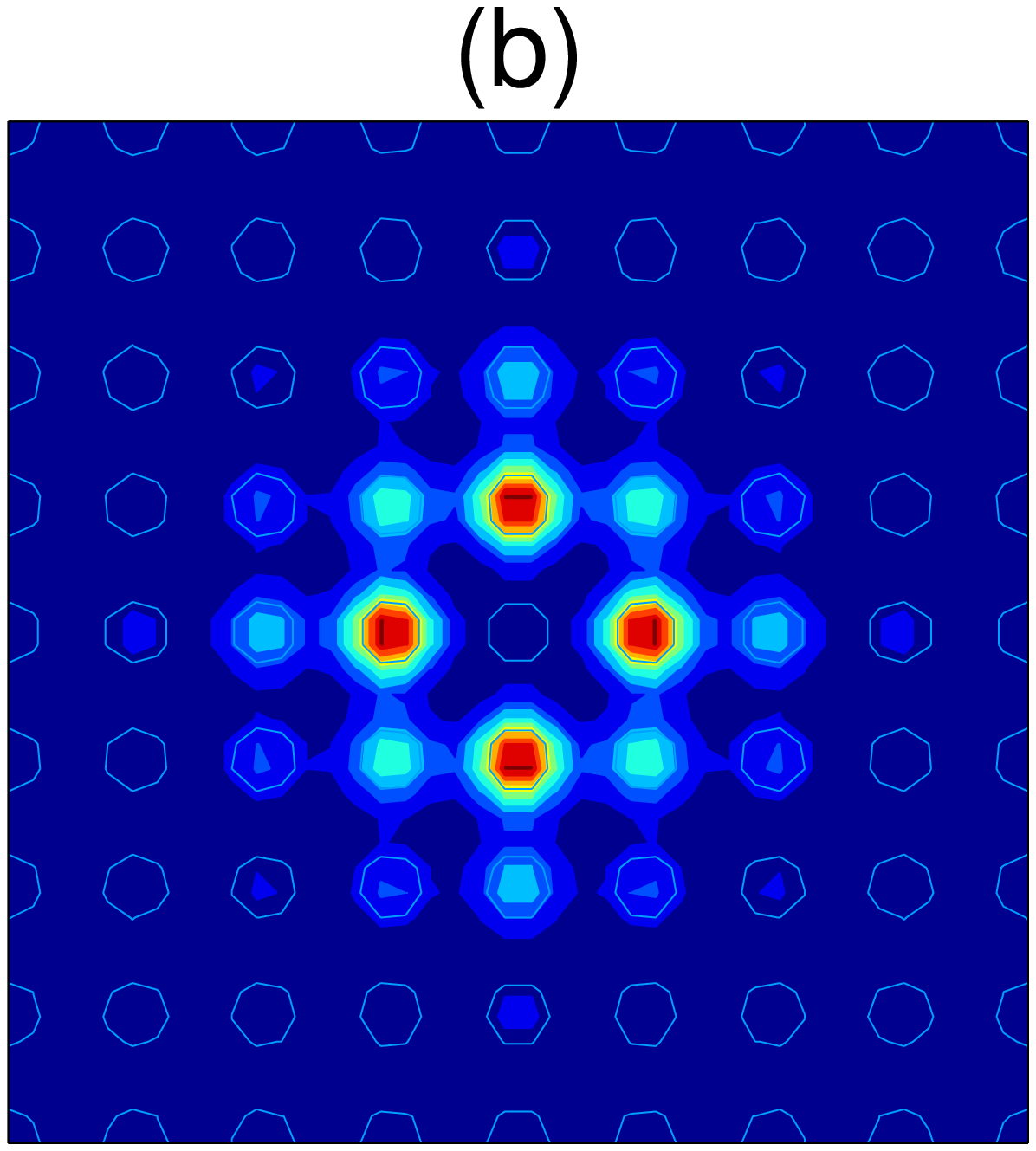}
\includegraphics[width=0.2\textwidth]{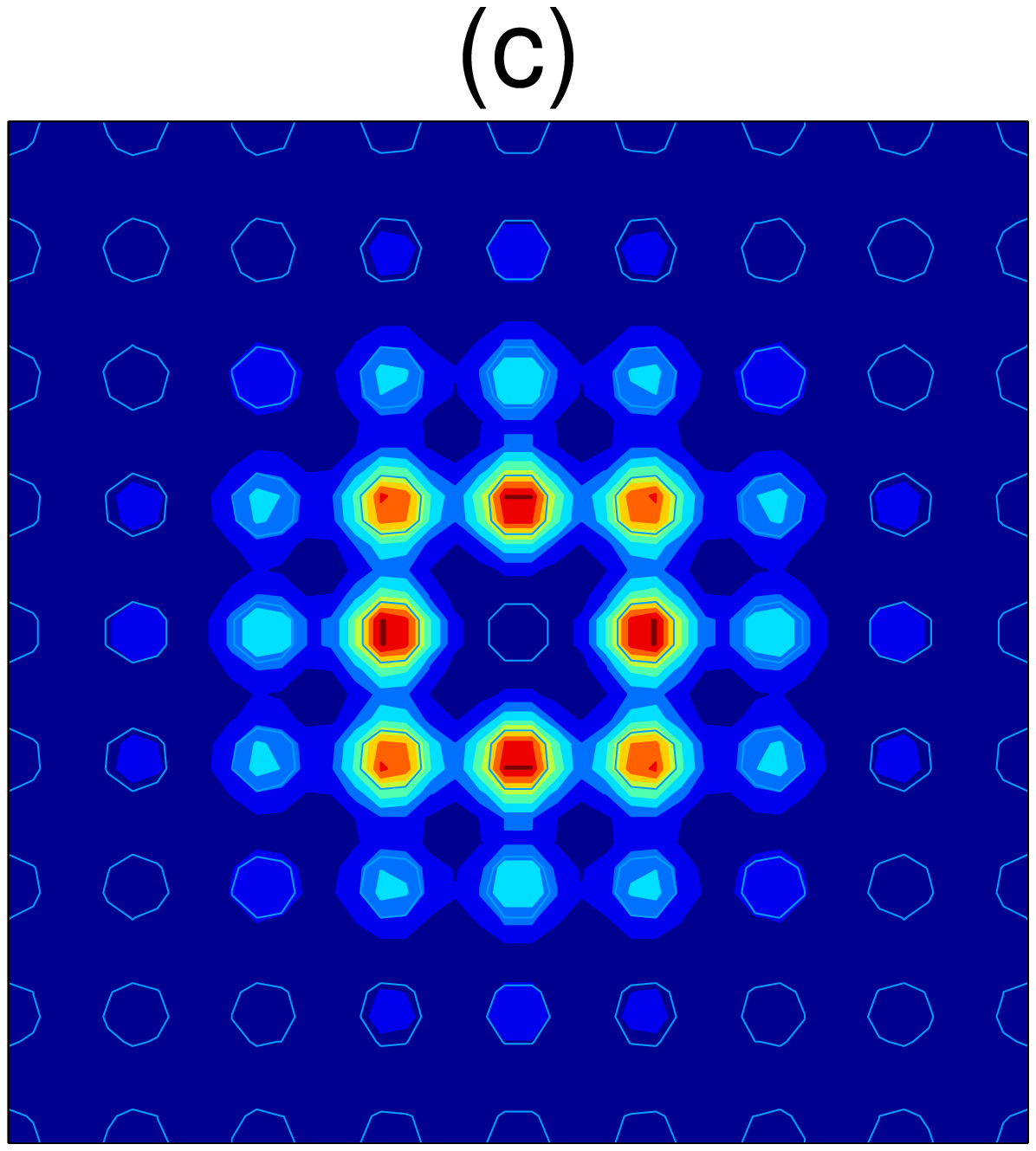}
\includegraphics[width=0.2\textwidth]{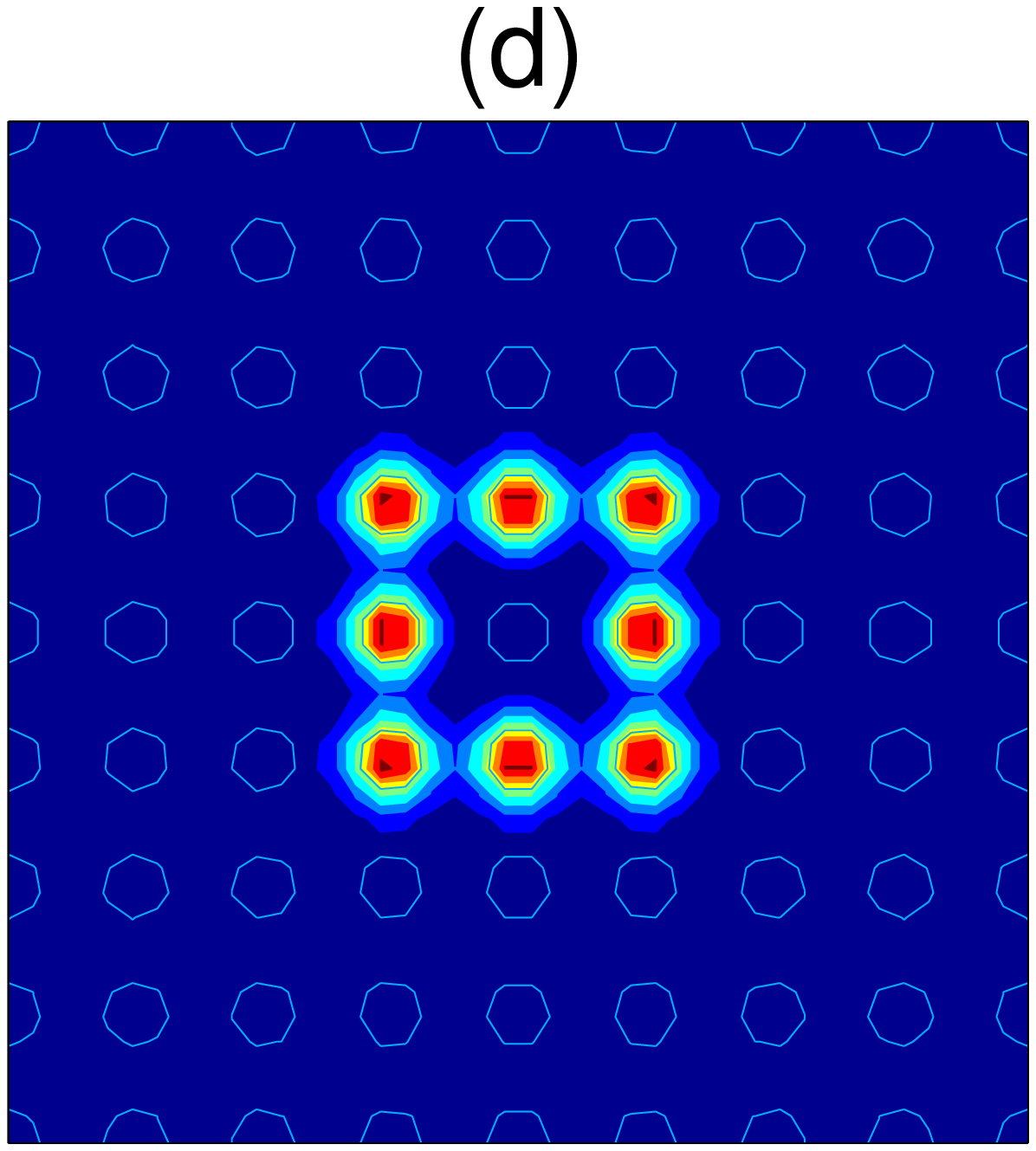}\\
\vspace{0.5cm}
\includegraphics[width=0.2\textwidth]{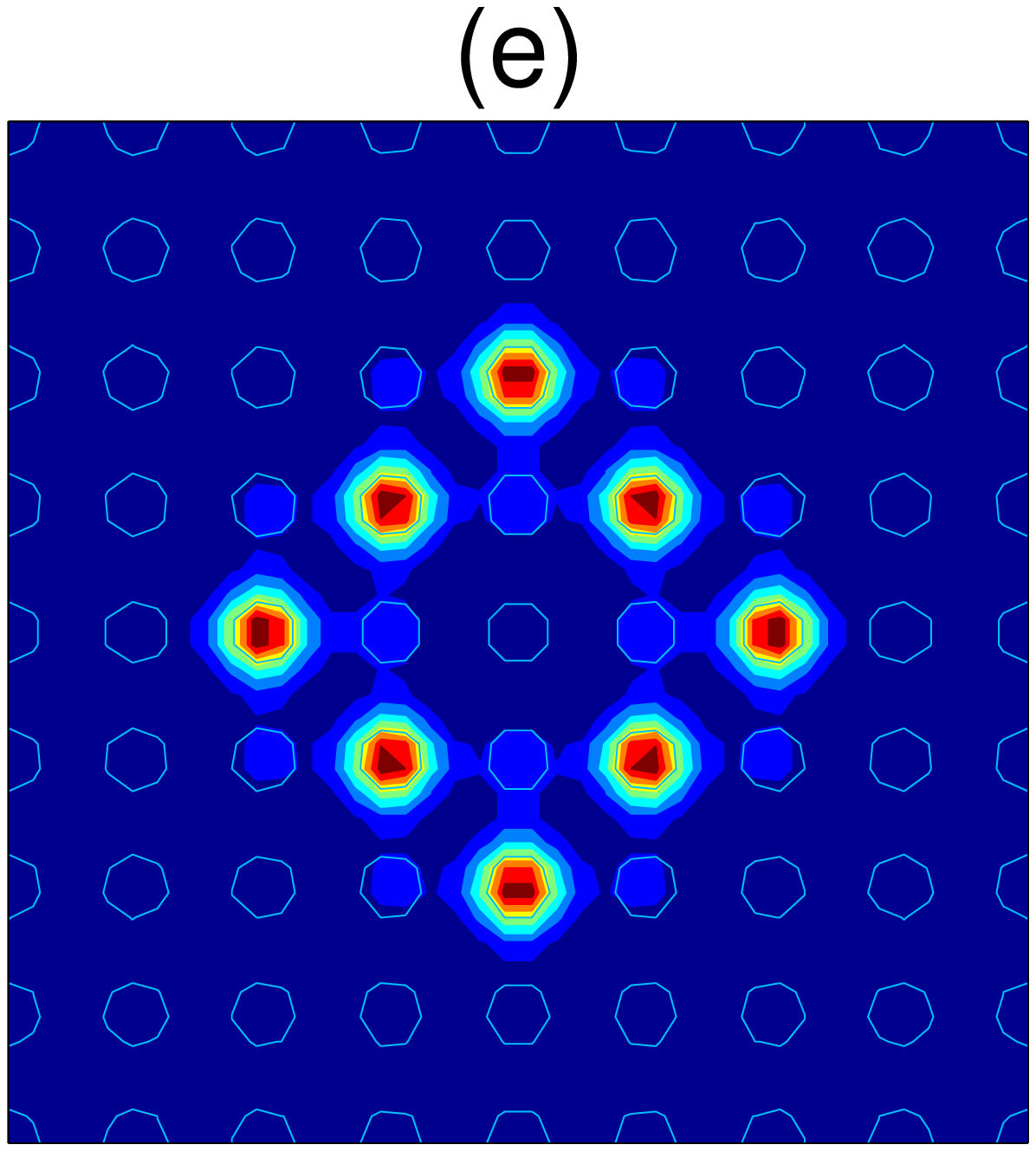}
\includegraphics[width=0.2\textwidth]{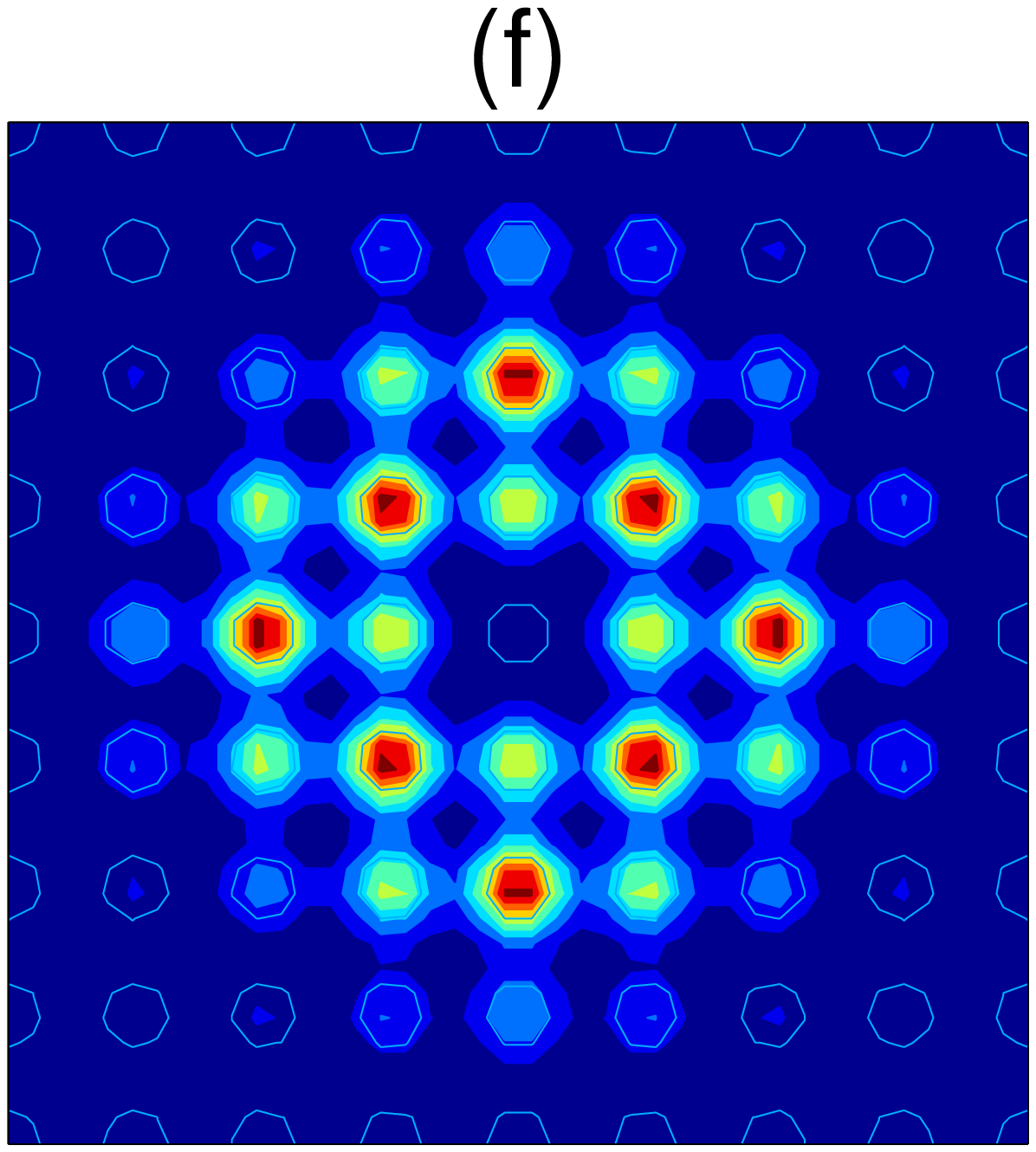}
\includegraphics[width=0.2\textwidth]{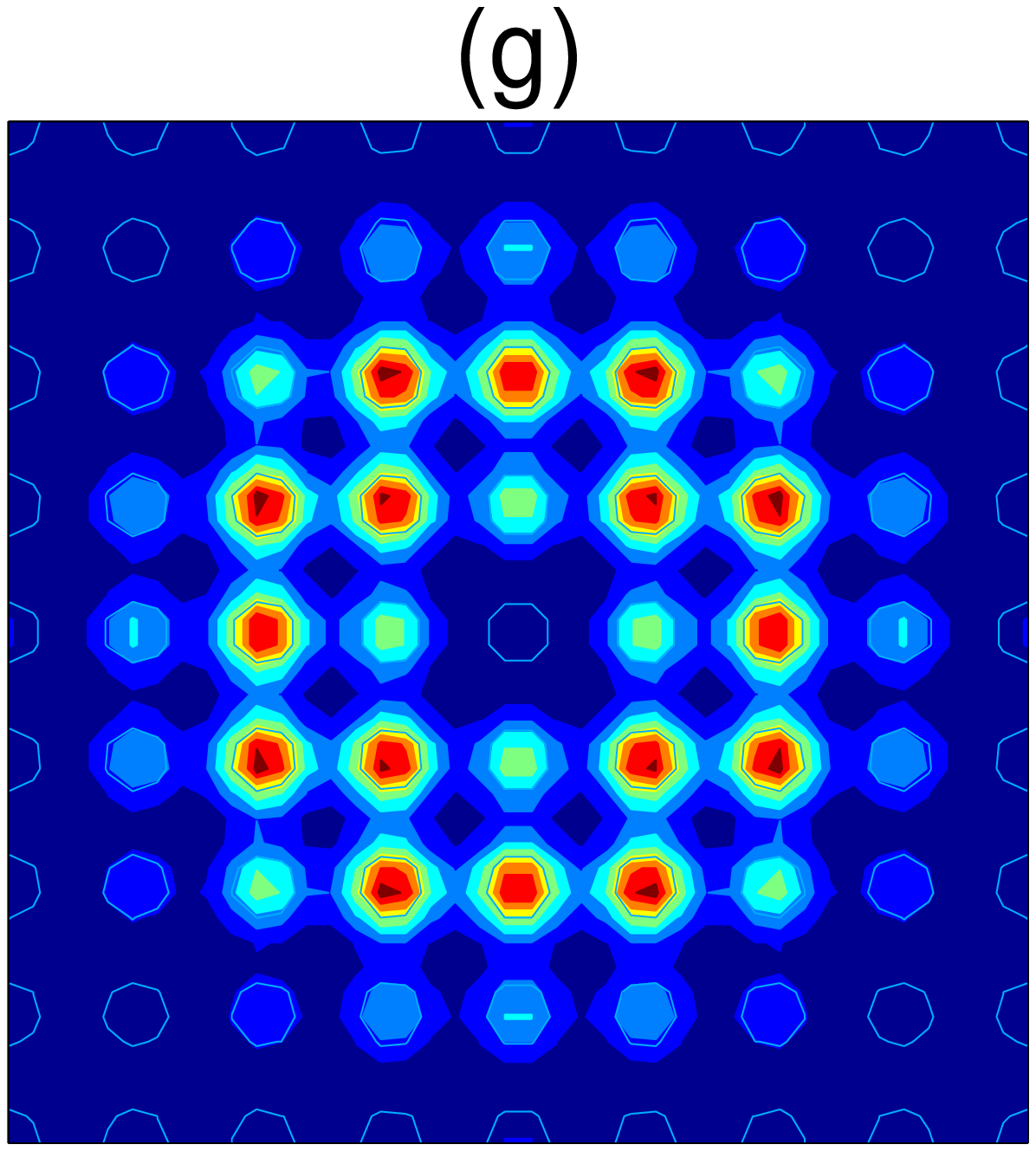}
\includegraphics[width=0.2\textwidth]{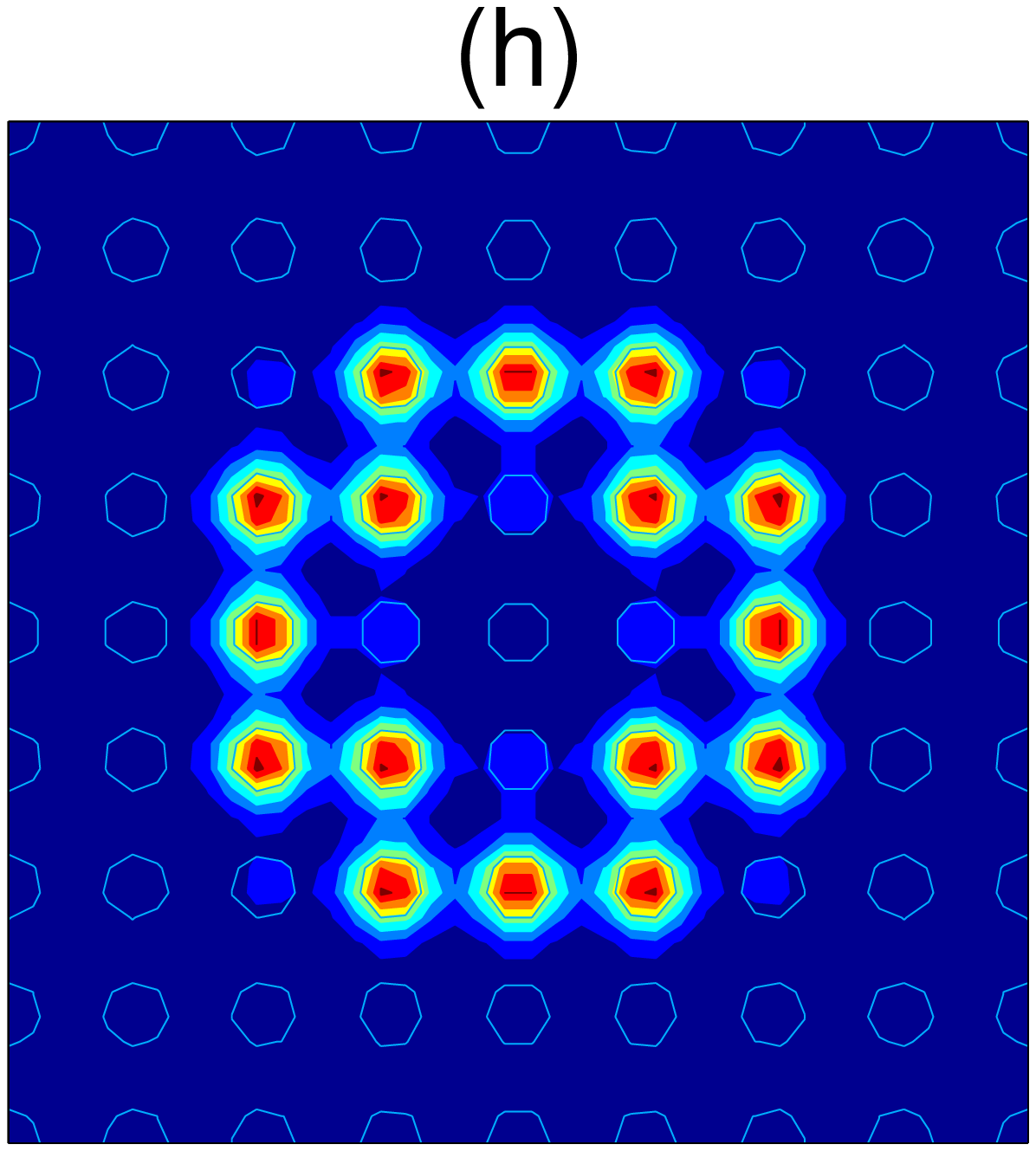}
\caption{(Color online) Vortex profiles ($|u|$) in the first (top)
and second (bottom) on-site solution families under focusing Kerr
nonlinearity at marked points in Fig. \ref{figure1} (a). The inset
in (a) is the typical phase structure of all these vortices. The
background circles represent the lattice sites (with high $V$
values), as in Figs. 3, \ref{figure6} and \ref{figure7}  as well.
 }\label{figure2}
\end{figure*}

First, we study vortex solitons supported by the focusing Kerr
nonlinearity, i.e. $\sigma=1$ in Eq.~(\ref{eq:three}). Two types of
vortex solitons --- on-site and off-site ones, will be sought.
On-site vortices are centered on a lattice site (lattice peak),
while off-site vortices are centered between lattice sites (lattice
minimum) \cite{Malomed,YangMuss,Chen_vortex,Segev_vortex}. For each
type of vortices, various families are found in the semi-infinite
gap. The power diagrams of the first two families of on-site and
off-site vortex solitons are shown in Fig. \ref{figure1} (a, b)
respectively. Here the power is defined by
$P=\int_{-\infty}^{\infty}\int_{-\infty}^{\infty}|u|^2dxdy$. A
distinctive feature about these power curves is that each curve has
a slanted U-shape. At each propagation constant $\mu$ away from the
band edge, one can find two vortex solitons with different powers
within each family. Another feature about these curves is that none
of them reaches the edge of the (first) Bloch band. This means that
none of these vortex families bifurcates from Bloch bands. These
features are in stark contrast with those for soliton families
reported in \cite{ZShi}, which do bifurcate from Bloch bands (in the
low-amplitude limit), and each propagation constant only corresponds
to a single soliton solution within each family. To find out what
types of vortex solitons are contained in each family, we display
them at four representative locations of each power curve in Fig. 2
(on-site) and Fig. 3 (off-site). First we examine the first on-site
vortex family whose power curve is the solid (blue) one in Fig.
1(a). At four marked positions `a-d' on this curve, the
corresponding vortex profiles are displayed in Fig. 2(a-d)
respectively. The phase structures of all these vortices are similar
to that of the ring vortex soliton in bulk media and shown in the
inset of Fig. 2(a). Winding around the center of the vortex, the
phase increases by $2\pi$, thus these vortices have charge one. The
intensity distributions of these vortices, on the other hand, differ
significantly at various locations of the power curve. On the lower
branch of the power curve, when $\mu$ is far away from the Bloch
band, the vortex soliton consists of four main humps residing in the
four lattice sites closest to the center site and forming a diamond
configuration [see Fig. \ref{figure2} (a)]. Hence this is the
familiar on-site vortex soliton which has been reported before
\cite{Malomed,Yang_NJP,Segev_vortex}. When $\mu$ increases toward
the band edge along the lower branch, the vortex's power as well as
its peak intensity decreases, and the vortex starts to develop tails
on the outside of its four main humps [see Fig. \ref{figure2} (b)].
As $\mu$ further increases beyond a cut-off value
$\mu^{(1)}_{c}$=4.107, which is close to but below the band edge
$\mu_{edge}$=4.126, vortices in this family can not be found
anymore. This phenomenon was first reported by one of the authors in
\cite{MussYang}, and it was puzzling at the time since we did not
expect a solution branch terminating abruptly away from edges of
Bloch bands. This puzzle is now resolved by Fig. 1(a). What happens
is that when $\mu$ reaches the cut-off value $\mu^{(1)}_{c}$, the
power curve turns around and enters the upper branch of the same
family [see Fig. 1(a)]. When passing through the turn-around, the
vortex undergoes rapid shape changes and evolves into eight main
humps in a square configuration surrounding the central lattice
site. Tails at the outside of this square are significant too. As
$\mu$ moves away from the Bloch band along the upper branch, the
tails gradually fade away, and the eight main humps intensity
totally dominate the vortex [see Fig. \ref{figure2} (d)]. Thus, in
this first family of on-site vortex solitons, the vortex undergoes
dramatic shape changes from a four-humped diamond-like structure
into an eight-humped square structure. This is quite unexpected and
was not known before.

\begin{figure*}
\centering
\includegraphics[width=0.2\textwidth]{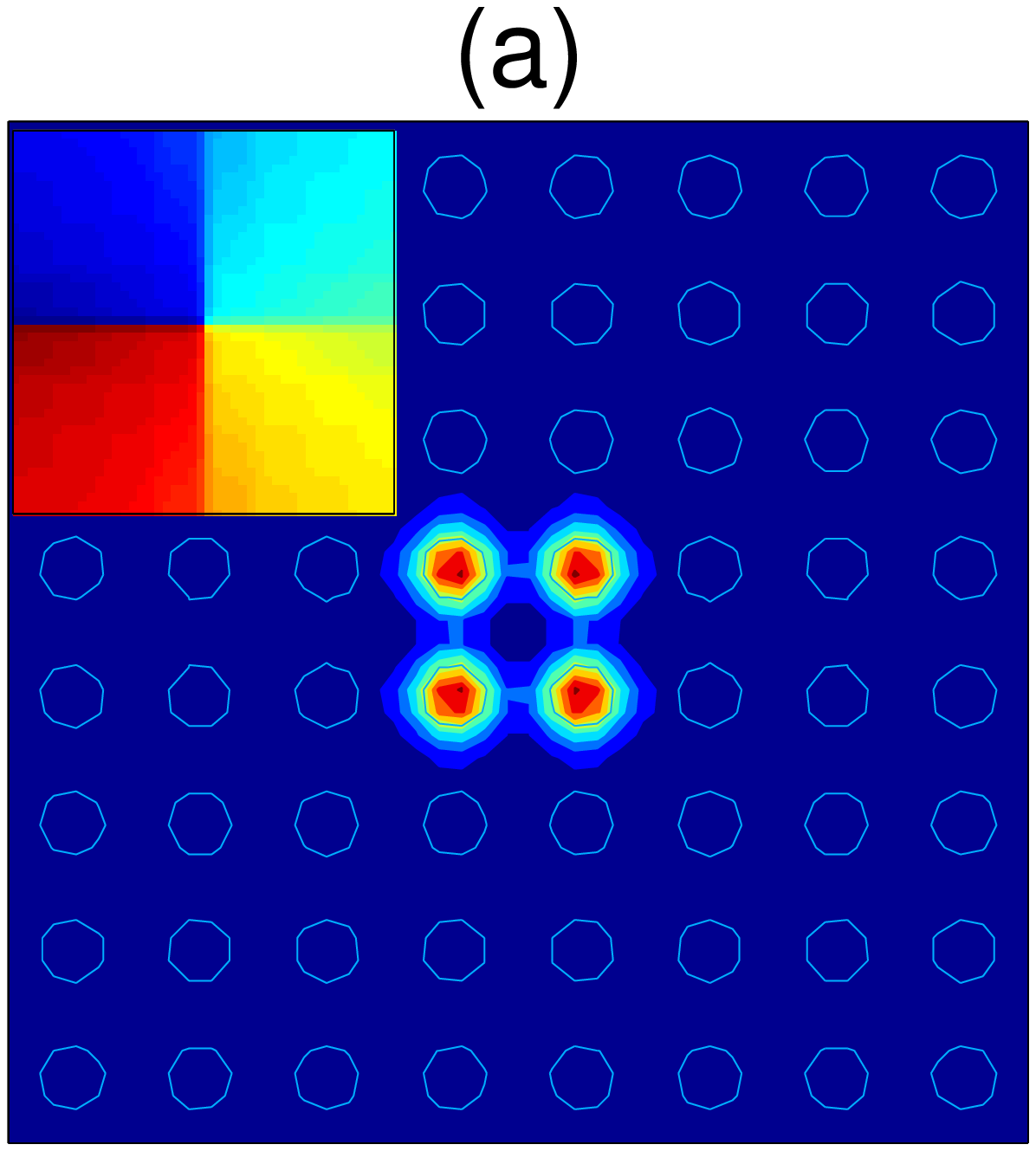}
\includegraphics[width=0.2\textwidth]{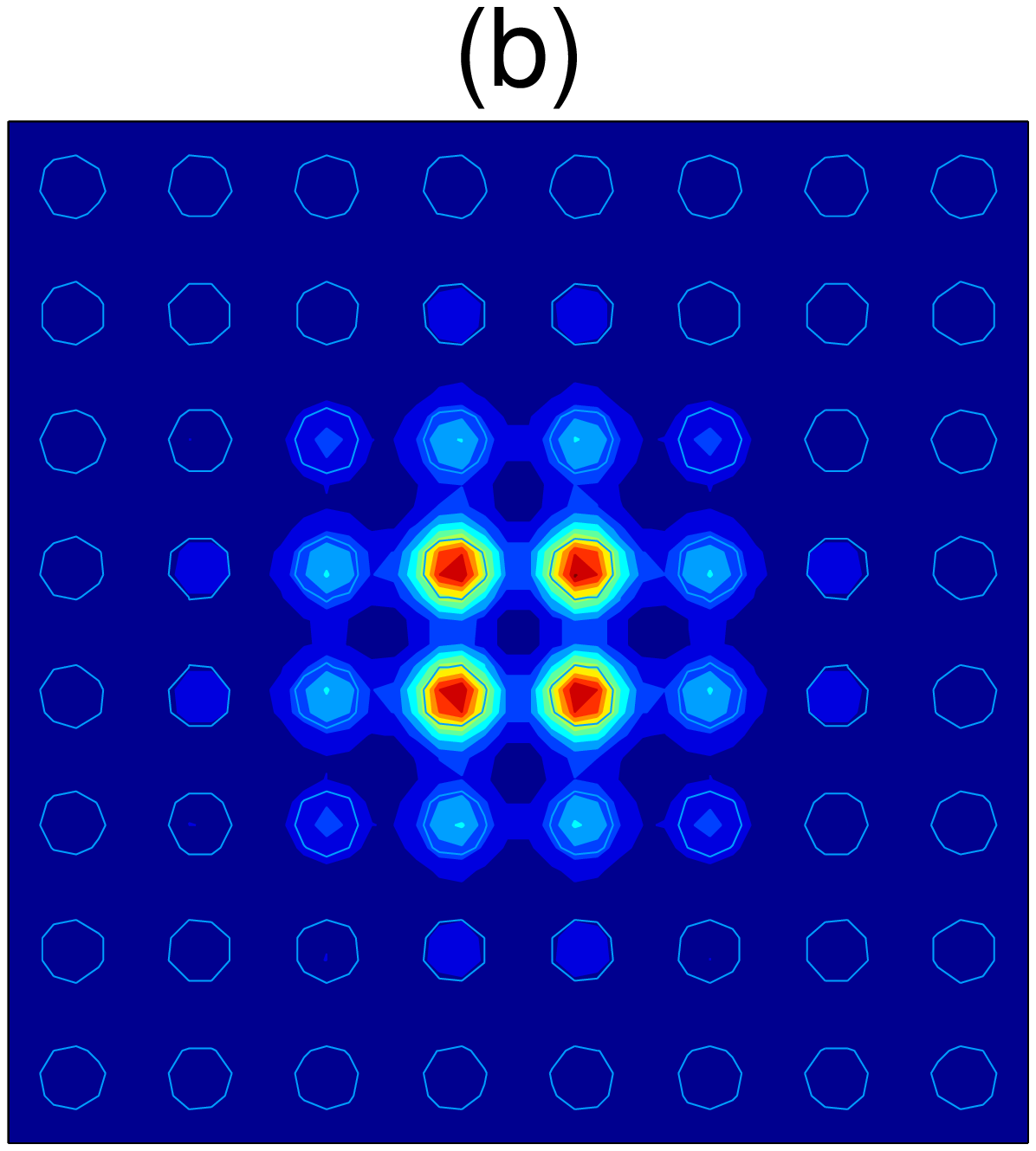}
\includegraphics[width=0.2\textwidth]{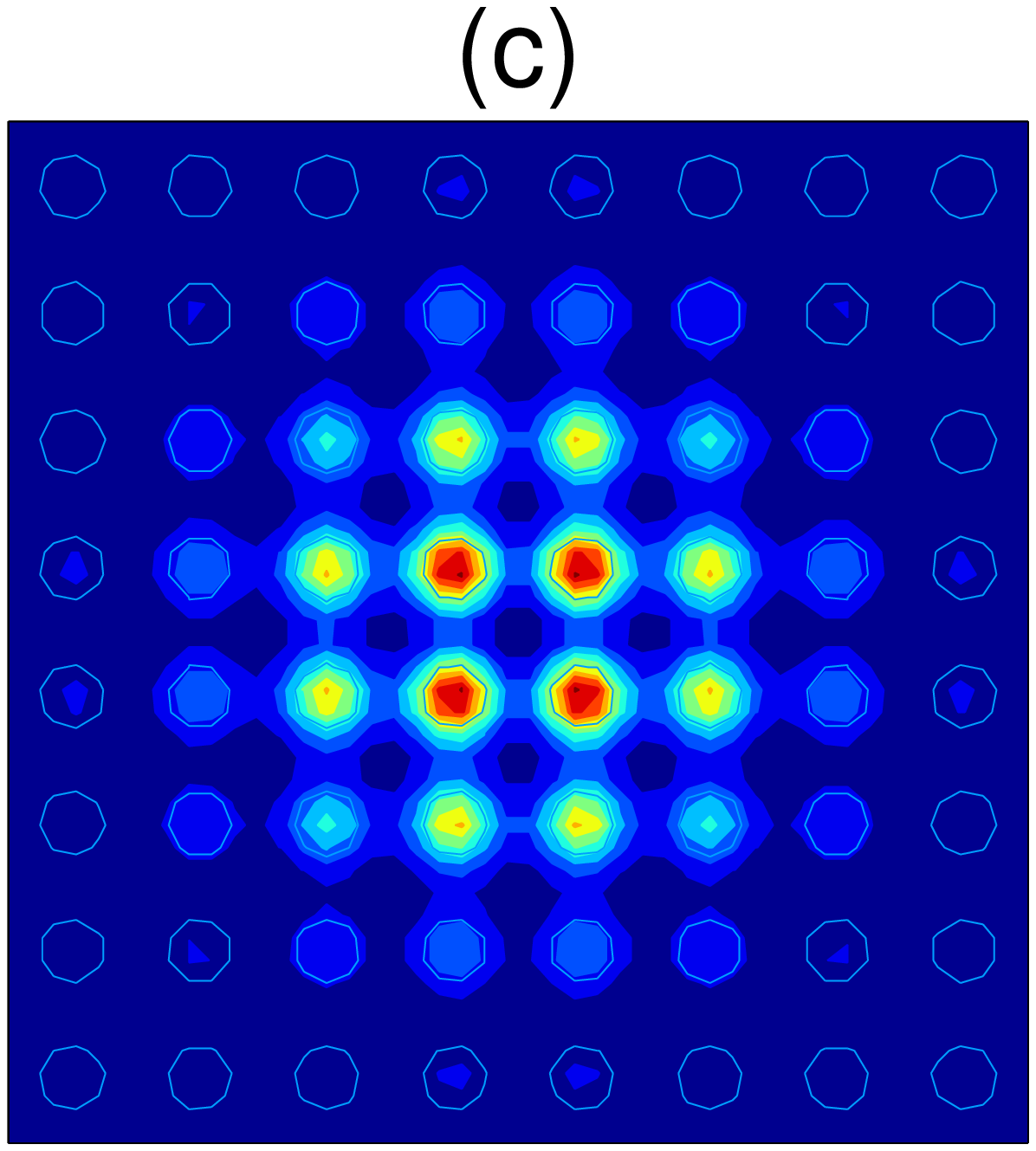}
\includegraphics[width=0.2\textwidth]{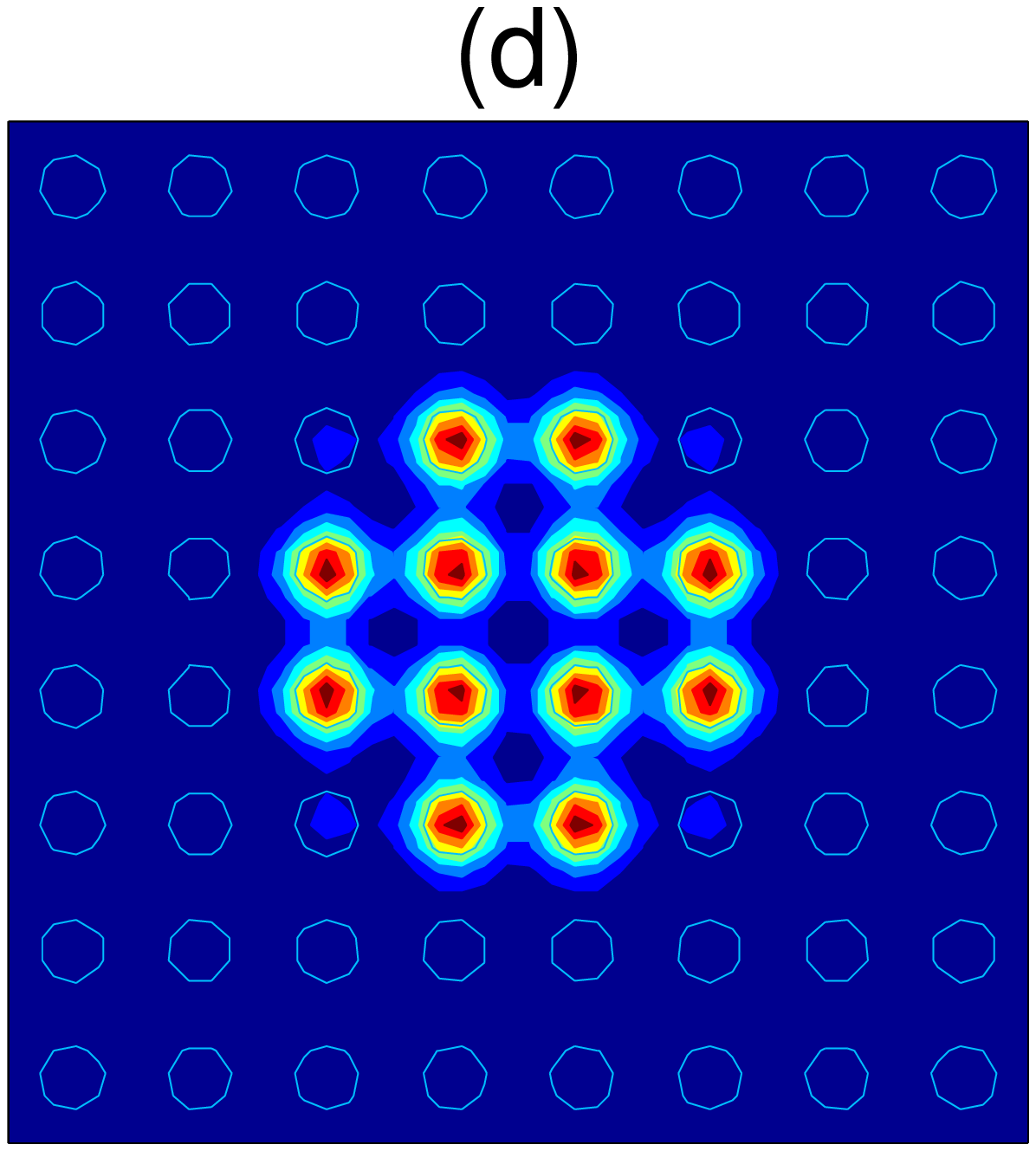}\\
\vspace{0.5cm}
\includegraphics[width=0.2\textwidth]{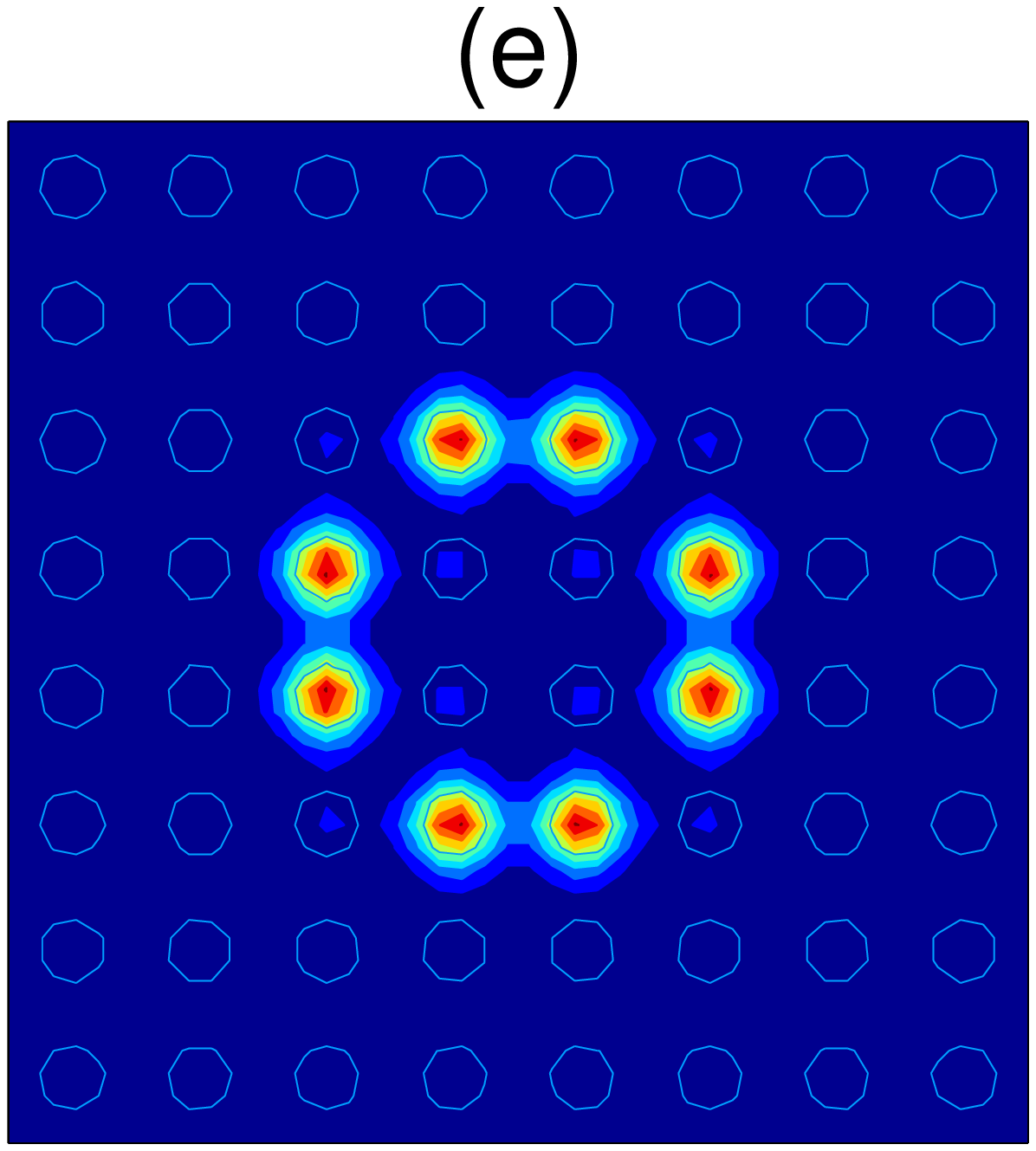}
\includegraphics[width=0.2\textwidth]{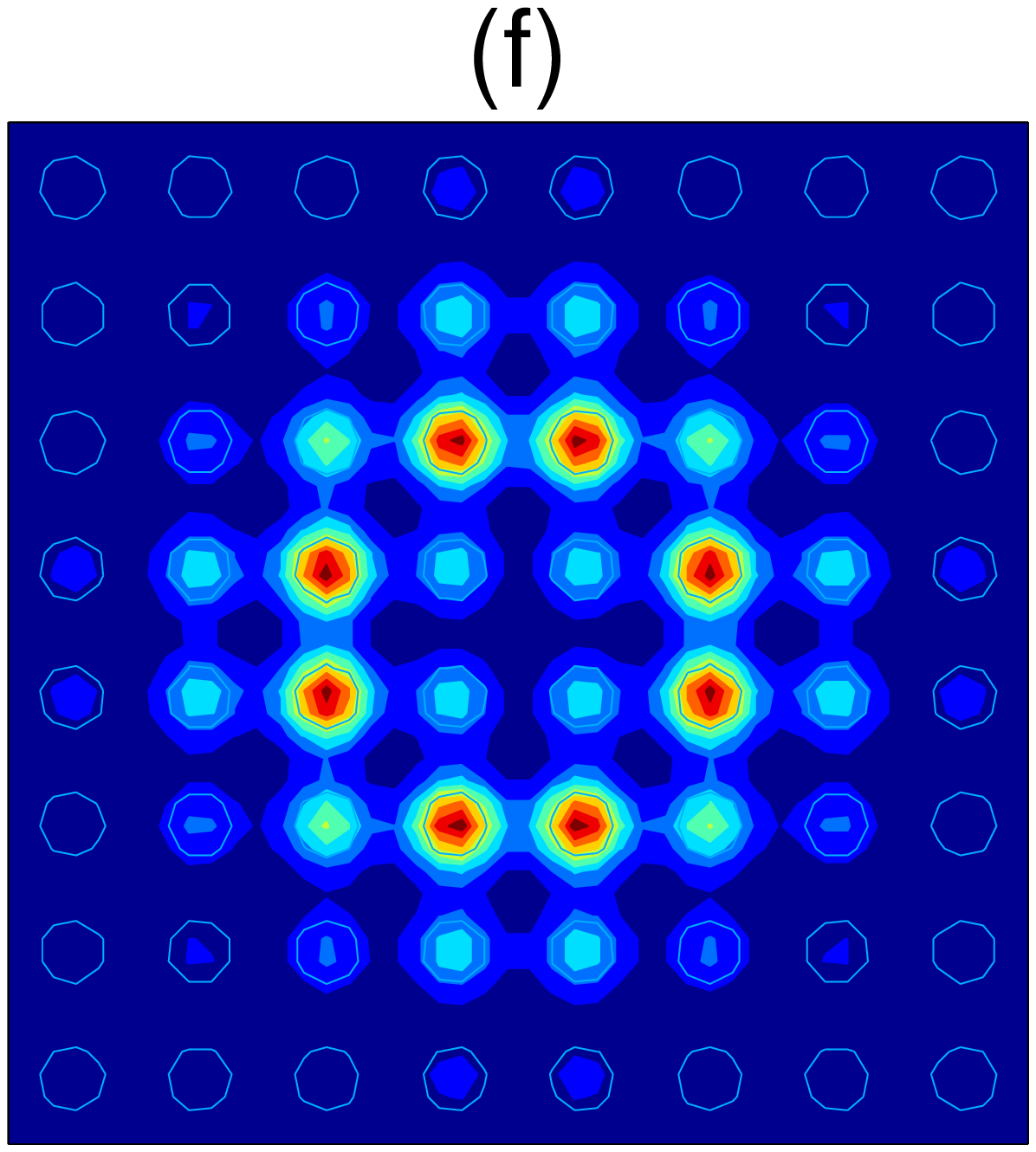}
\includegraphics[width=0.2\textwidth]{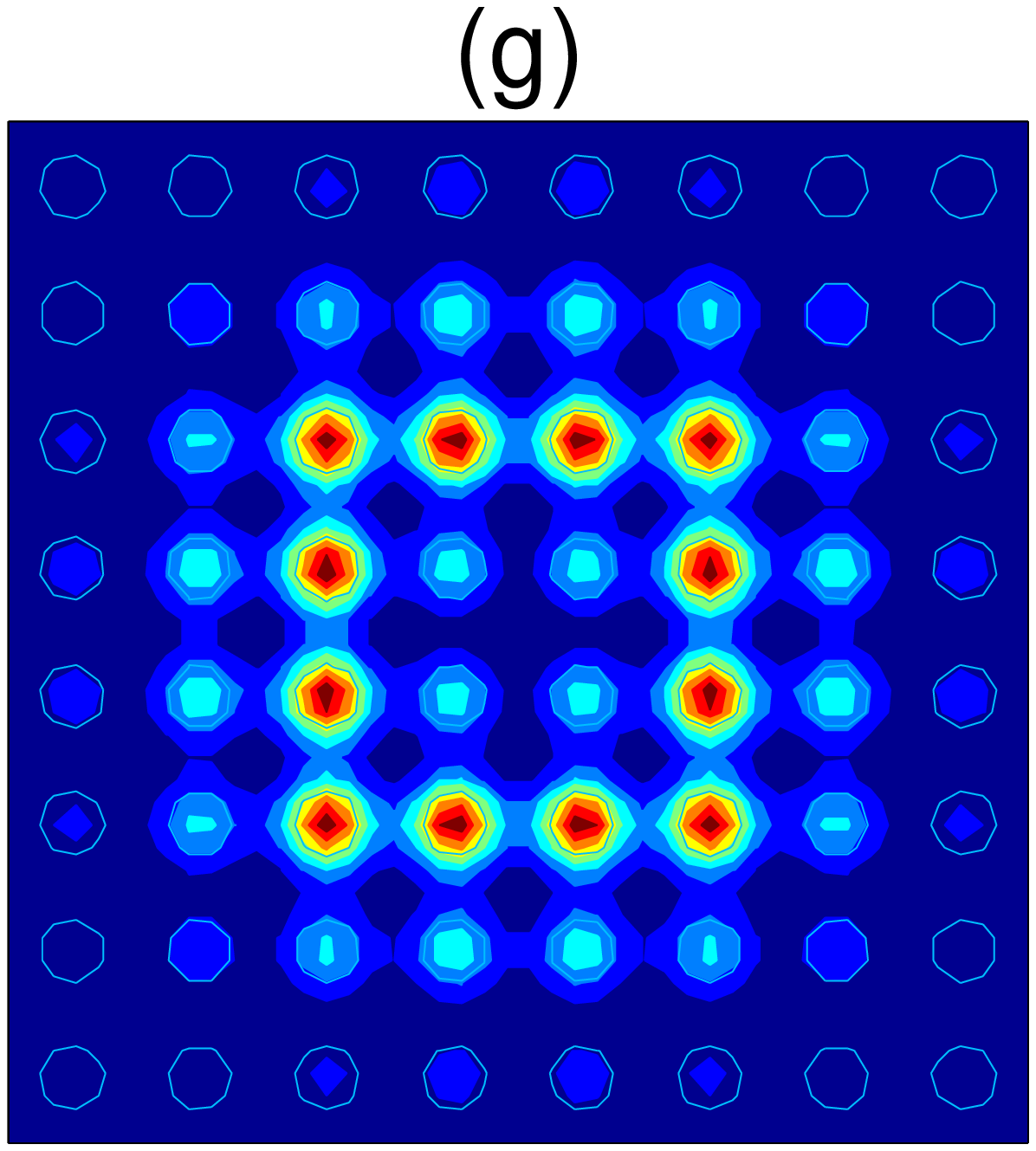}
\includegraphics[width=0.2\textwidth]{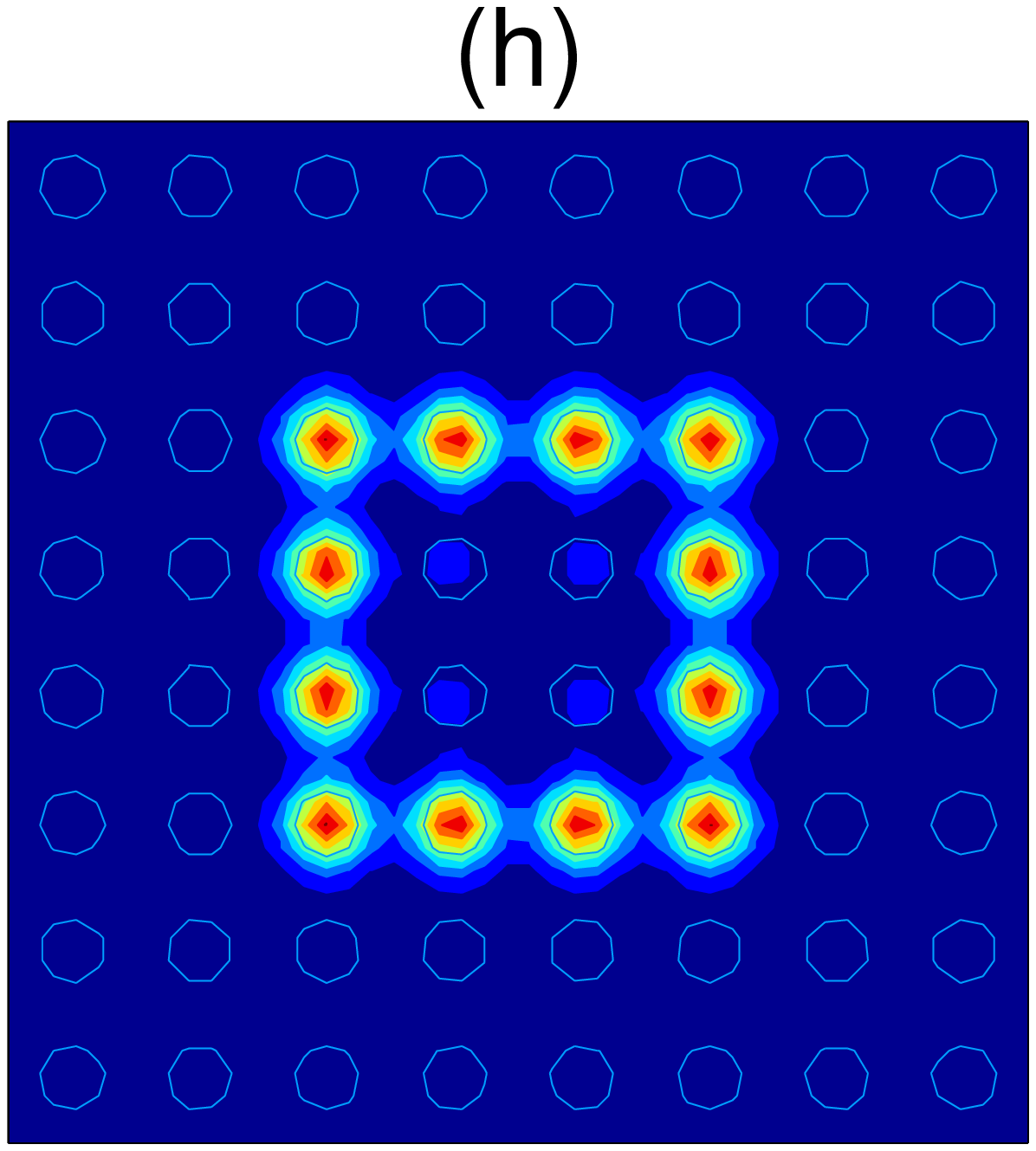}
\caption{(Color online) Vortex profiles ($|u|$) in the first (top)
and second (bottom) off-site solution families under focusing Kerr
nonlinearity at marked points in Fig. \ref{figure1} (b). The inset
in (a) is the typical phase structure of all these vortices.
}\label{figure3}
\end{figure*}

Regarding the second on-site family, vortex solitons at four
representative positions `e-h' of the power curve are displayed in
Fig. \ref{figure2} (e-h). The phase structures of these vortices are
similar to those in the first family [see inset of Fig. 2(e)], i.e.
these vortices also have charge one. But the intensity patterns in
this second family are quite different from those in the first
family. Specifically, in the lower branch of the second family, away
from the Bloch band, the vortex contains eight main humps in a
larger-diamond configuration [see Fig. 2(e)]. As $\mu$ moves
continuously from the lower branch to the upper one, the vortex
continuously evolves from this eight-humped diamond configuration to
a sixteen-humped configuration resembling the surrounded wall of a
castle [see Fig. 2(e-h)]. This dramatic shape change within the same
solution family occurs for all vortex families, and it is one of the
main findings of this article.

For off-site vortex solitons in the semi-infinite gap, various
solution families are found as well. In the first off-site family,
representative solution profiles are displayed in Fig. 3 (a-d).
These vortices also have a simple $2\pi$ phase winding structure
around the vortex center, thus having charge one [see inset in Fig.
3(a)]. Regarding their intensity patterns, when $\mu$ is far away
from the Bloch band on the lower branch, the vortex has four main
humps occupying four adjacent lattice sites in a compact square
configuration. This is the familiar off-site vortex soliton reported
in \cite{YangMuss,MussYang,Chen_vortex,Segev_vortex}. As $\mu$ moves
from the lower branch to the upper one, the vortex changes from the
four-humped square configuration to a twelve-humped compact cross
configuration [see Fig. 3(a-d)]. In the second off-site family,
lower-branch vortex solitons away from the Bloch band have eight
main humps arranged in an octagon configuration [see Fig. 3(e)]. As
$\mu$ moves to the upper branch, however, the vortex becomes a
square with twelve main humps on its perimeter [see Fig. 3(h)].
Notice from Fig. \ref{figure1} (b) that the upper branches of these
two off-site families have almost the same power for the same
propagation constant. This is because on these two upper branches,
vortices of both families have the same number of main humps, and
nearby humps are all aligned along the lattice direction and
separated by one lattice spacing [see Fig. 3(d,h)]. One may recall
that for on-site vortices, the upper branch of the first family and
the lower branch of the second family also have the same number of
humps [see Fig. 2(d, e)]. But adjacent humps in the latter case are
aligned along diagonal directions of the lattice and have larger
spatial separations than in the former case. This caused noticeable
power differences between these two branches as can be seen in Fig.
\ref{figure1} (a).

From Fig. 1, we see that both on-site and off-site vortices can
exist close to the Bloch band [see points `c, g' in Fig. 1(a) for
example]. We find that such vortices near the Bloch band have lower
amplitudes. For instance, the vortex at point `c' in Fig. 1(a) has
amplitude 0.55, and the one at point `g' has amplitude 0.48.
Comparatively, vortex amplitudes far away from the Bloch band are
much higher. Additionally, vortices close to the Bloch band have
long tails, and they resemble the Bloch wave of the band edge
modulated by a ring-vortex envelope [see Fig. 2(c, g) for instance].
This seems to suggest that vortex solitons near Bloch bands can be
treated analytically by asymptotic methods. However, such asymptotic
analysis encounters subtle problems for vortex solitons, as we will
explain below. Low-amplitude Bloch-wave packets in Eq.
(\ref{eq:one}) have been analyzed in \cite{ZShi}. Near the band edge
$\mu_{edge}$ in Fig. 1 and under focusing nonlinearity, the
leading-order asymptotic solution is $u(x,y)=\epsilon A(X, Y)p(x,
y)$, $\mu=\mu_{edge}-\epsilon^2$. Here, $p(x,y)$ is the Bloch wave
at the band edge $\mu_{edge}$ (with $\Gamma$-point symmetry),
$X=\epsilon x, Y=\epsilon y$, $\epsilon \ll 1$, and the envelope
function $A(X,Y)$ satisfies the equation
\begin{equation} \label{A}
D_1(A_{XX}+A_{YY})-A+\alpha_0 |A|^2 A=0,
\end{equation}
where $D_1>0$ is the second-order dispersion coefficient at the band
edge, and $\alpha_0 > 0$ is a constant. The envelope equation
(\ref{A}) admits a ring-vortex solution $A=f(R)e^{i\Theta}$, where
$(R, \Theta)$ is the polar coordinates of the $(X, Y)$ plane. Due to
some additional constraints on the solution $u(x, y)$, this ring
envelope can only be centered at certain locations at or between
lattice sites \cite{ZShi}. Centering the ring-vortex envelope at a
lattice site, the corresponding solution $u(x,y)$ would then be a
low-amplitude on-site vortex soliton which resembles the ones such
as Fig. 2 (c, g). Centering the ring-vortex envelope between a
lattice site, we would get a low-amplitude off-site vortex soliton
which resembles the ones in Fig. 3 (c, f). However, a contradiction
between this asymptotic analysis and numerical results is that,
according to the asymptotic analysis, these vortices should exist
continuously as $\mu$ approaches the band edge $\mu_{edge}$ (i.e.
$\epsilon\to 0$), but the numerical results in Fig. 1 indicate that
true vortex solitons actually do not approach the band edge for each
vortex family. Apparently, the asymptotic analysis above is not
entirely correct for vortex solitons, and it must be revised in
order to explain the true vortex behaviors in Figs. 1-3. How this
can be done is still unclear yet. Note that these vortex behaviors
are similar to multi-packet solitary waves in the fifth-order
Korteweg de Vries equation \cite{Akylas}, so the analytical
treatment of \cite{Akylas} might be useful for the present problem.
This remains to be seen.

Fig. \ref{figure1} just shows the first two families of on-site and
off-site vortex solitons. We have also found other vortex families
in Eq. (\ref{eq:three}) whose power curves have similar slanted
U-shapes and are higher than those in Fig. \ref{figure1}. Vortices
in these higher families contain more intensity humps which are
located further away from the center of the vortex, but their phase
structures remain similar to those of the first two families (i.e.
they all have charge one). It can be inferred that infinite families
of such vortex solitons exist in the semi-infinite gap, and all of
them do not bifurcate from the edge of the first Bloch band. Within
each higher family, vortex shapes also undergo drastic changes as
$\mu$ moves from the lower branch to the upper one.

The vortex solitons studied above are symmetric about the vortex
center. Asymmetric vortices also exist in a square lattice
\cite{asymmetric_vortex}. We find that the power curves of
asymmetric vortices also have slanted U-shapes as in Fig. 1, and
they do not bifurcate from Bloch bands either. We also find that
asymmetric vortices reported in \cite{asymmetric_vortex} belong to
the lower branches of these power curves. As $\mu$ moves to the
upper branches, shapes of asymmetric vortices become very different
and more complex.

\begin{figure*}
\centering
\includegraphics[width=0.45\textwidth]{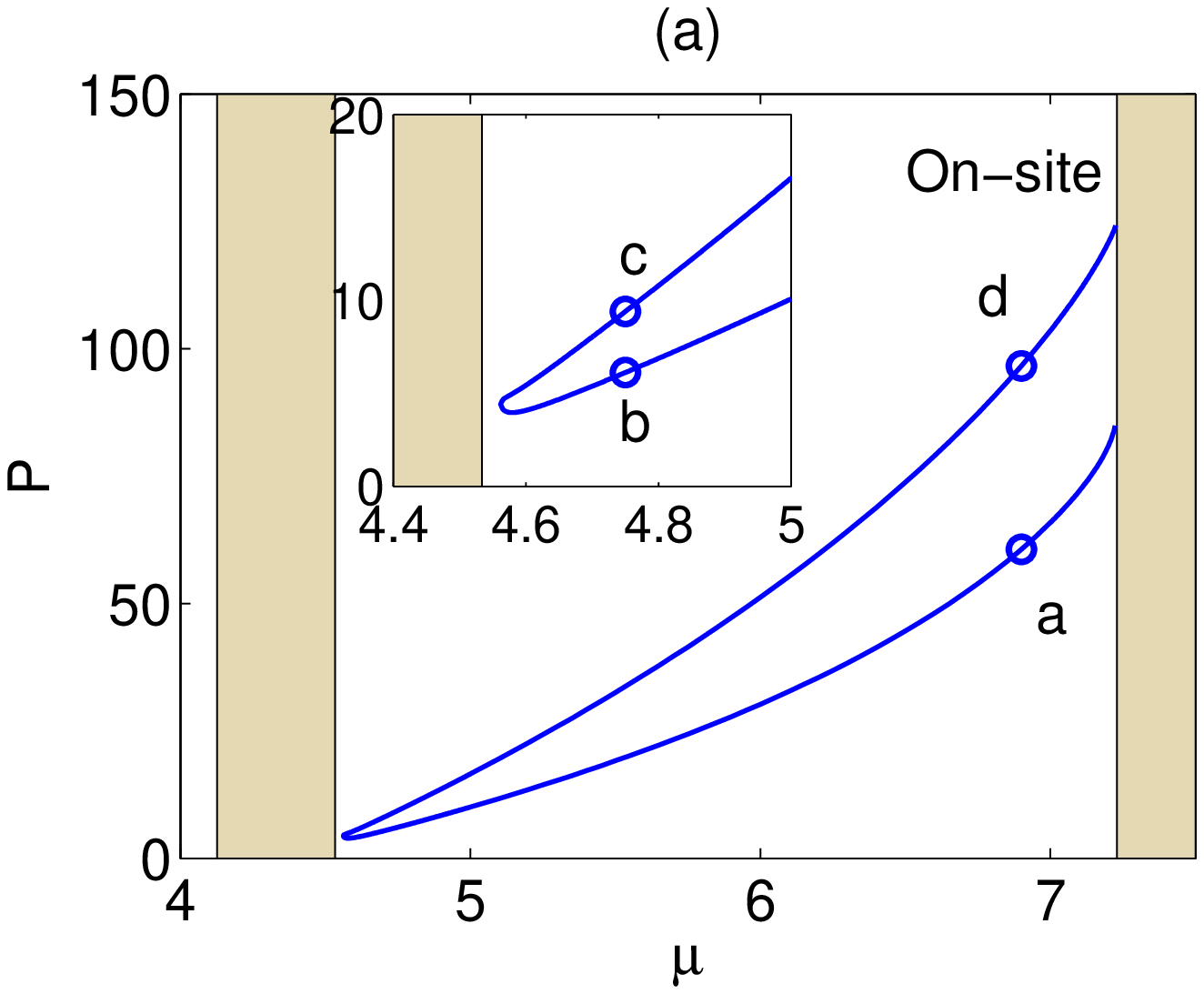}
\includegraphics[width=0.45\textwidth]{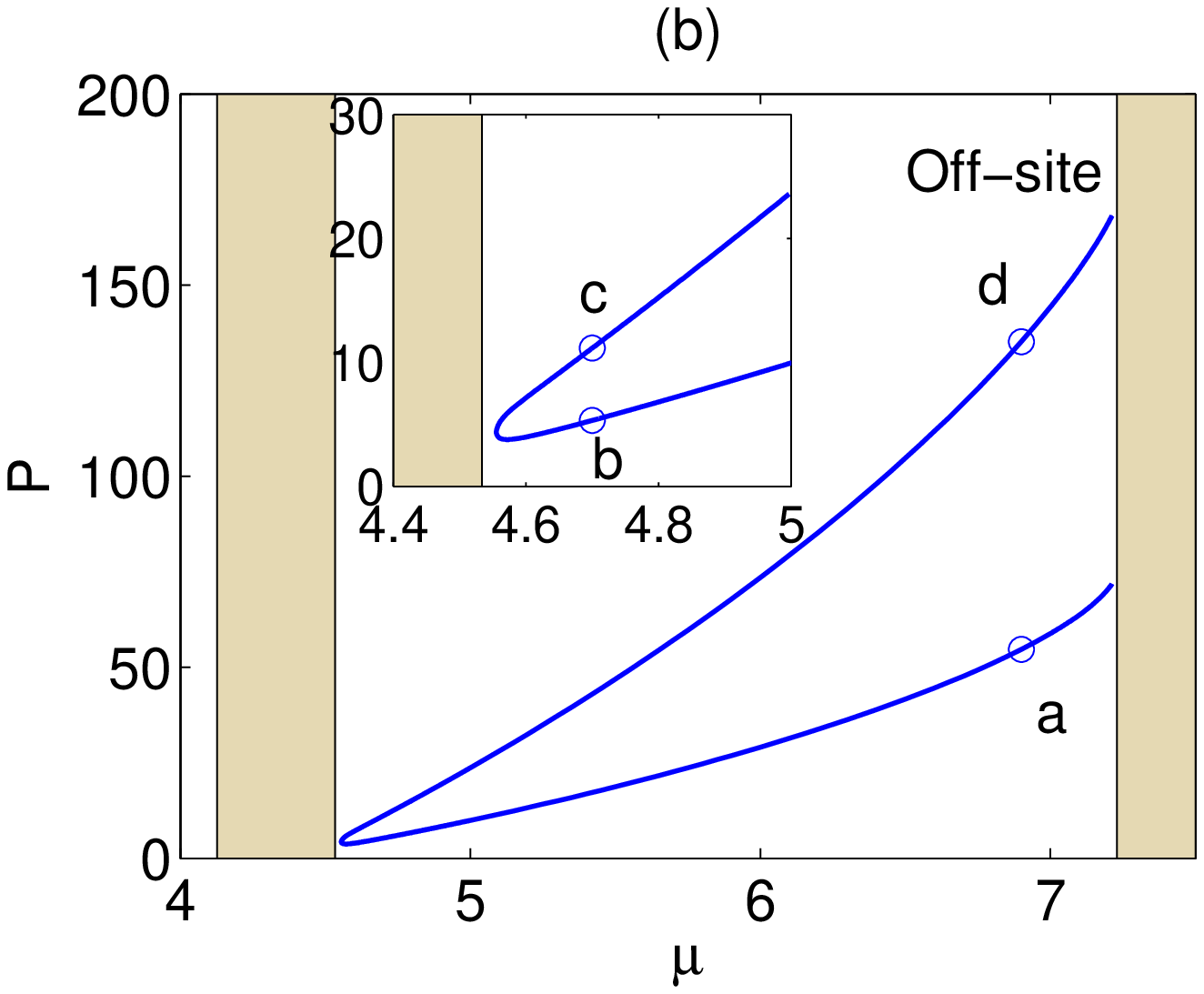}
\caption{Power diagrams of the first family of on-site (a) and
off-site (b) vortex solitons in the first band gap under defocusing
Kerr nonlinearity. The insets zoom in on the graphs near the band
edge. Vortex profiles at the marked points are shown in Fig.
\ref{figure6} (on-site) and Fig. \ref{figure7} (off-site)
respectively. Shaded: the first two Bloch bands.}\label{figure5}
\end{figure*}

\begin{figure*}
\centering
\includegraphics[width=0.2\textwidth]{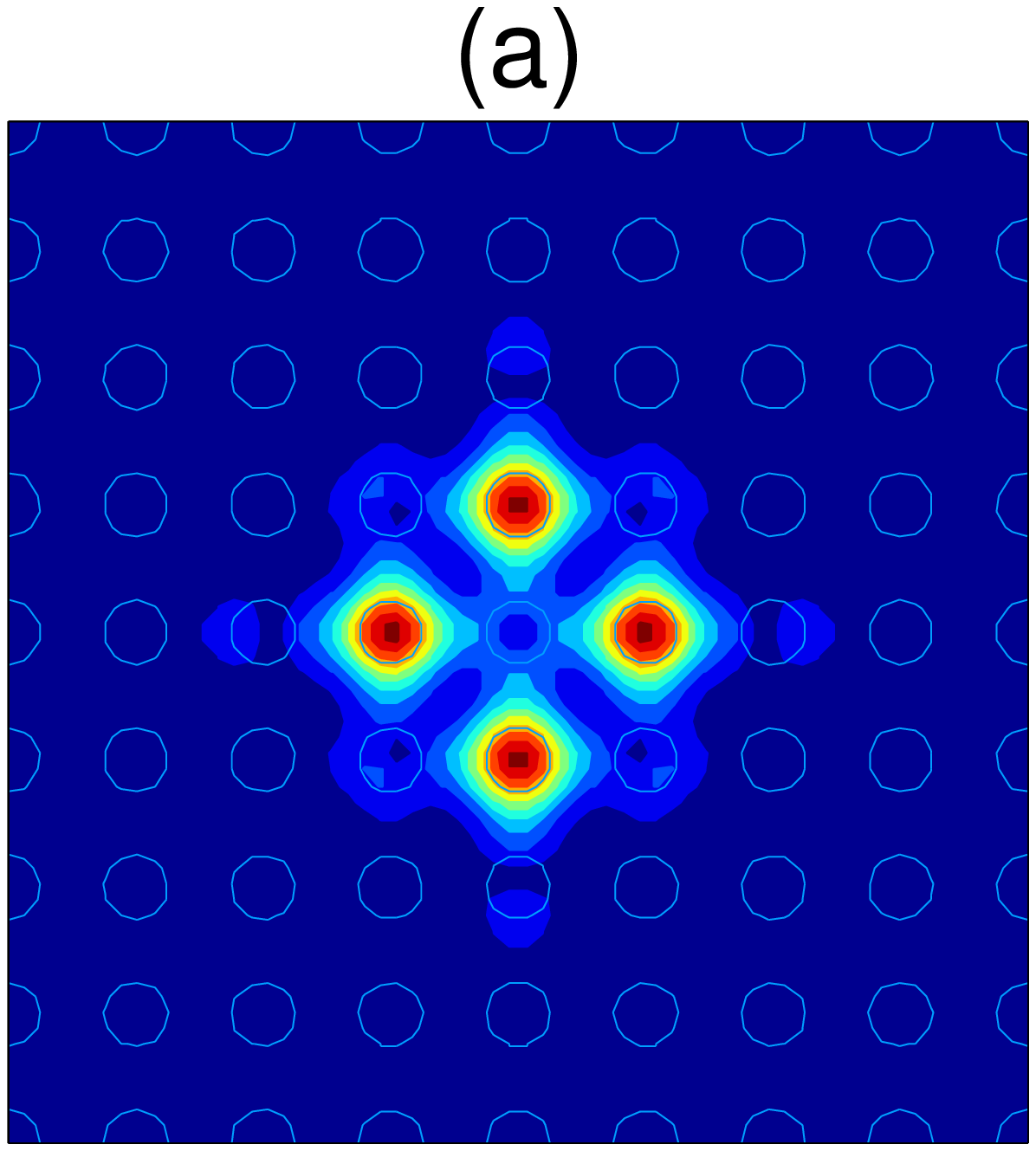}
\includegraphics[width=0.2\textwidth]{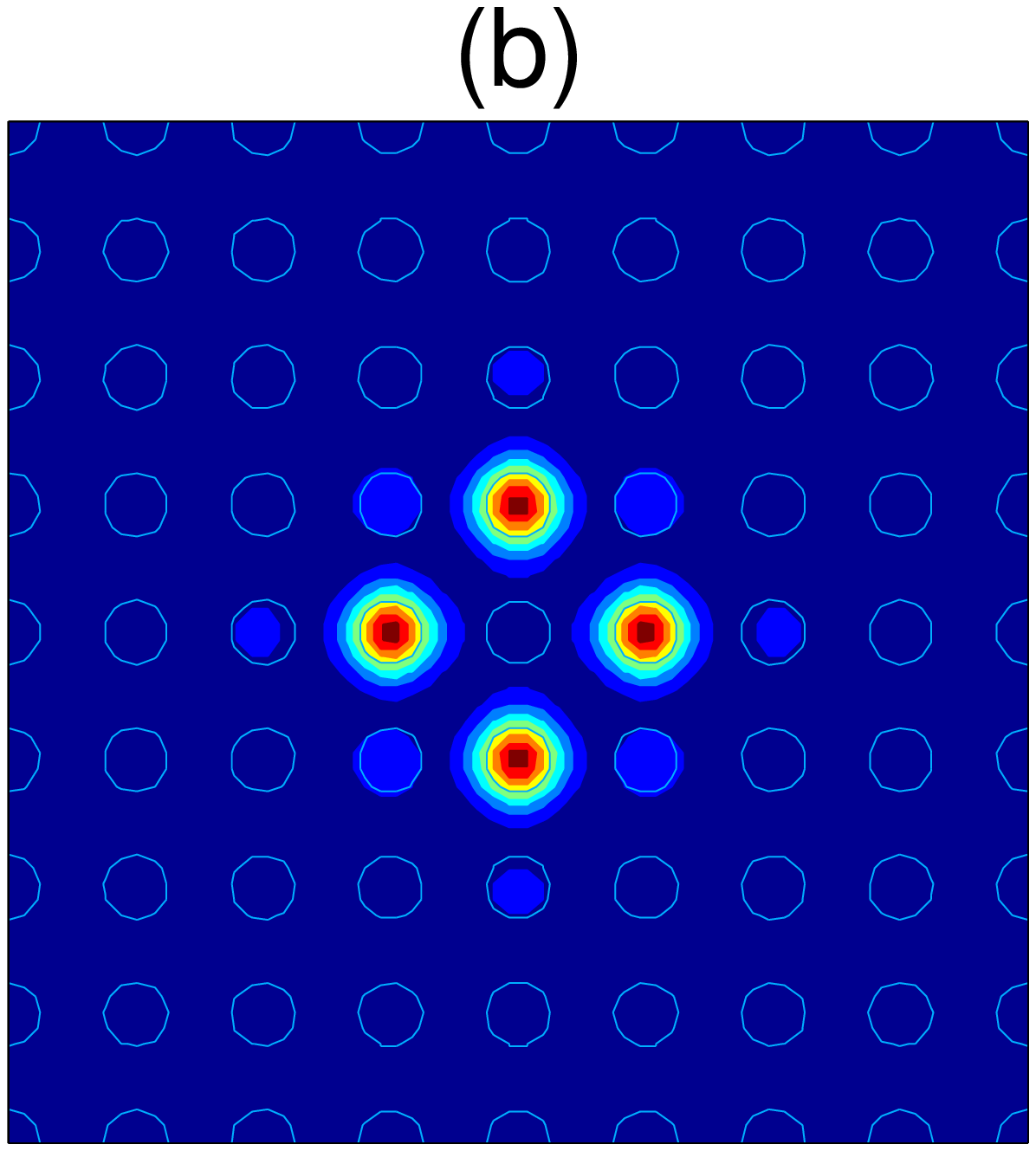}
\includegraphics[width=0.2\textwidth]{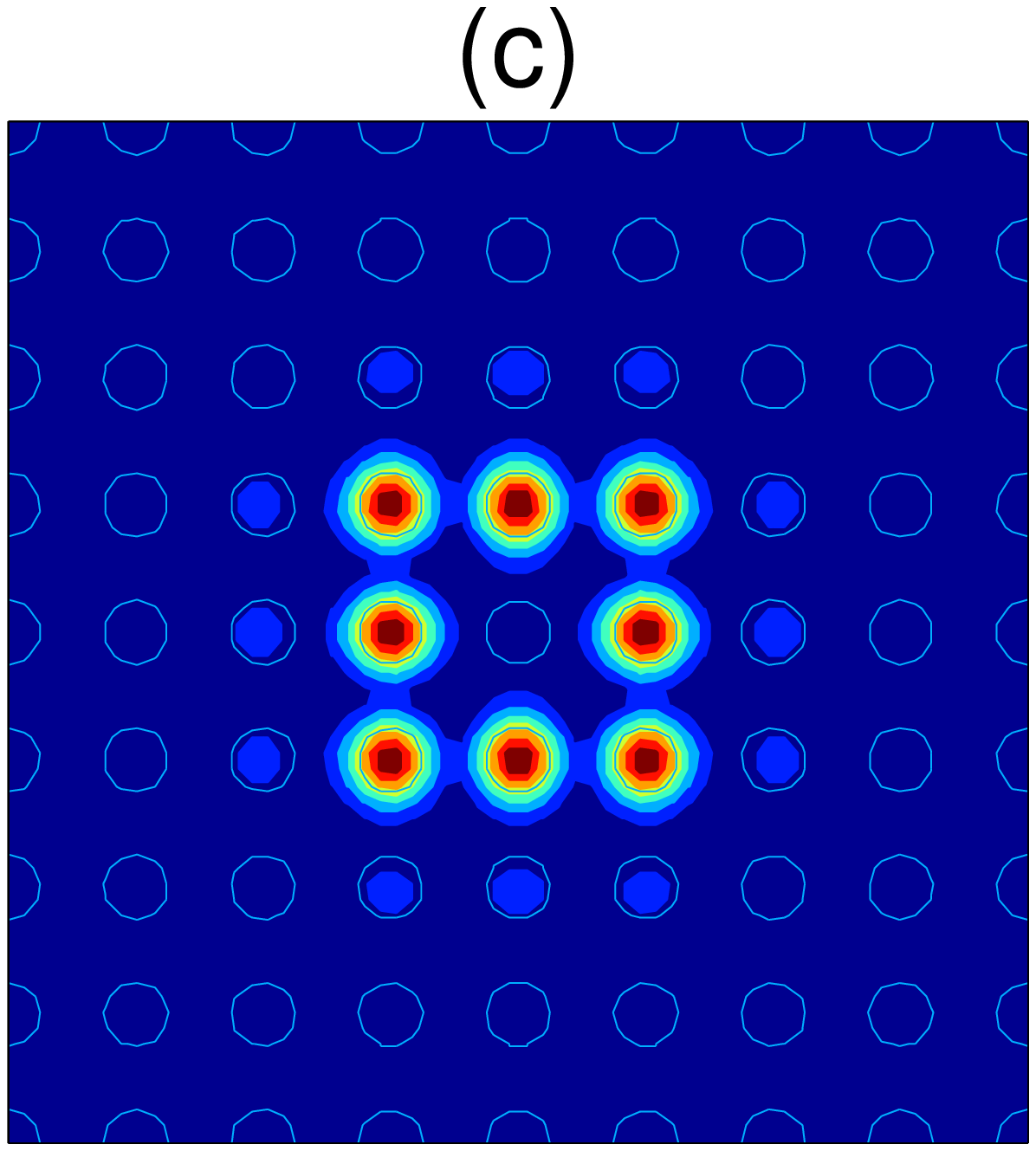}
\includegraphics[width=0.2\textwidth]{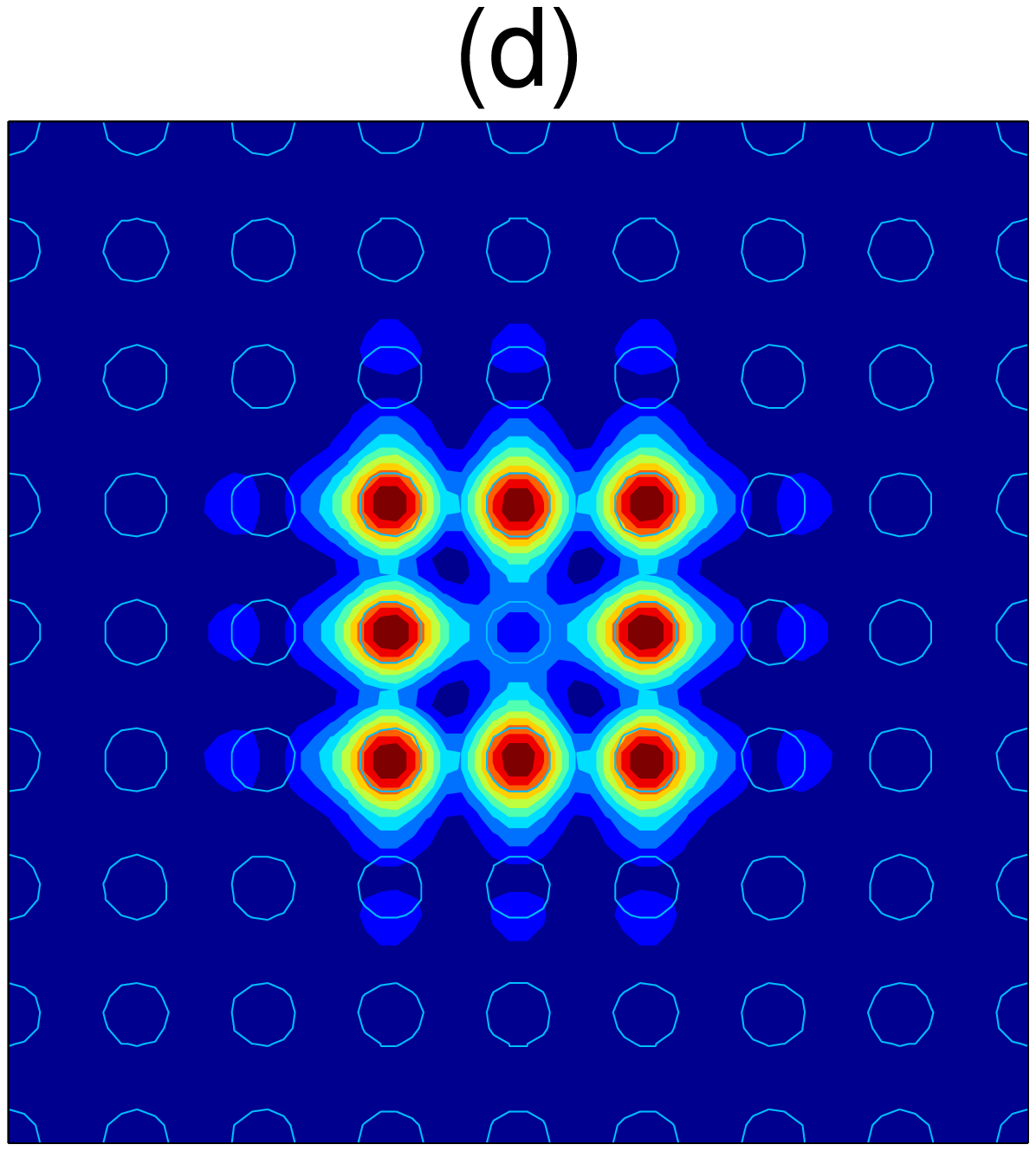}\\
\vspace{0.5cm} \hspace{0.83cm}
\includegraphics[width=0.2\textwidth]{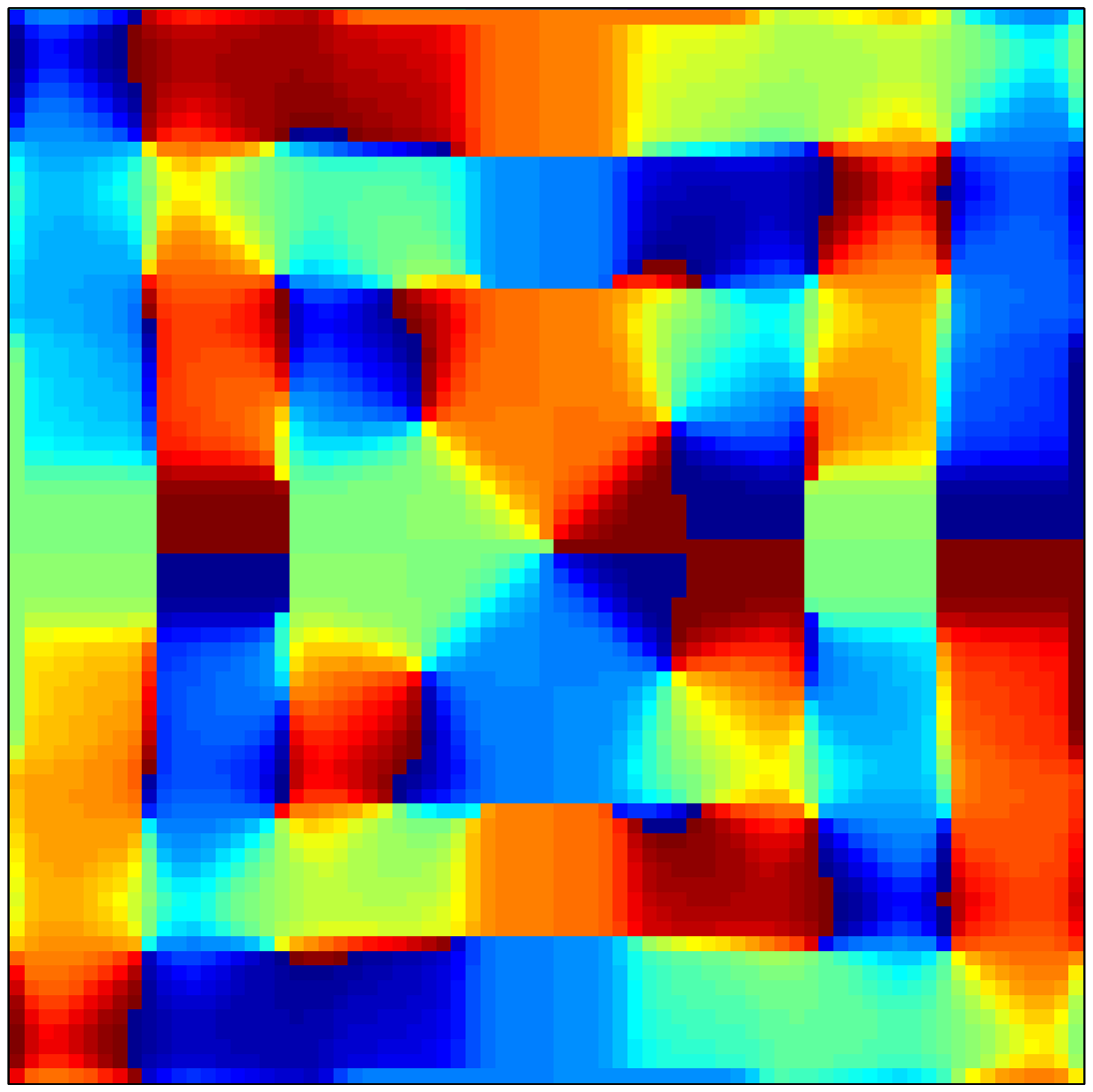}
\includegraphics[width=0.2\textwidth]{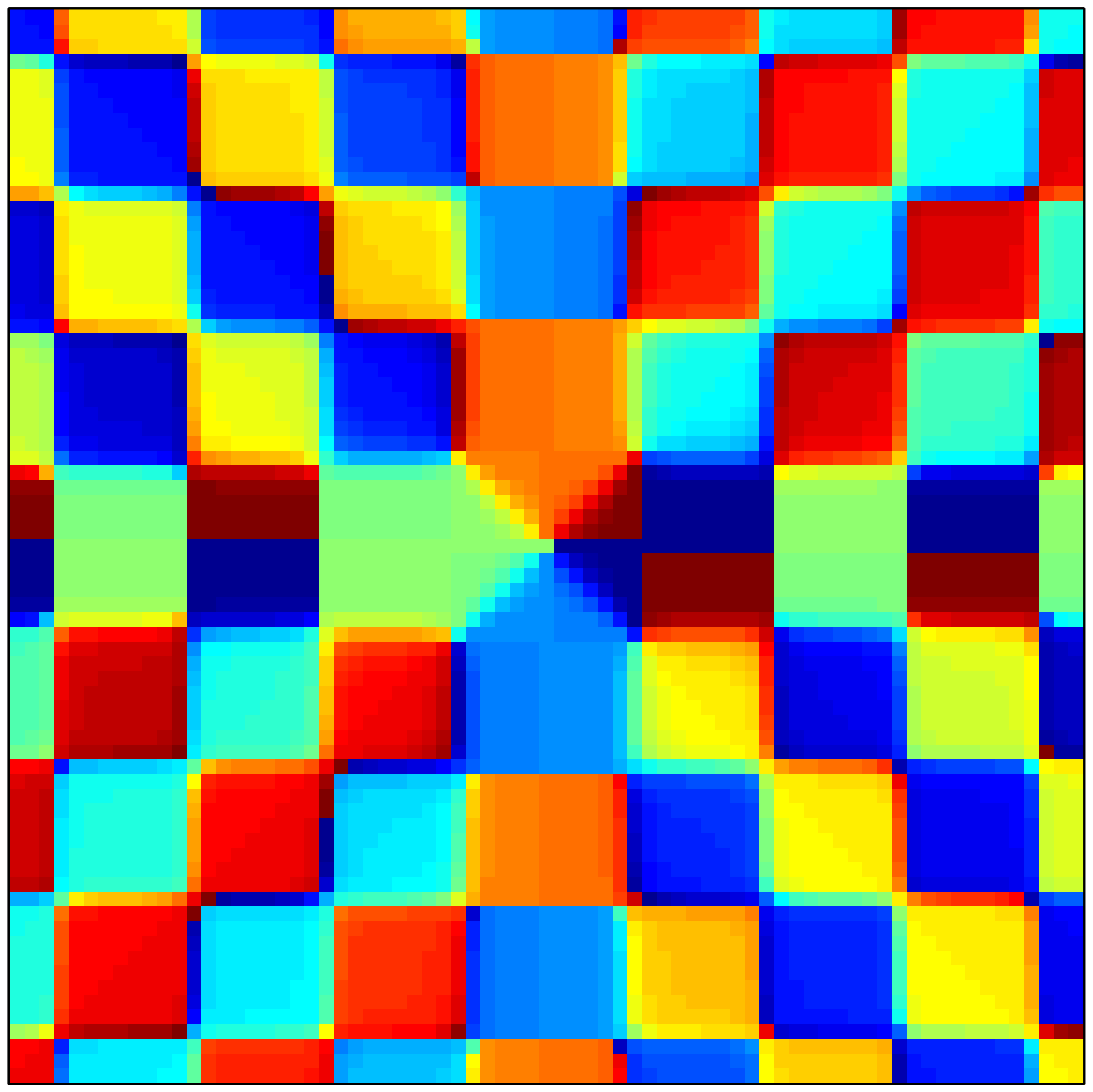}
\includegraphics[width=0.2\textwidth]{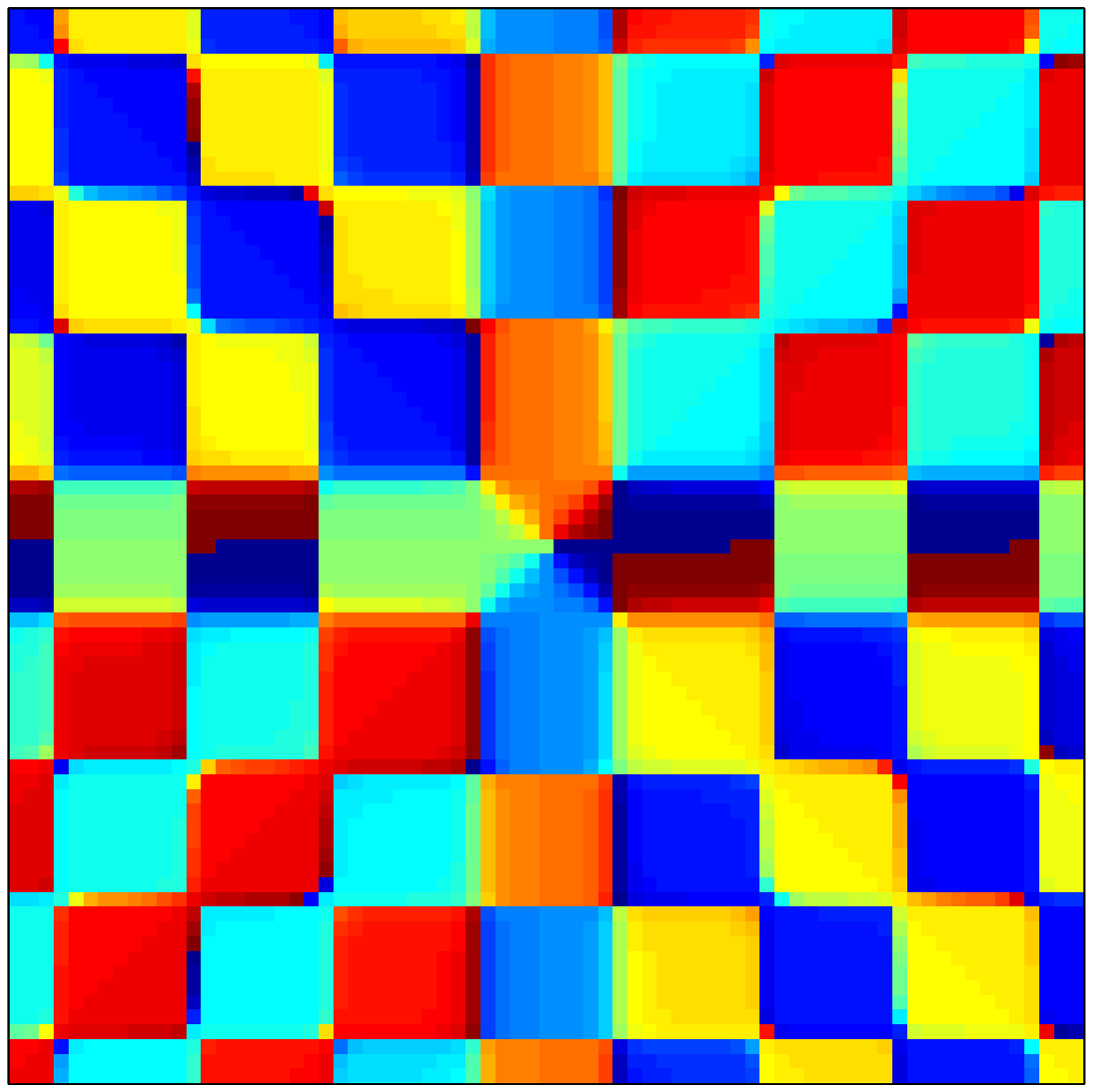}
\includegraphics[width=0.2\textwidth]{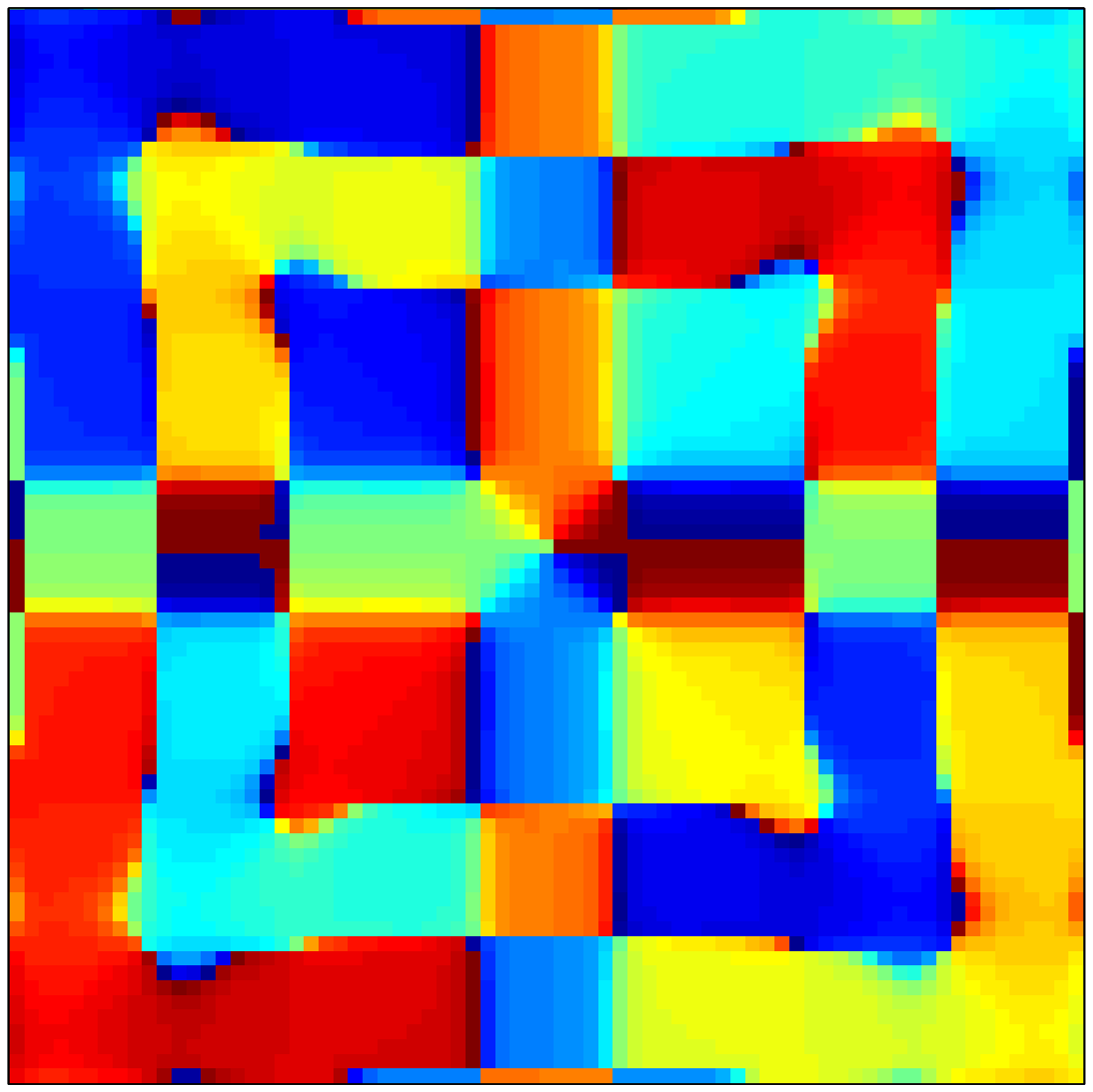}
\includegraphics[width=0.05\textwidth]{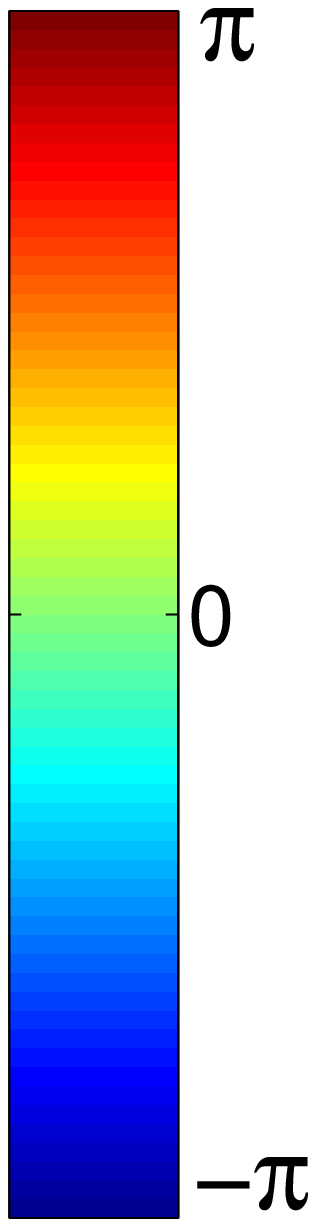}
\caption{(Color online) On-site gap vortex profiles ($|u|$, top row)
and phase structures (bottom row) corresponding to the marked points
in Fig. \ref{figure5} (a) under defocusing Kerr nonlinearity. Bottom
right: color bar of the phase figures.}\label{figure6}
\end{figure*}

\begin{figure*}
\centering
\includegraphics[width=0.2\textwidth]{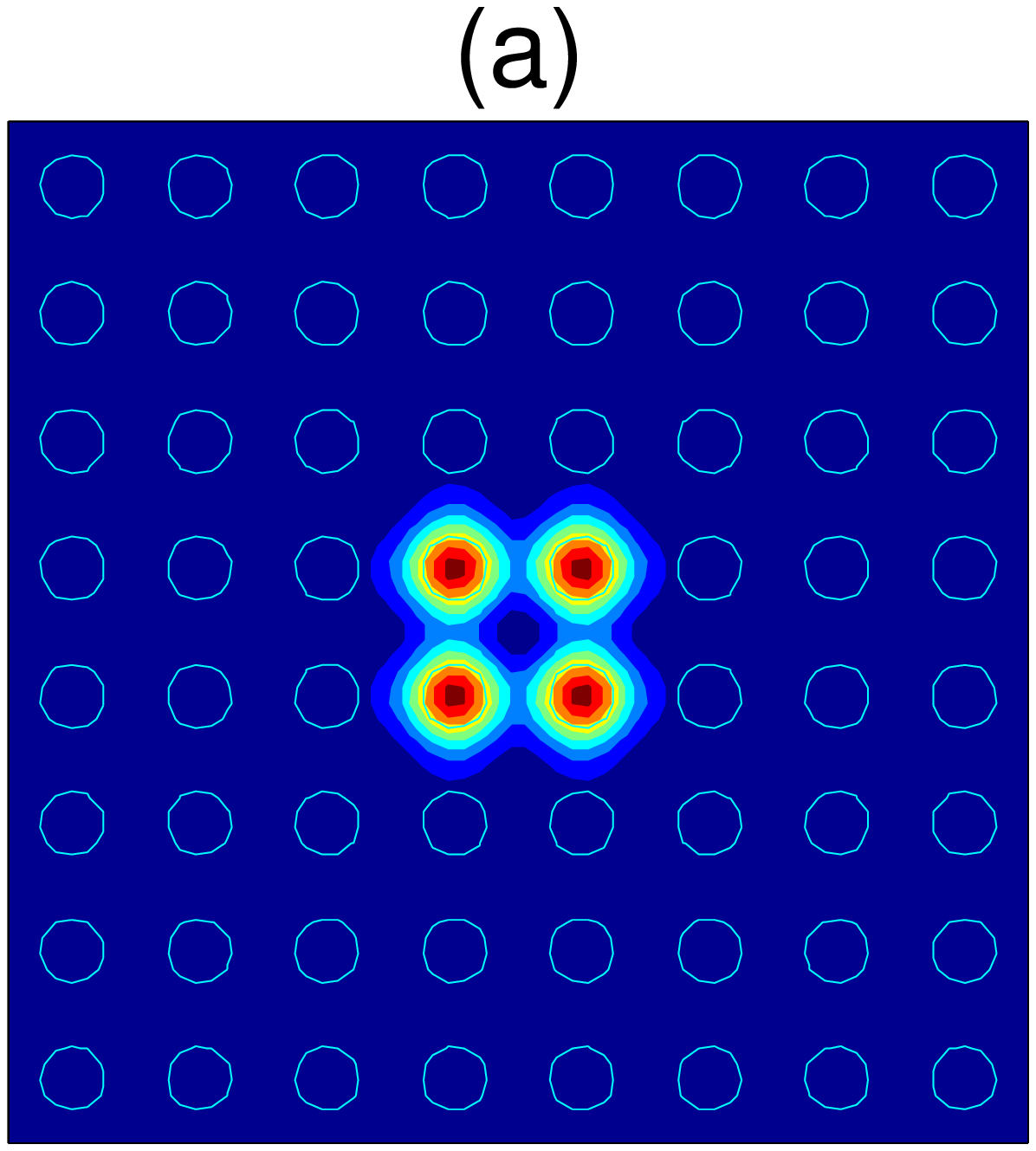}
\includegraphics[width=0.2\textwidth]{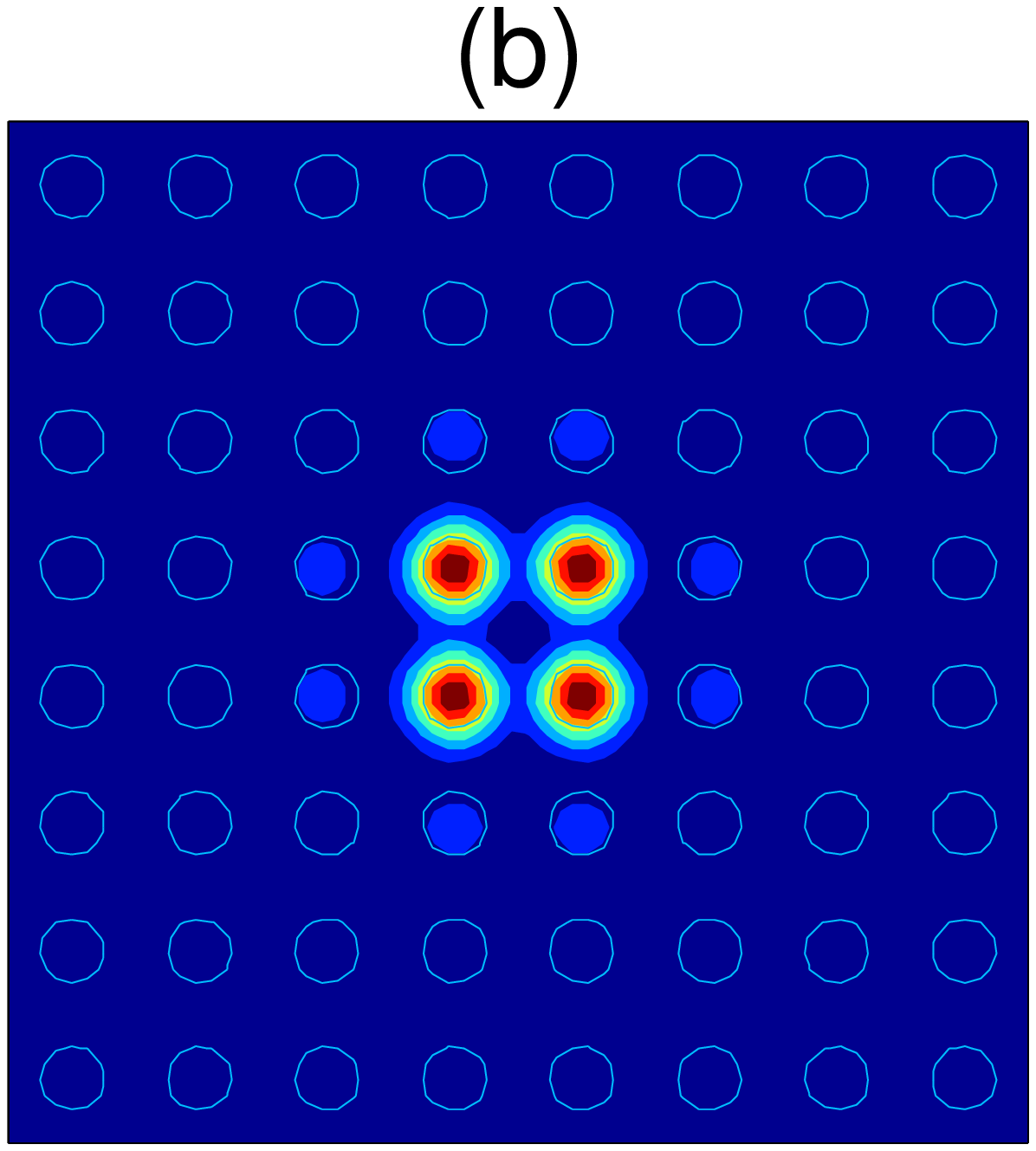}
\includegraphics[width=0.2\textwidth]{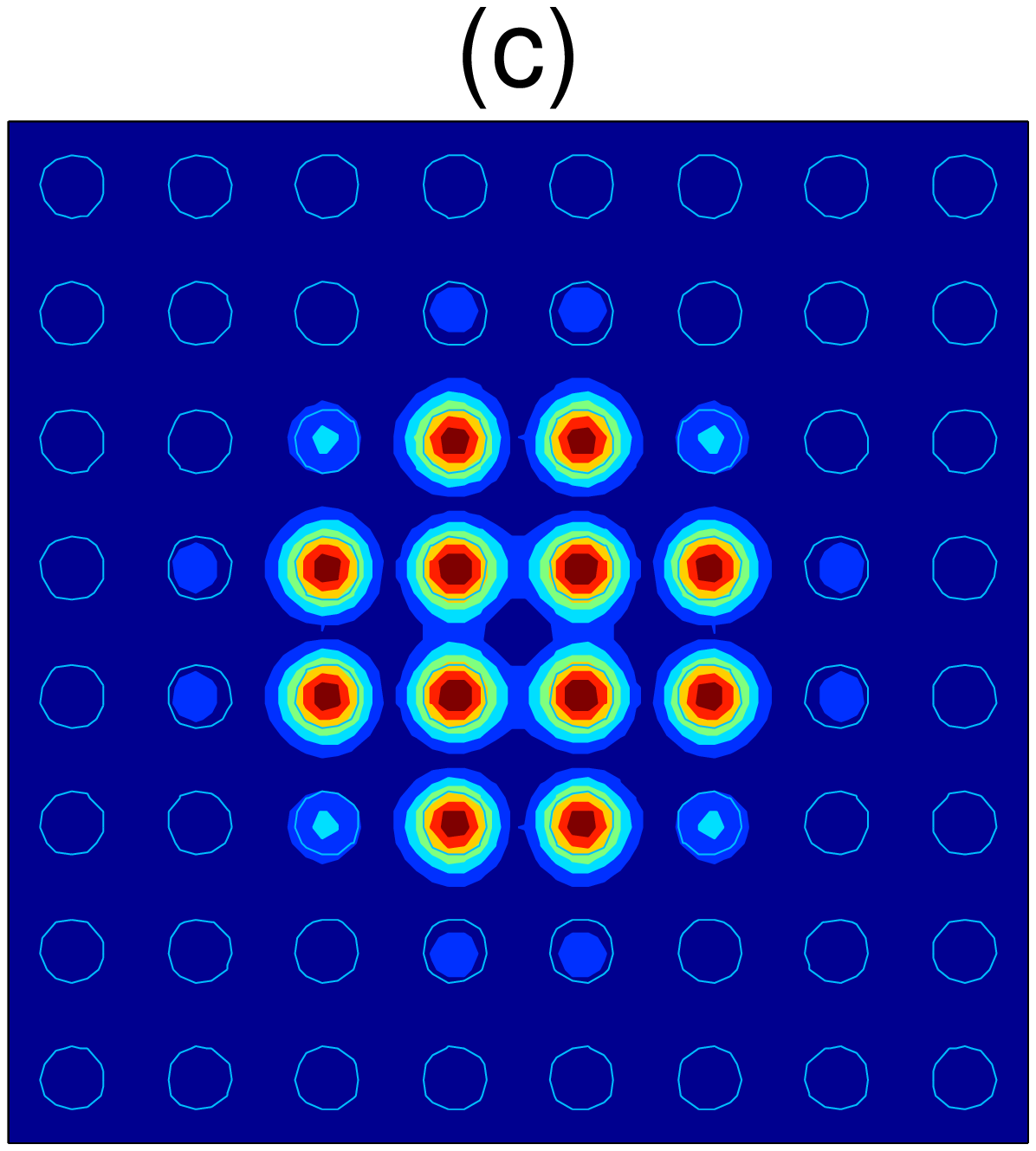}
\includegraphics[width=0.2\textwidth]{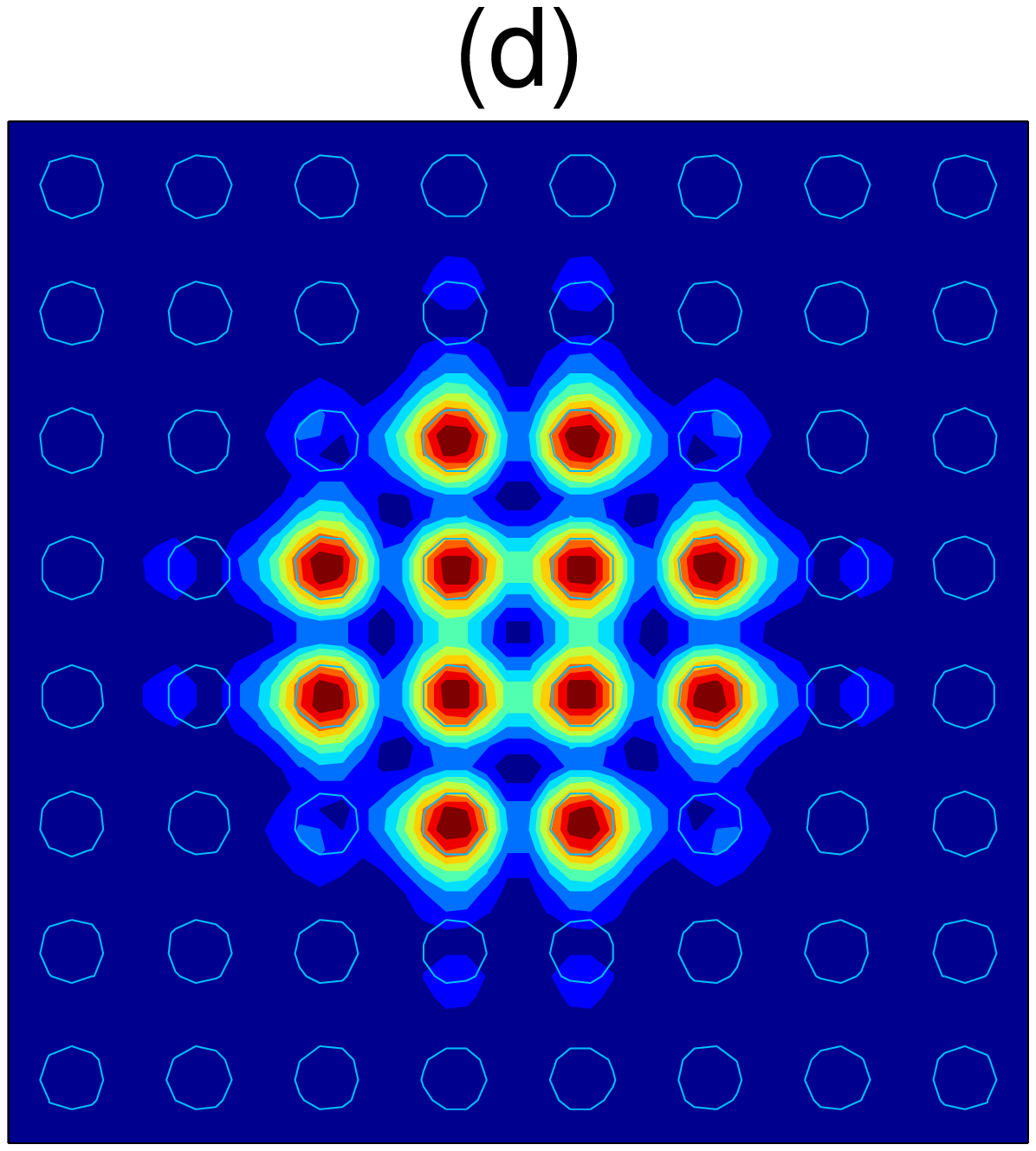}\\
\vspace{0.5cm} \hspace{0.83cm}
\includegraphics[width=0.2\textwidth]{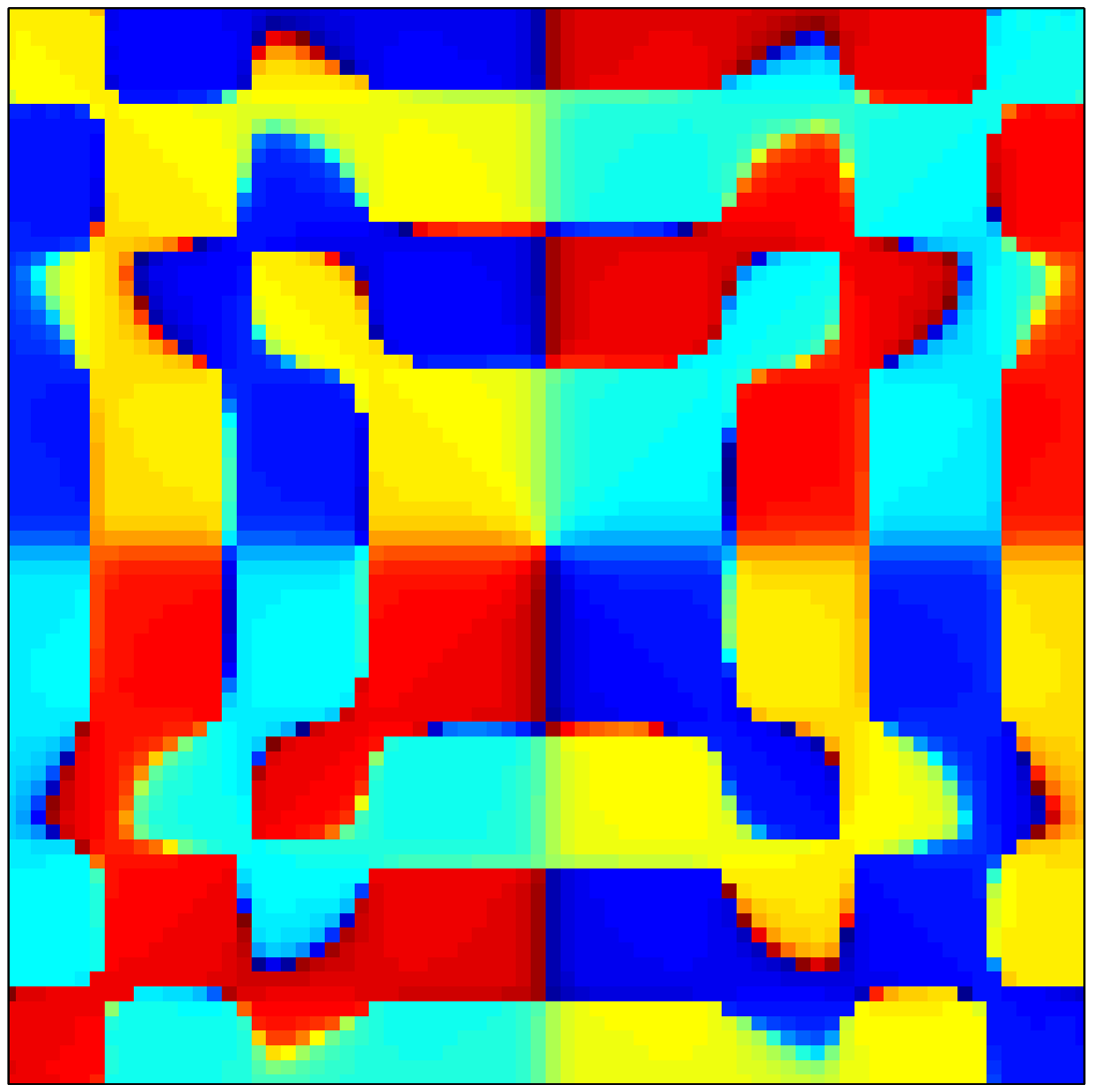}
\includegraphics[width=0.2\textwidth]{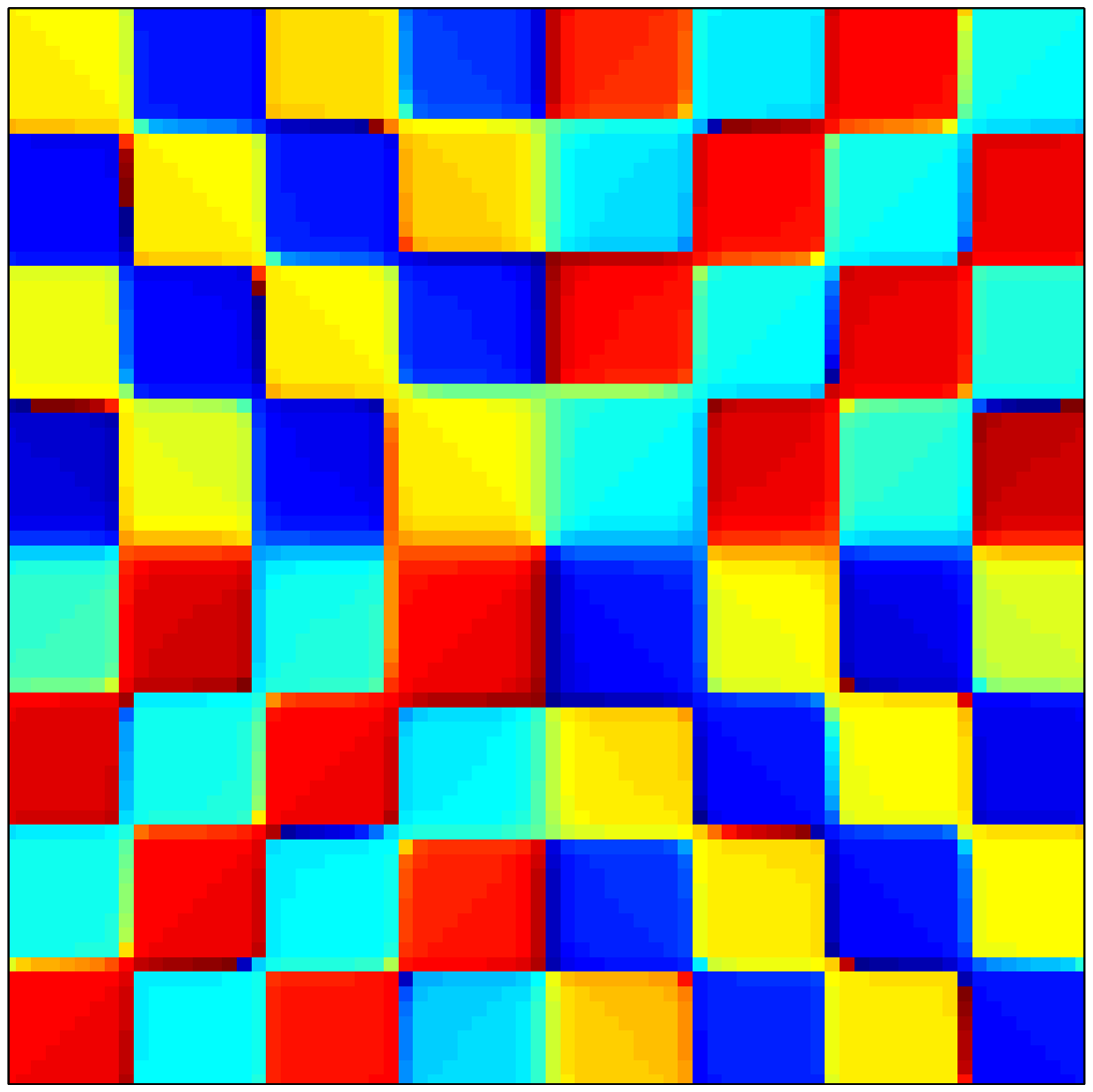}
\includegraphics[width=0.2\textwidth]{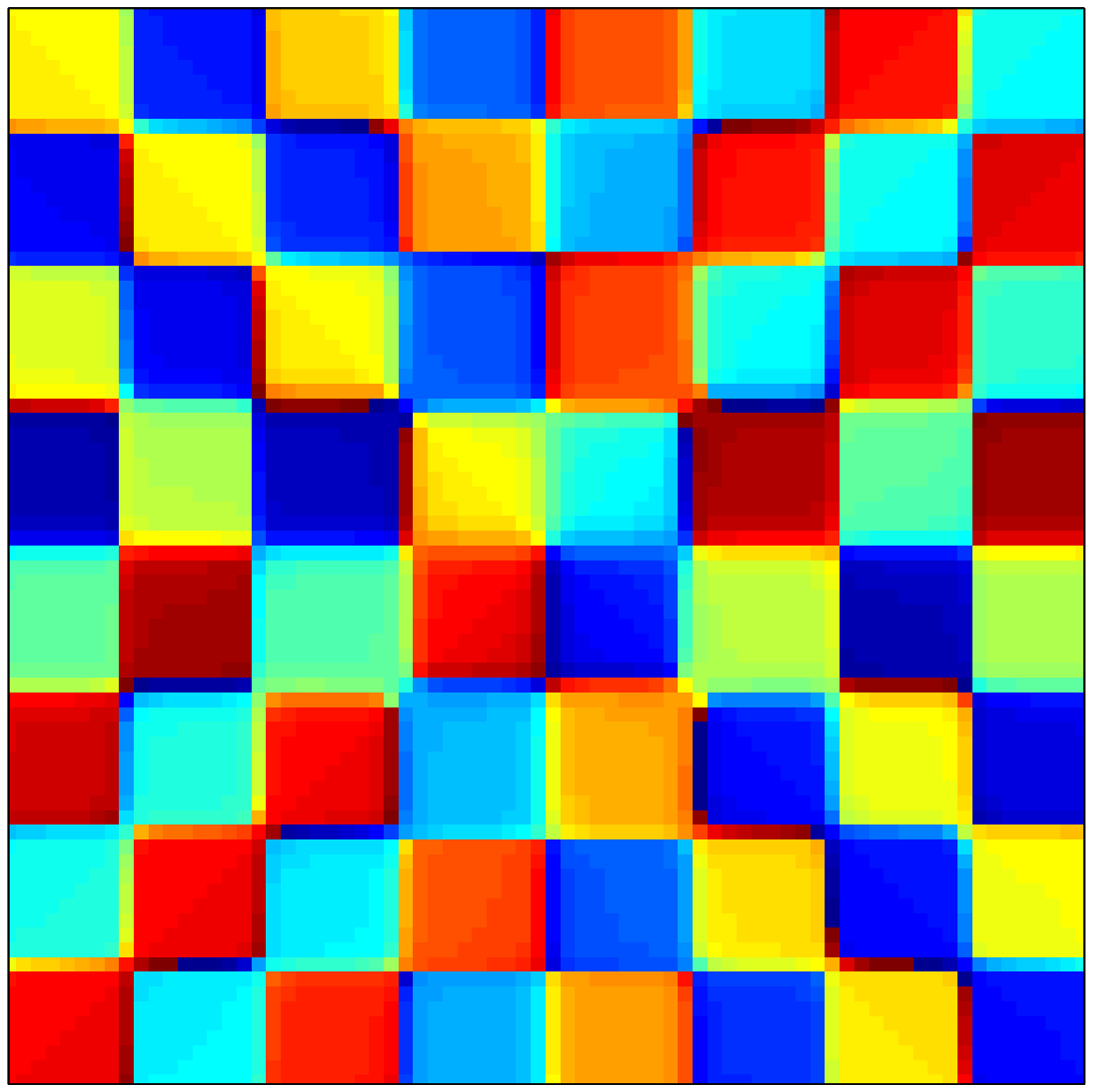}
\includegraphics[width=0.2\textwidth]{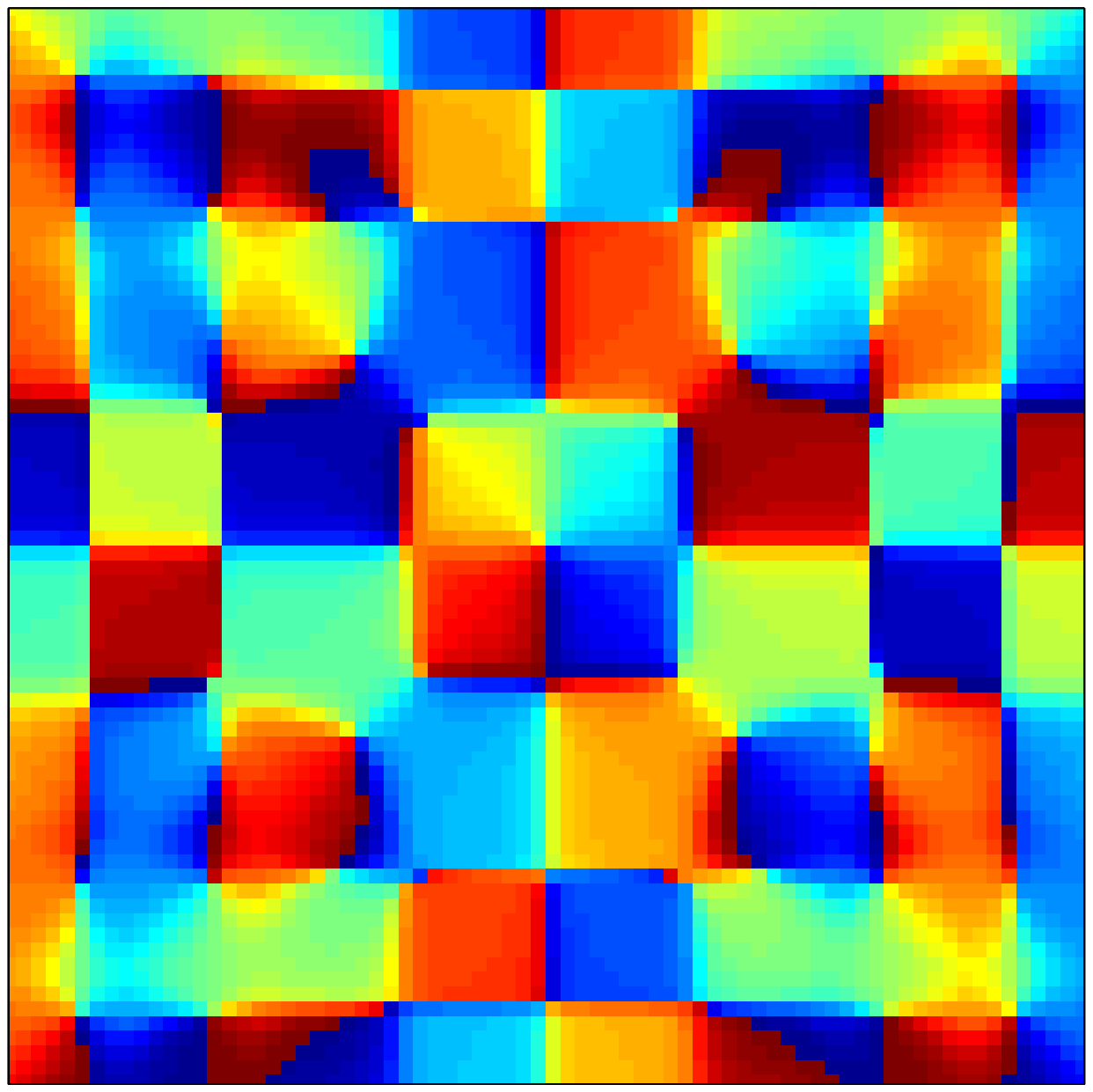}
\includegraphics[width=0.05\textwidth]{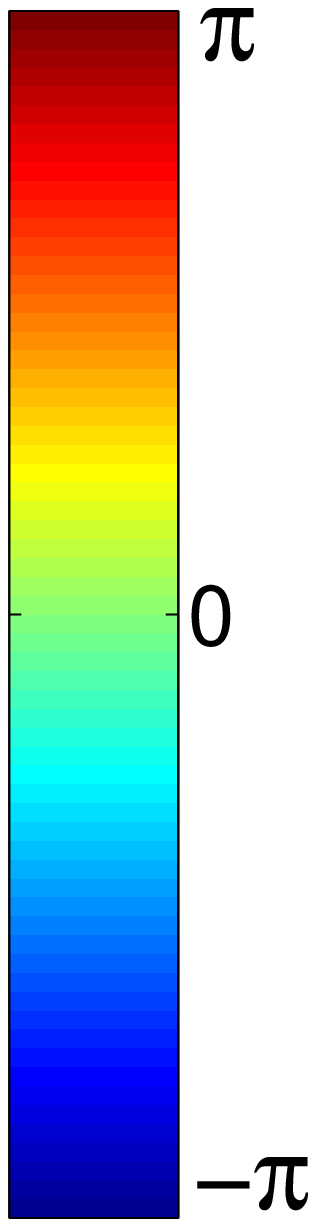}
\caption{(Color online) Off-site gap vortex soliton profiles ($|u|$,
top row) and phase structures (bottom row) corresponding to the
marked points in Fig. \ref{figure5} (b) under defocusing Kerr
nonlinearity. Bottom right: color bar of the phase
figures.}\label{figure7}
\end{figure*}

\subsection{The Case of Defocusing Kerr Nonlinearity}
In this subsection, we examine vortex solitons under defocusing Kerr
nonlinearity, i.e. $\sigma=-1$ in Eq.~(\ref{eq:three}). These
vortices exist as infinite families in the first band gap. Fig.
\ref{figure5} shows the power curves of the first on-site and
off-site gap-vortex families. It is seen that these power curves
also have a U-shape, but slanted in an opposite direction from the
focusing case (see Fig. 1). Spatial profiles and phase structures of
these gap vortices at the marked points in Fig. \ref{figure5} are
displayed in Fig. \ref{figure6} (on-site) and Fig. \ref{figure7}
(off-site). We see that the shapes of these first-family defocusing
gap vortices ($|u|$) resemble those of the first-family focusing
vortices. Specifically, for the first on-site defocusing family, as
$\mu$ moves from the lower branch to the upper one, the shape of gap
vortices changes from a four-humped diamond configuration to an
eight-humped square configuration (see Fig. \ref{figure6}),
analogous to the focusing case (see Fig. 2). For the first off-site
defocusing family, as $\mu$ moves from the lower branch to the upper
one, gap vortices change from a four-humped square configuration to
a twelve-humped cross configuration (see Fig. \ref{figure7}), also
analogous to the focusing case (see Fig. 3). The most important
difference between defocusing vortices and focusing vortices lies in
their phase structures. The phase fields of focusing vortices have a
simple $2\pi$-winding structure like in ring vortices of bulk media
[see insets in Figs. 2(a) and 3(a)]. However, phase structures of
defocusing gap vortices are much more complex (see Figs.
\ref{figure6} and \ref{figure7}). In the central region of gap
vortices, the phase field has a simple $2\pi$-winding structure
around the vortex center. But in the outer region, vortex phases at
adjacent lattice sites sometimes have $\pi$ difference [see Figs.
\ref{figure6}(b,c) and \ref{figure7}(b,c)], but some other times do
not [see Figs. \ref{figure6}(d) and \ref{figure7}(a)]. These
complicated phase structures make the concept of topological charges
ambiguous and not well defined for gap vortices. Note that gap
vortices reported earlier in \cite{EAOstrovskaya2,EAOstrovskaya}
correspond to the lower branches of our gap vortices in Fig.
\ref{figure5}.

From Fig. \ref{figure5}, we see that gap vortices can exist quite
close to the right edge of the first Bloch band. At this band edge,
the second-order dispersion coefficient $D_1$ is negative. Thus
according to the envelope equation for this band edge, which is
similar to Eq. (\ref{A}) with only the sign of the $A$ term
reversed, ring-vortex solutions $A=f(R)e^{i\Theta}$ exist under
defocusing nonlinearity (where $\alpha_0<0$). Hence for $\mu$ close
to this band edge, one may construct the analytical gap-vortex
solution $u(x,y)=\epsilon A(X, Y)p(x, y)$, where $p(x,y)$ is the
Bloch wave at this right edge (with $M$-point symmetry). This
analytical gap-vortex solution qualitatively reproduces the
intensity distribution, and more importantly, the intricate phase
structure, of the true solution near the right edge of the first
band [see Figs. \ref{figure6}(b) and \ref{figure7}(b) for instance].
In particular, it explains the simple $2\pi$ winding of the phase
near the vortex center, and the $\pi$ phase difference between
lattice sites in the far field (due to the $M$-point symmetry of the
Bloch wave at this band edge). Of course, this analytical gap-vortex
solution can not explain why true gap vortices exist as infinite
families, and each family does not continuously reach the band edge
--- the same difficulty we have seen with the focusing nonlinearity
in the previous subsection.

\section{Vortex Solitons Under Saturable Nonlinearity}
In photorefractive crystals where many lattice-soliton experiments
were performed in recent years, the nonlinearity is saturable
\cite{Christodoulides_PRE}. Even though wave phenomena under
saturable and Kerr nonlinearities have many features in common,
significant differences exist as well. In this section, we examine
vortex solitons under saturable nonlinearity.

A simple theoretical model for solitons in photorefractive crystals
imprinted with a photonic lattice is
\cite{Christodoulides_PRE,Yang_NJP}
\begin{equation}
u_{xx}+u_{yy}-\frac{E_0}{1+I(x,y)+|u|^2}u=-\mu u, \label{eq:four}
\end{equation}
where $E_0$ is the applied dc field,
\begin{equation} \label{Ixy}
I(x,y)=I_0\cos^2(x)\cos^2(y)
\end{equation}
is a periodic square-lattice function, $I_0$ is the peak intensity
of this lattice, and $\mu$ is the propagation constant. Here all
variables have been non-dimensionalized \cite{Yang_NJP}. When the
soliton amplitude is low ($|u|\ll 1$), the saturable nonlinearity
becomes the same as the Kerr nonlinearity. Therefore, we can expect
that near Bloch bands, the vortex behavior under saturable
nonlinearity should be similar to that under Kerr nonlinearity. In
particular, they can not bifurcate from Bloch bands either. Below we
numerically investigate vortex solitons in the semi-infinite gap in
Eq. (\ref{eq:four}) under focusing nonlinearlty ($E_0>0$). Without
loss of generality, we choose parameters $I_0=3$ and $E_0=6$. Vortex
solitons in the first band gap under defocusing saturable
nonlinearity will be briefly described in the end of this section.

\begin{figure*}
\centering
\includegraphics[width=0.45\textwidth]{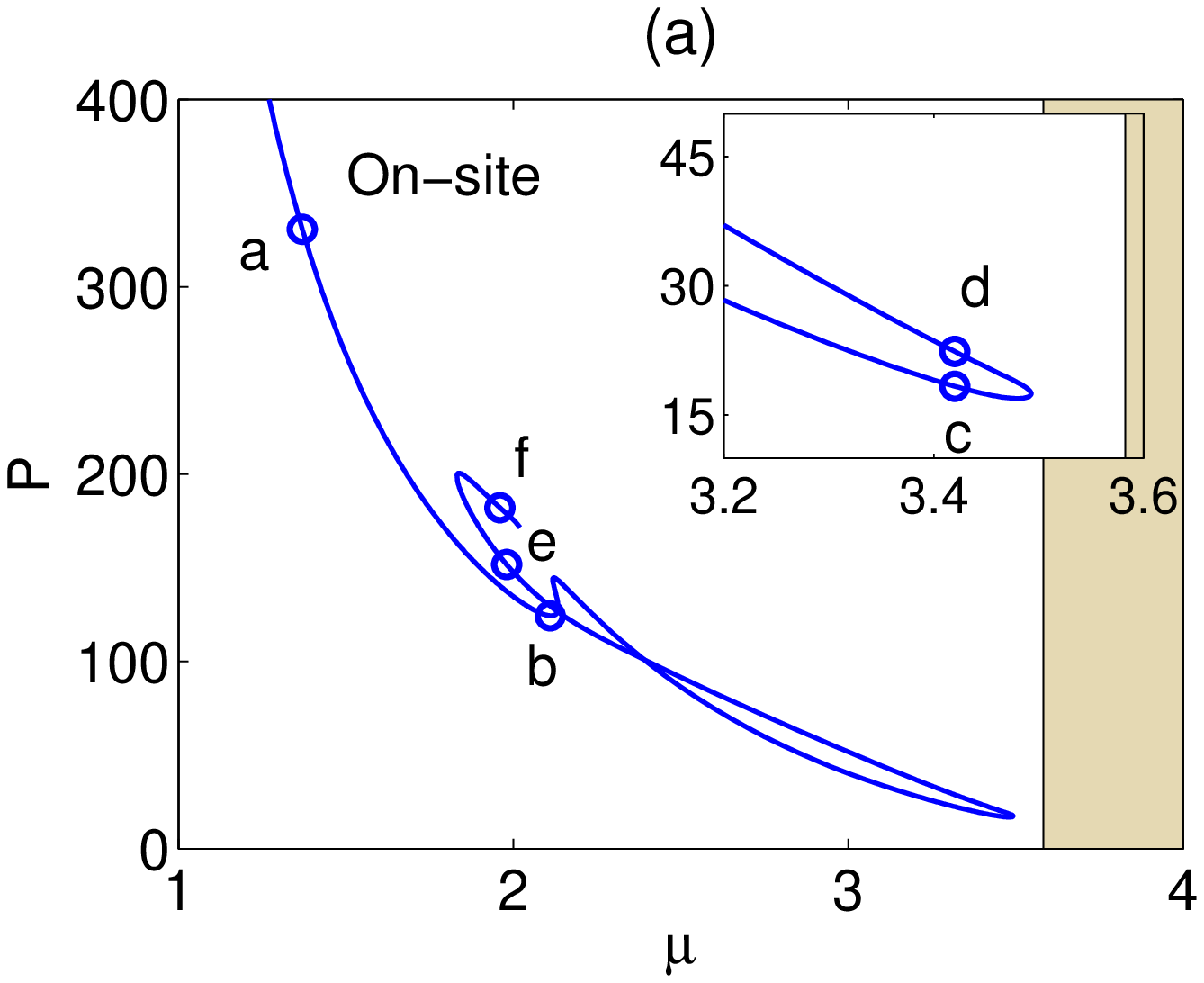}
\includegraphics[width=0.45\textwidth]{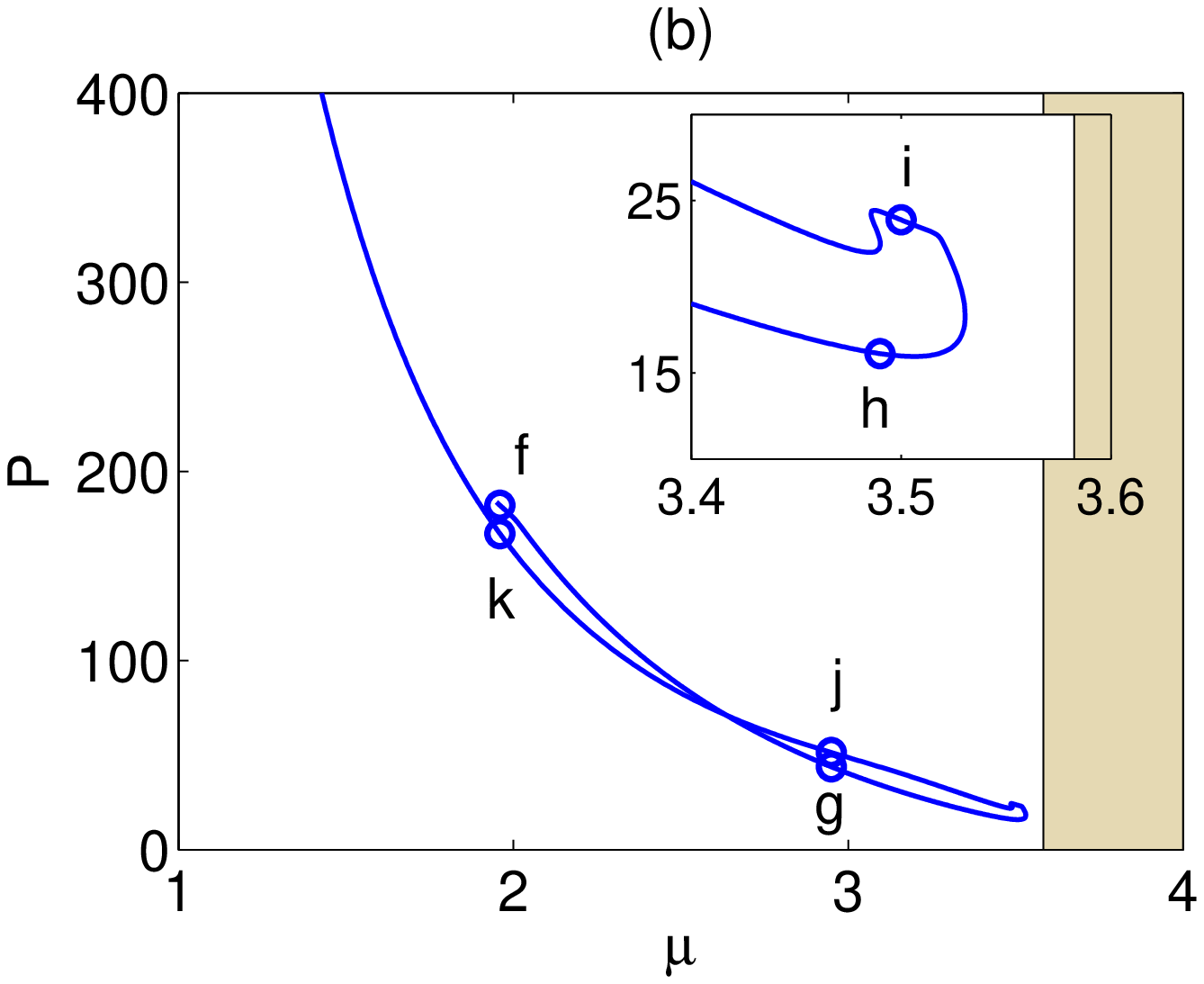}
\caption{Power diagrams of an on-site vortex family in the
semi-infinite gap under focusing saturable nonlinearity. The power
curve after point `f' in (a) is shown in (b). Vortex profiles at the
marked points are shown in Fig. \ref{figure9}. The insets zoom in on
the graphs near the band edge. }\label{figure8}
\end{figure*}

\begin{figure*}
\centering
\includegraphics[width=0.15\textwidth]{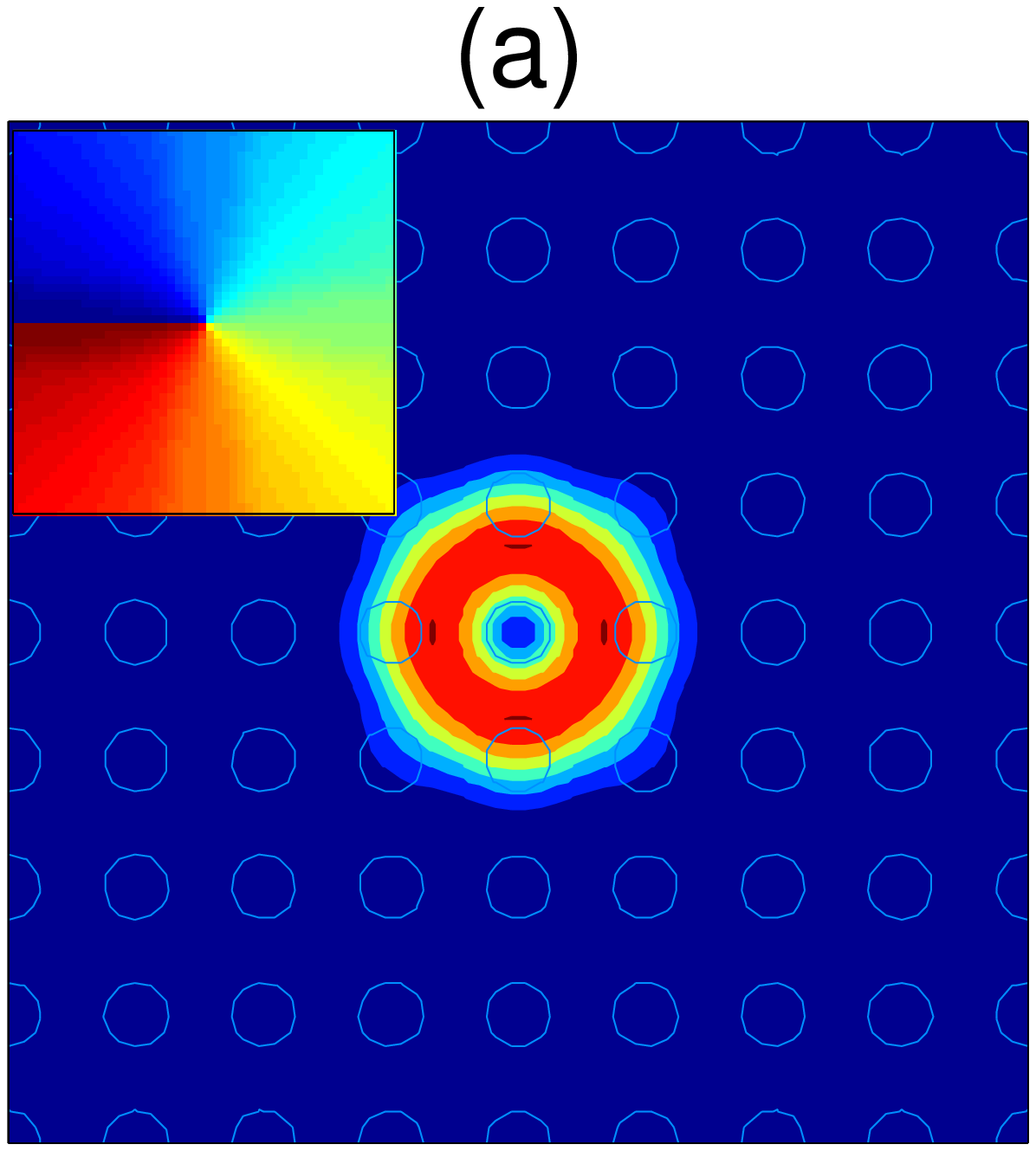}
\includegraphics[width=0.15\textwidth]{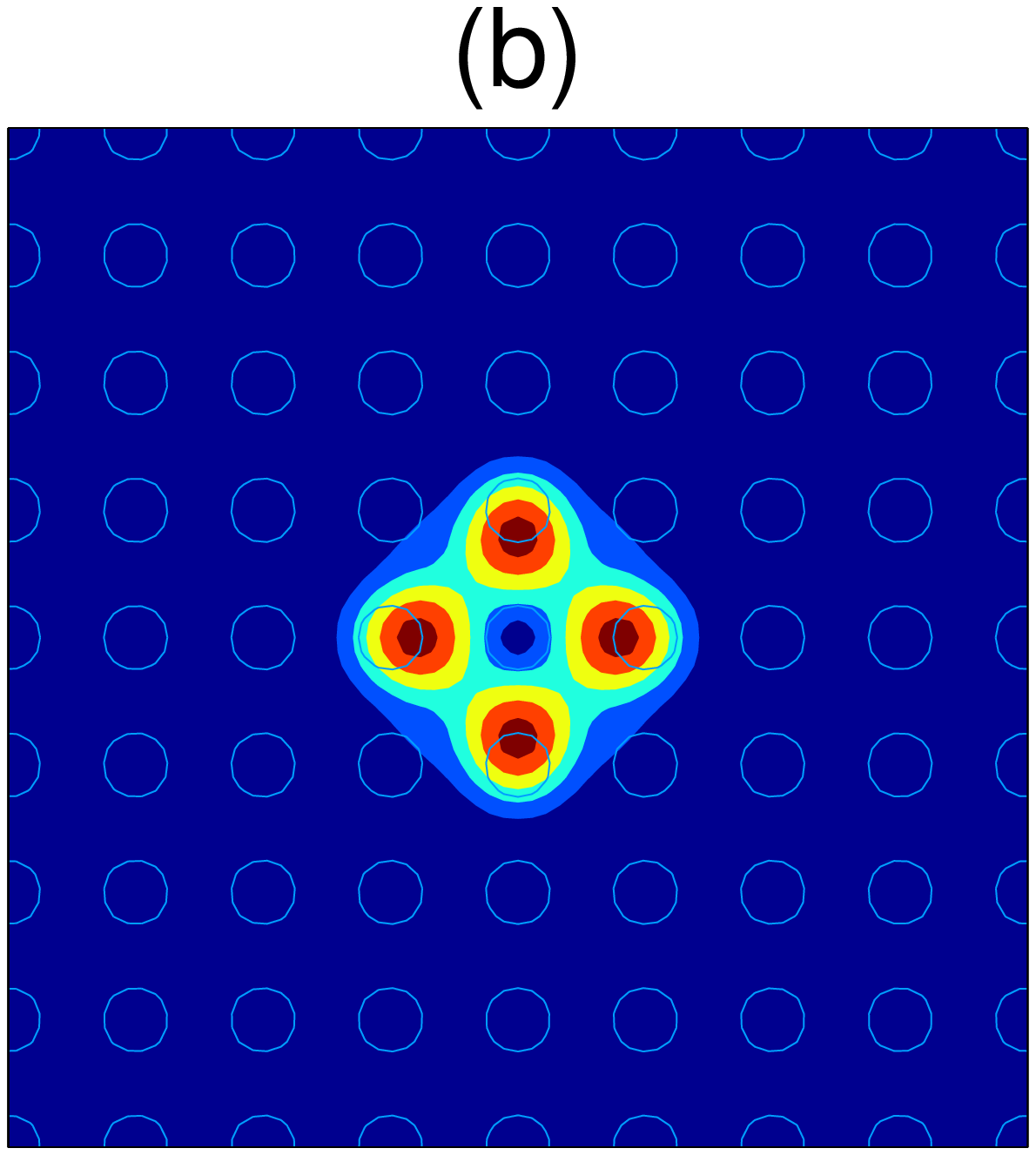}
\includegraphics[width=0.15\textwidth]{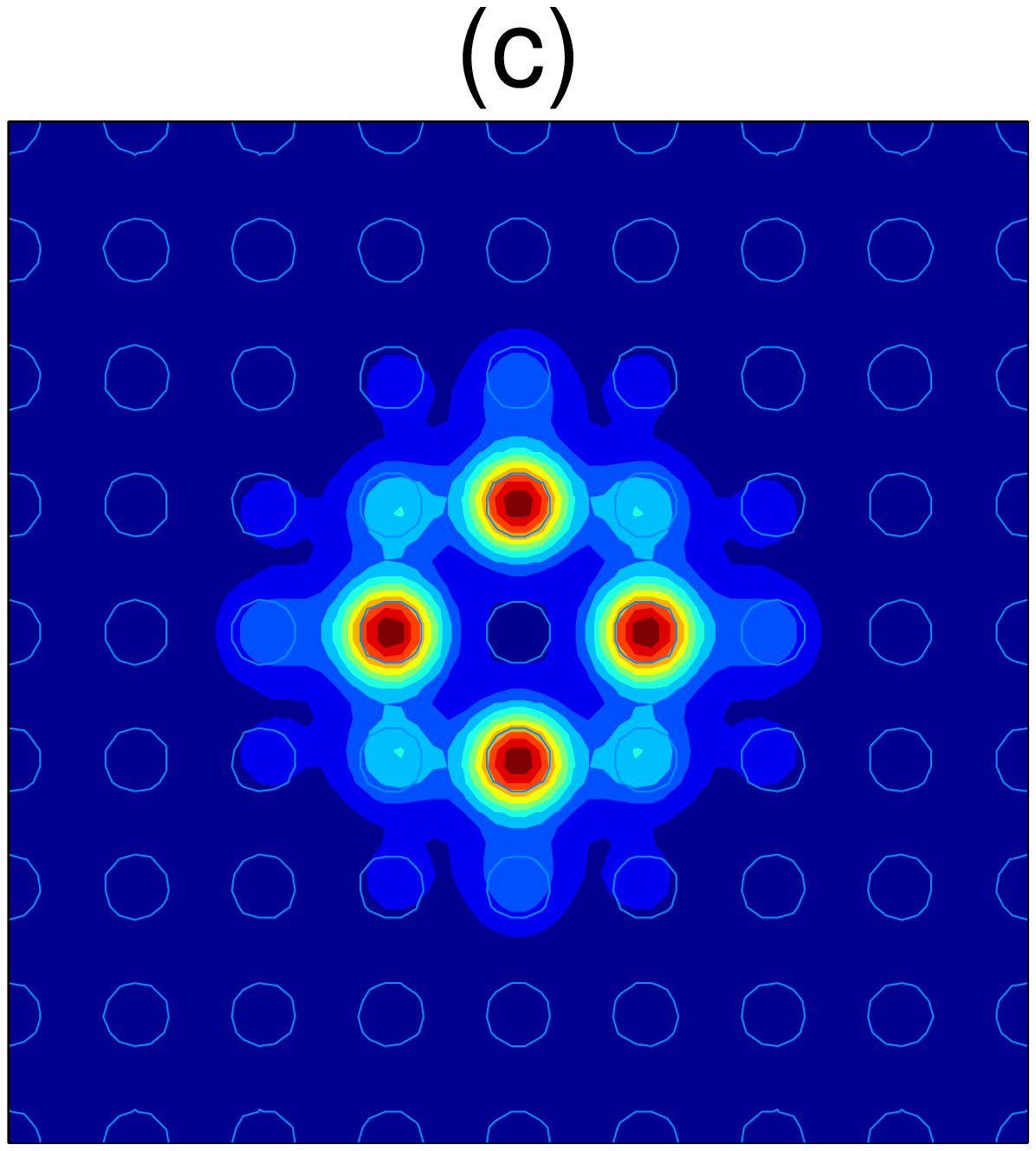}
\includegraphics[width=0.15\textwidth]{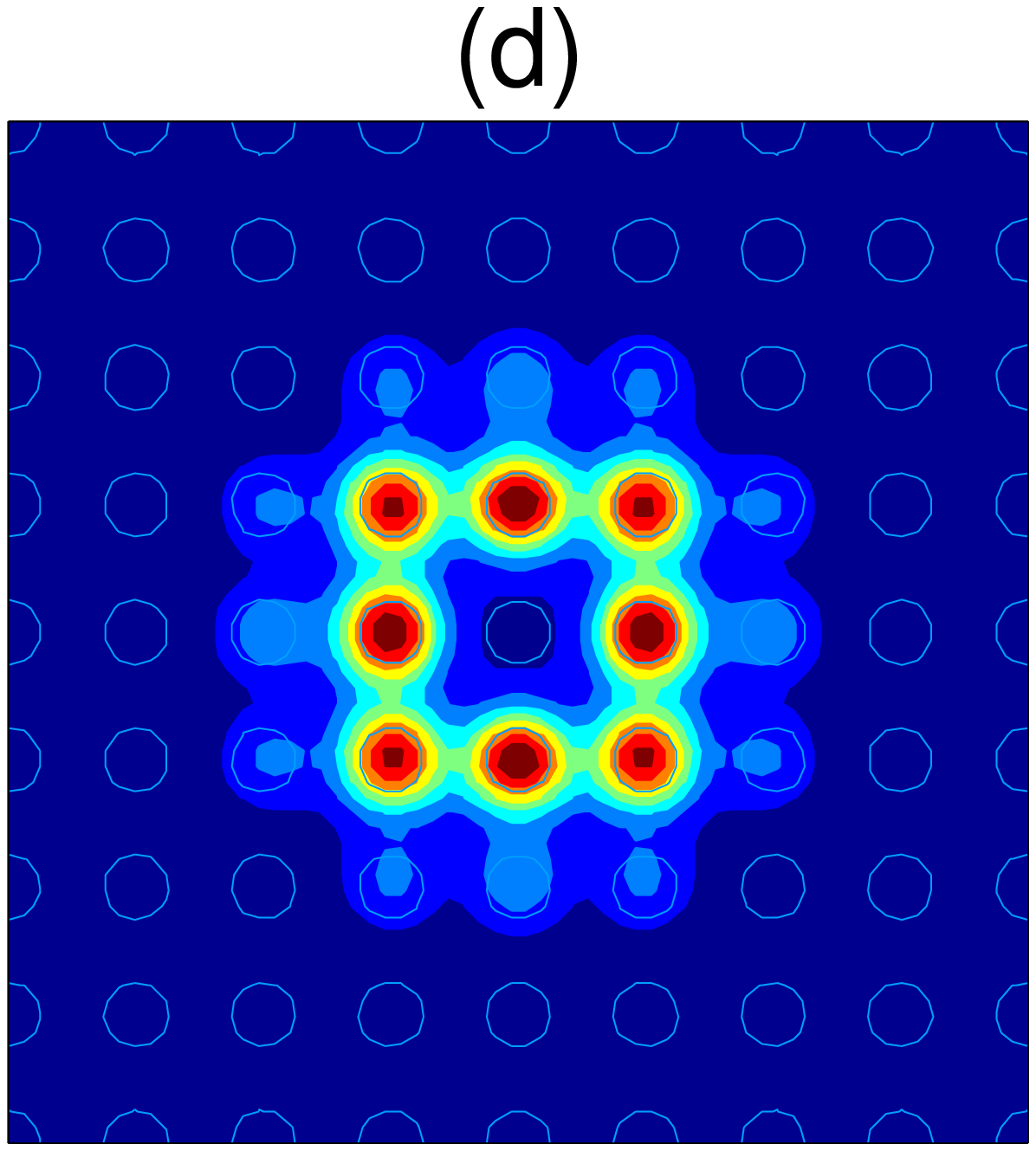}
\includegraphics[width=0.15\textwidth]{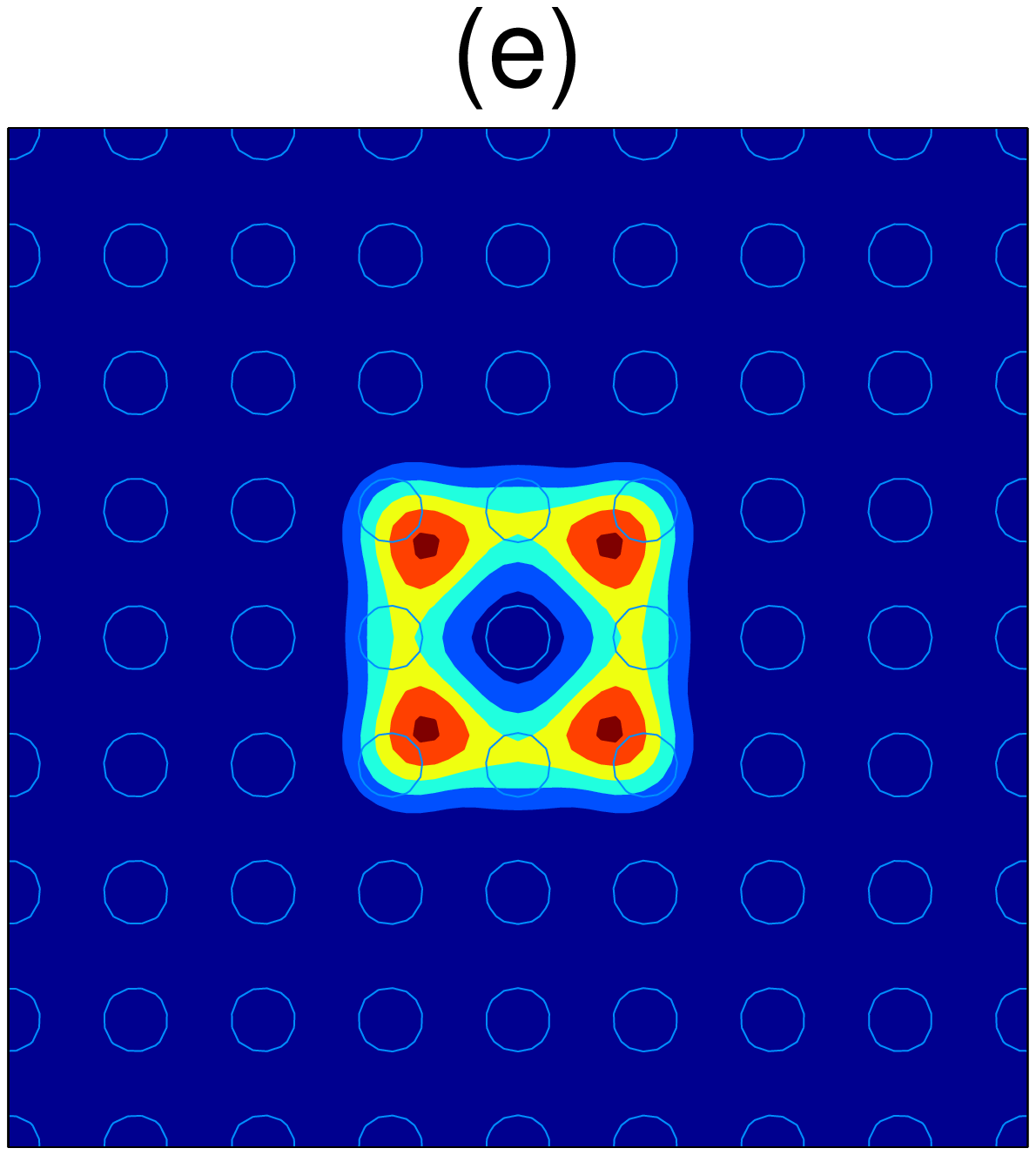}
\includegraphics[width=0.15\textwidth]{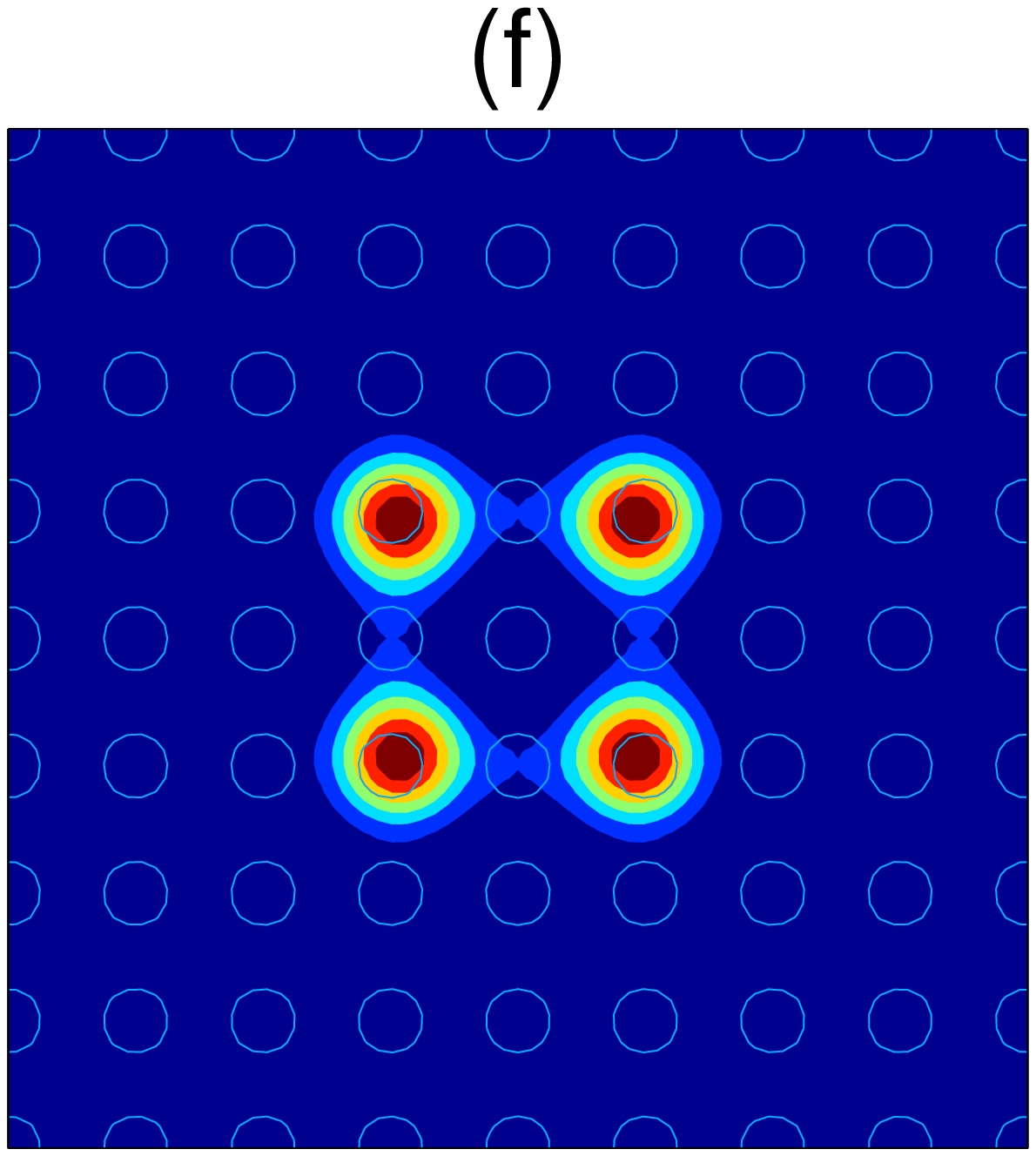}\\
\vspace{0.3cm}
\includegraphics[width=0.15\textwidth]{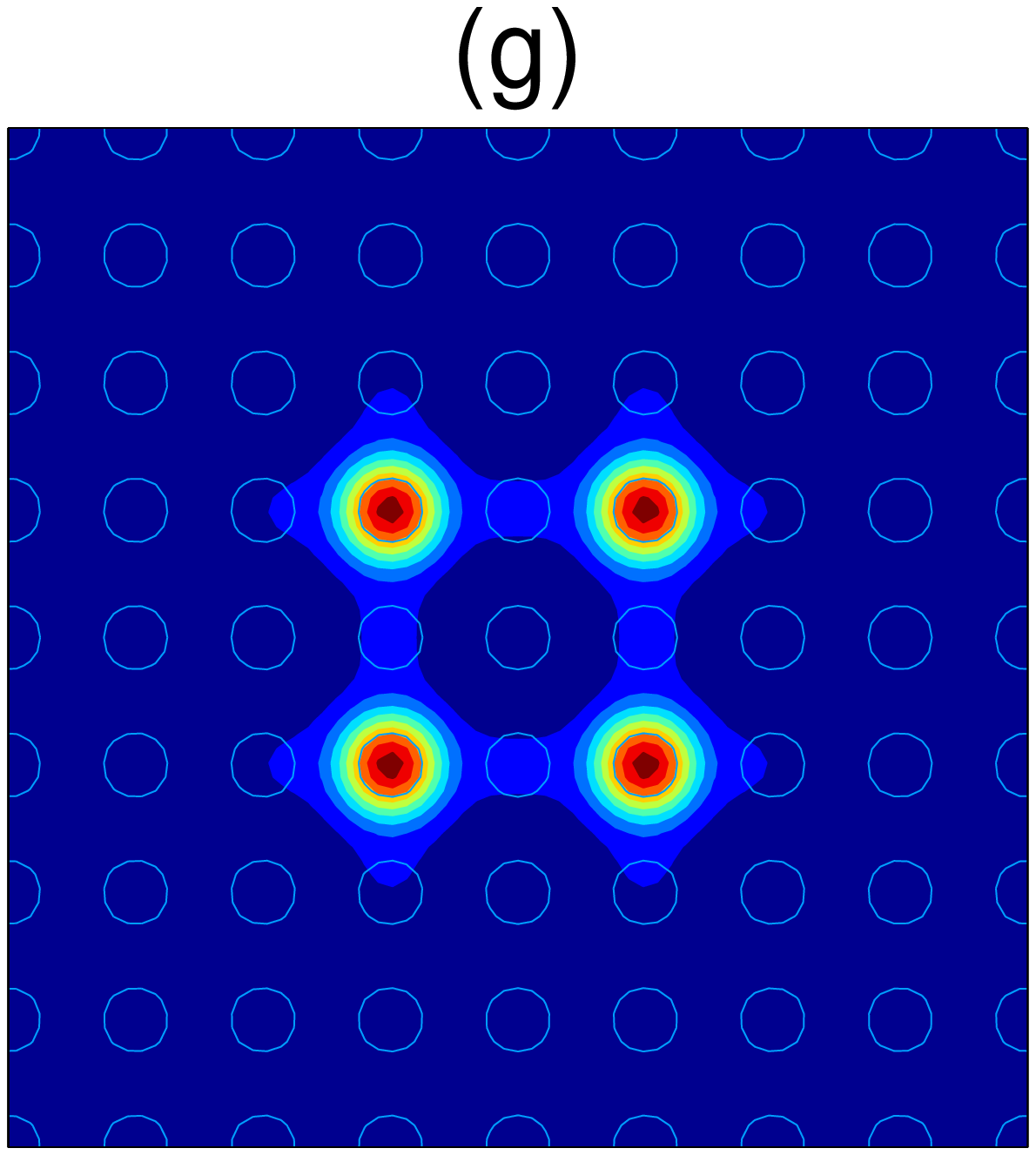}
\includegraphics[width=0.15\textwidth]{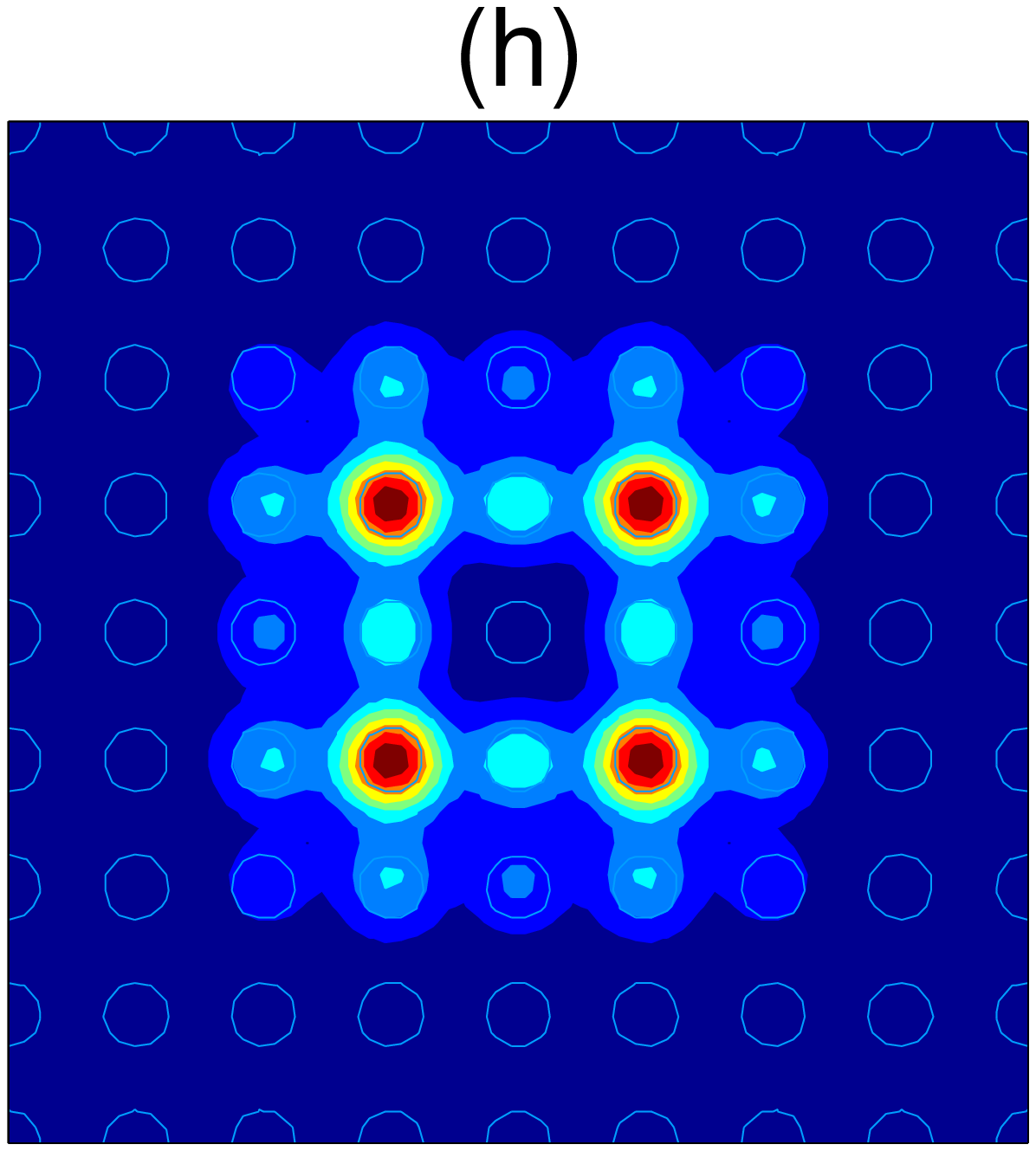}
\includegraphics[width=0.15\textwidth]{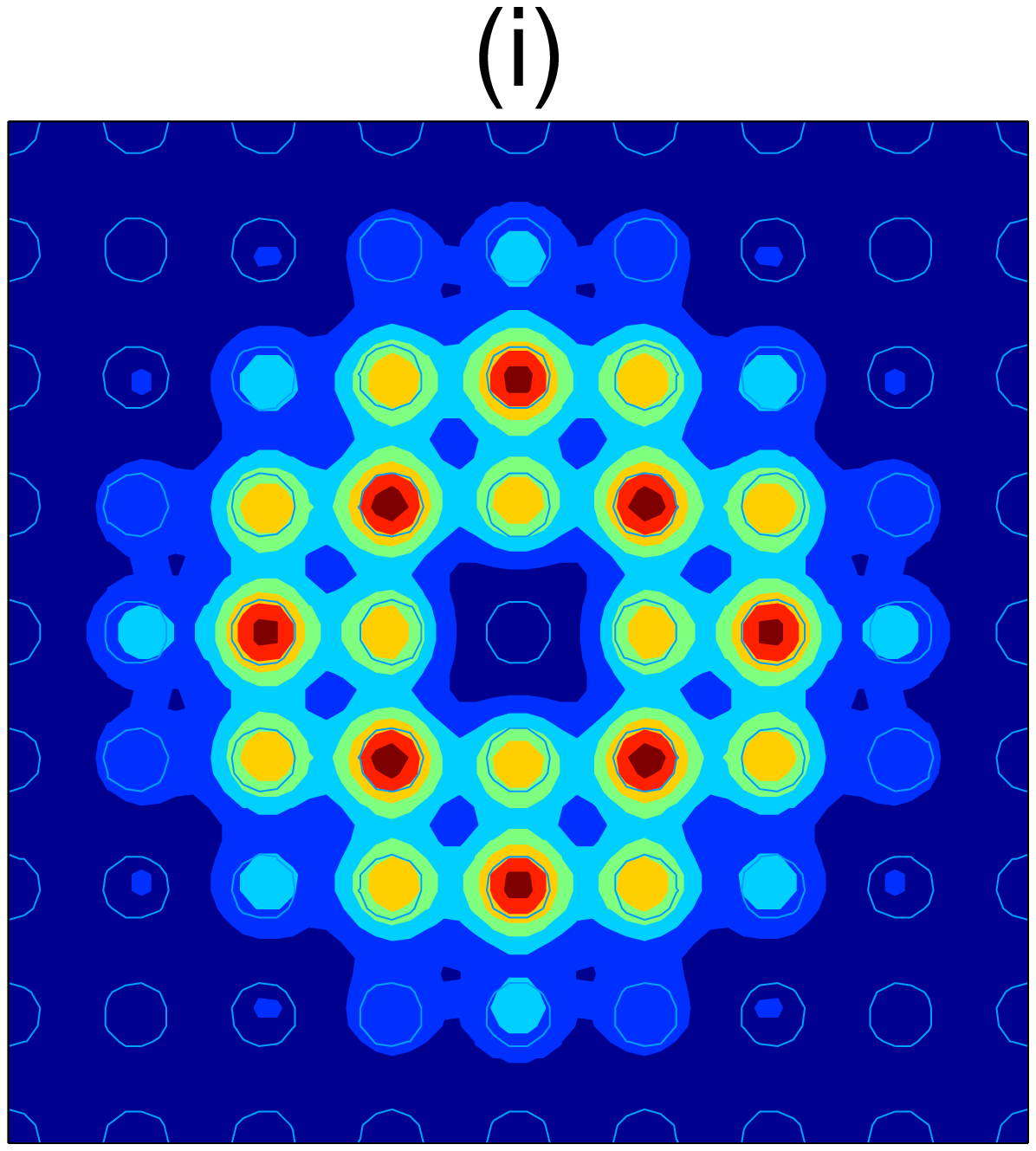}
\includegraphics[width=0.15\textwidth]{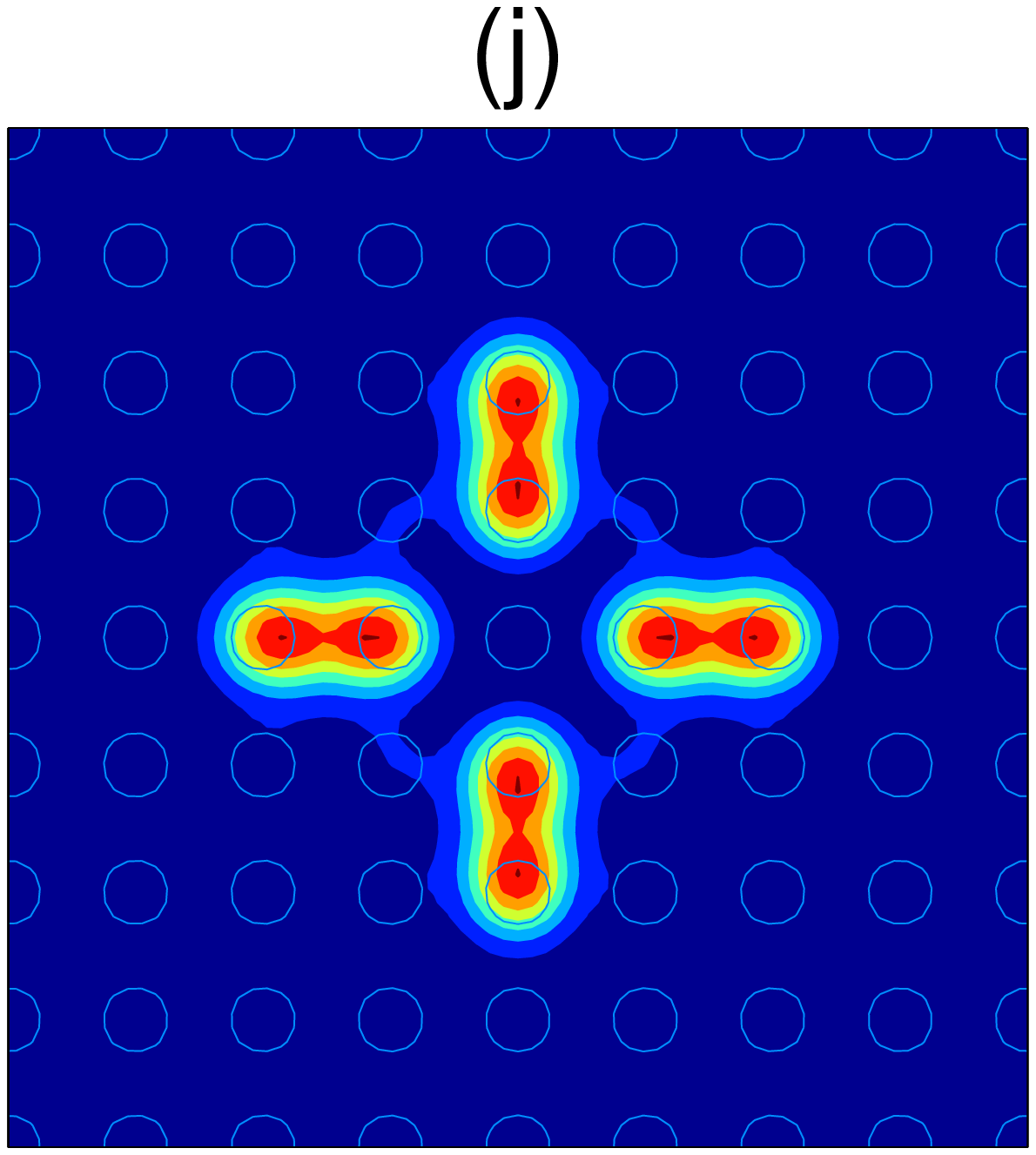}
\includegraphics[width=0.15\textwidth]{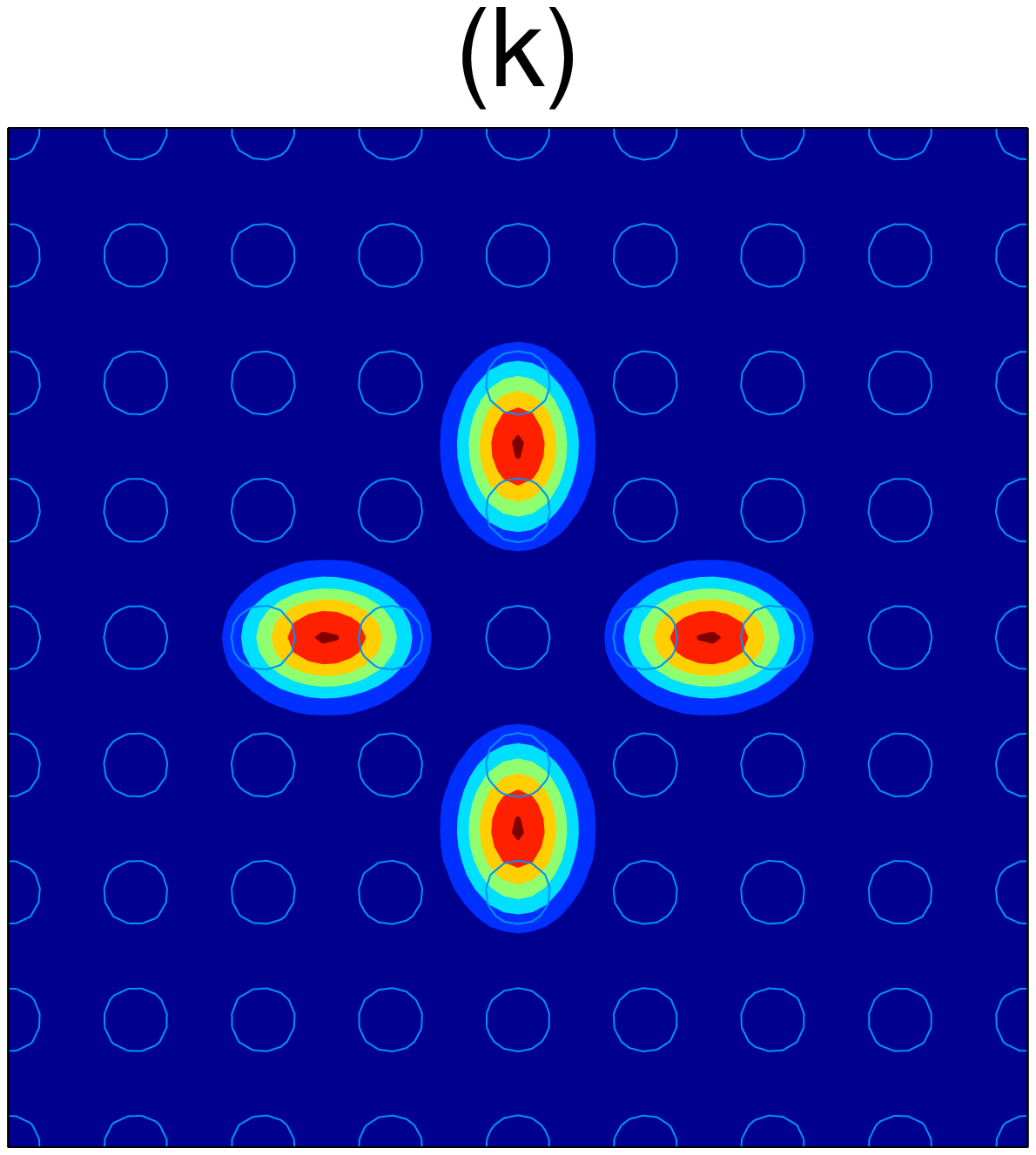}
\includegraphics[width=0.15\textwidth]{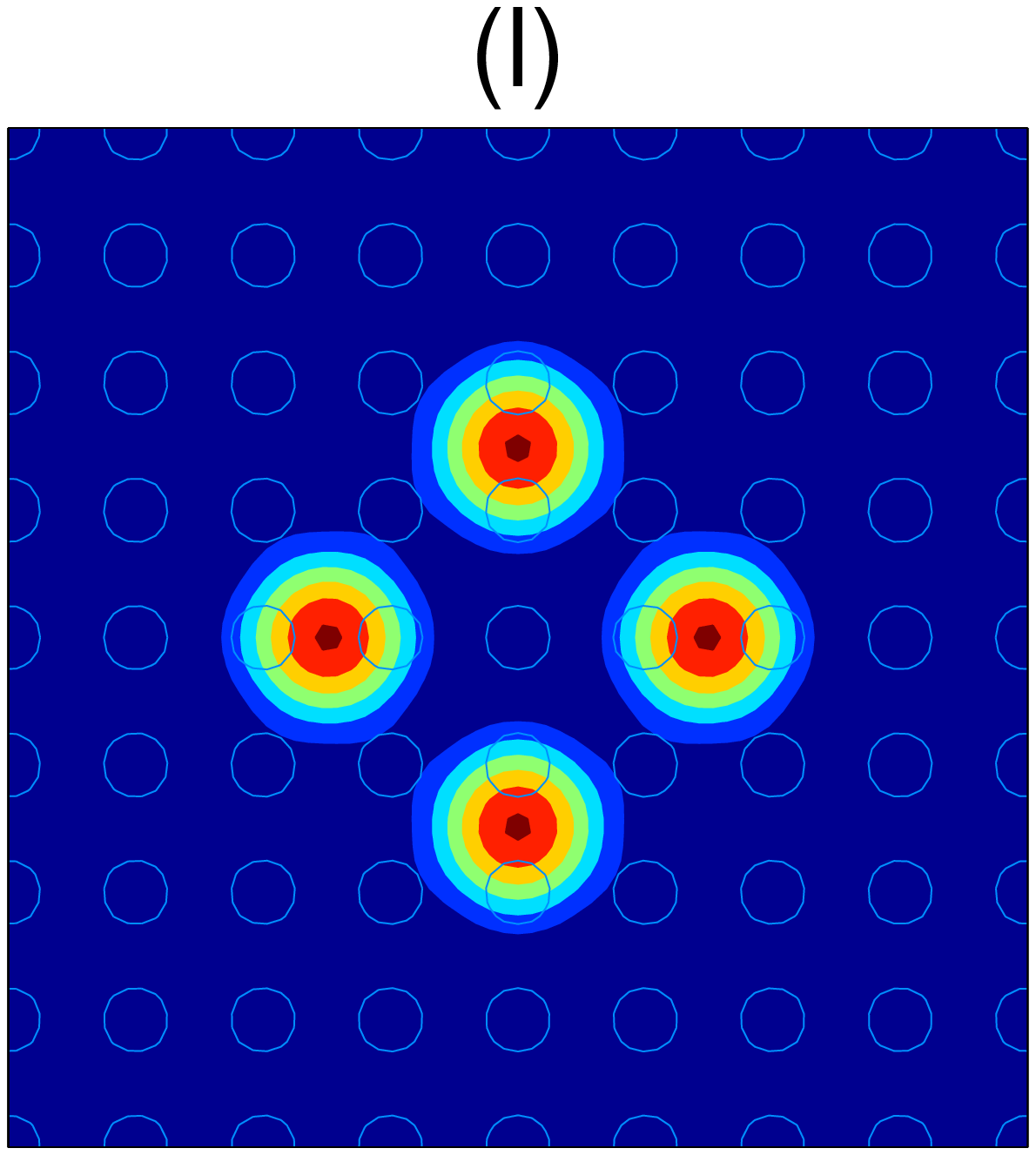}
\caption{(Color online) On-site vortex profiles ($|u|$)
corresponding to the marked points in Fig. \ref{figure8} under
focusing saturable nonlinearity. (l) shows the profile of the vortex
at $\mu$=1 (far from the band edge). The inset in (a) is the typical
phase structure of all these vortices. The background circles
represent the lattice sites (with high $I$ values), as in Fig.
\ref{figure11}.}\label{figure9}
\end{figure*}

First, we examine on-site vortex solitons under focusing saturable
nonlinearity. An infinite number of such vortex families exist in
the semi-infinite gap, and one of them is reported below. The power
curve of this family is displayed in Fig. \ref{figure8}. We see that
this curve is quite complicated. It has a number of turning points
(twists), and winds up and down in an unexpected way (a somewhat
similar power curve was reported for 1D lattice soltons under
cubic-quintic nonlinearities in \cite{JWang}). All vortices in this
family have simple $2\pi$ phase winding structures, thus having
charge one [see inset of Fig. \ref{figure9}(a)]. Their intensity
structures along the power curve are displayed in Fig.
\ref{figure9}. It is seen that far from the band edge (point `a' in
Fig. \ref{figure8}), the power is very high, and the vortex has a
ring profile [see Fig. \ref{figure9}(a)] . This ring profile can be
easily understood since the vortex here has high intensities and
dominates over the lattice, thus it effectively becomes a ring
vortex in a lattice-free (bulk) media. As the power and intensity
decrease (when $\mu$ increases), the lattice effect starts to
appear, and the vortex reshapes itself by distributing its main
power to the lattice sites. This leads to a four-humped on-site
vortex as seen in Fig. \ref{figure9}(b, c). This vortex has been
theoretically analyzed in \cite{Yang_NJP} and experimentally
observed in \cite{Segev_vortex}. This reshaping process, however, is
a delicate one as is evidenced by the two turning points near `b' in
the power curve of Fig. \ref{figure8} (a). As the power curve gets
close to the first band (see points `c, d'), the vortex starts to
spread out, and its amplitude becomes low. Vortices here closely
resemble those of the first on-site Kerr family near the first band
[see Fig. \ref{figure2} (b, c)]. Therefore, the power curve of
saturable vortices can not reach the first band either, and has to
turn around before reaching the band edge. After the turning point,
the power starts to increase (as $\mu$ moves away from the Bloch
band), and the vortex reshapes itself again. This time, after a
complex reshaping process, the eight-humped square vortex at point
`d' becomes a four-humped square vortex with its humps residing near
the four diagonal lattice sites around the vortex center [see Fig.
\ref{figure9} (e, f)]. Strangely, in this reshaping process, the
power curve turns around again at $\mu$=1.832 [between `e' and `f'
in Fig. \ref{figure8} (a)], and starts to decrease toward the first
band [see `g' in Fig. \ref{figure8} (b)]. Near the first band, the
power curve turns around again, and moves back into the
semi-infinite gap [see inset in Fig. \ref{figure8} (b)]. Near the
turning point, the vortex spreads out again [see Fig. \ref{figure9}
(h,i)]. Shortly after turning back, the vortex exhibits another
complex reshaping process evidenced by a twist in its power curve
[see the inset in Fig. \ref{figure8} (b) near point `i'].
Afterwards, the power curve moves into the semi-infinite gap
smoothly, and the vortex ultimately becomes four fundamental
solitons with $\pi/2$ phase delay between each other. Fig.
\ref{figure9} (l) shows such a vortex when the propagation constant
$\mu$=1, which is far away from the Bloch band. The complexity of
the power curve as well as the intricate reshaping process of
vortices exhibited in this saturable-vortex family is very
remarkable and is rarely seen in other wave systems.

\begin{figure}
\centering
\includegraphics[width=0.45\textwidth]{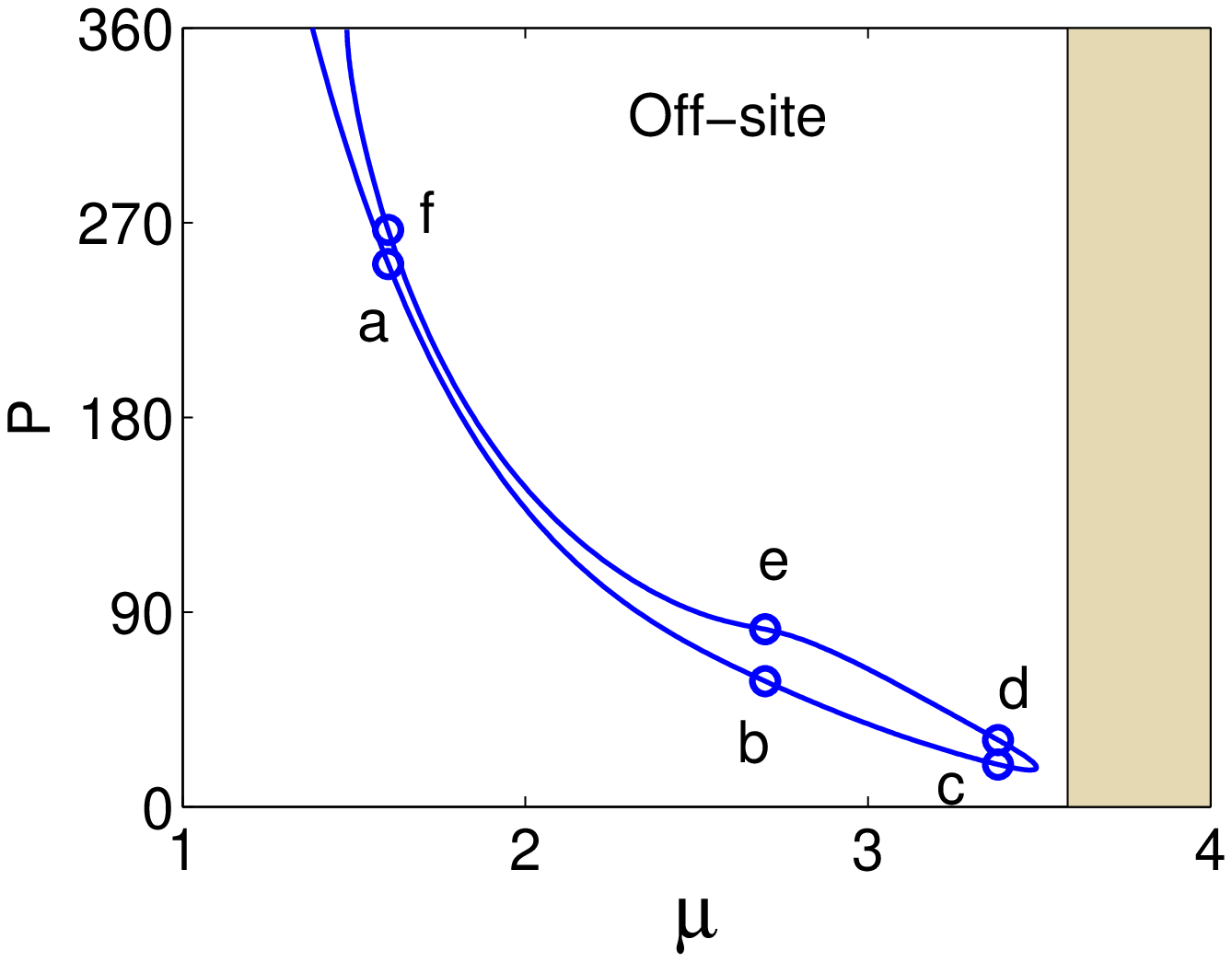}
\caption{The power diagram of an off-site vortex family in the
semi-infinite gap under focusing saturable nonlinearity. Vortex
profiles at the marked points are shown in Fig. \ref{figure11}.
}\label{figure10}
\end{figure}

\begin{figure}
\centering
\includegraphics[width=0.15\textwidth]{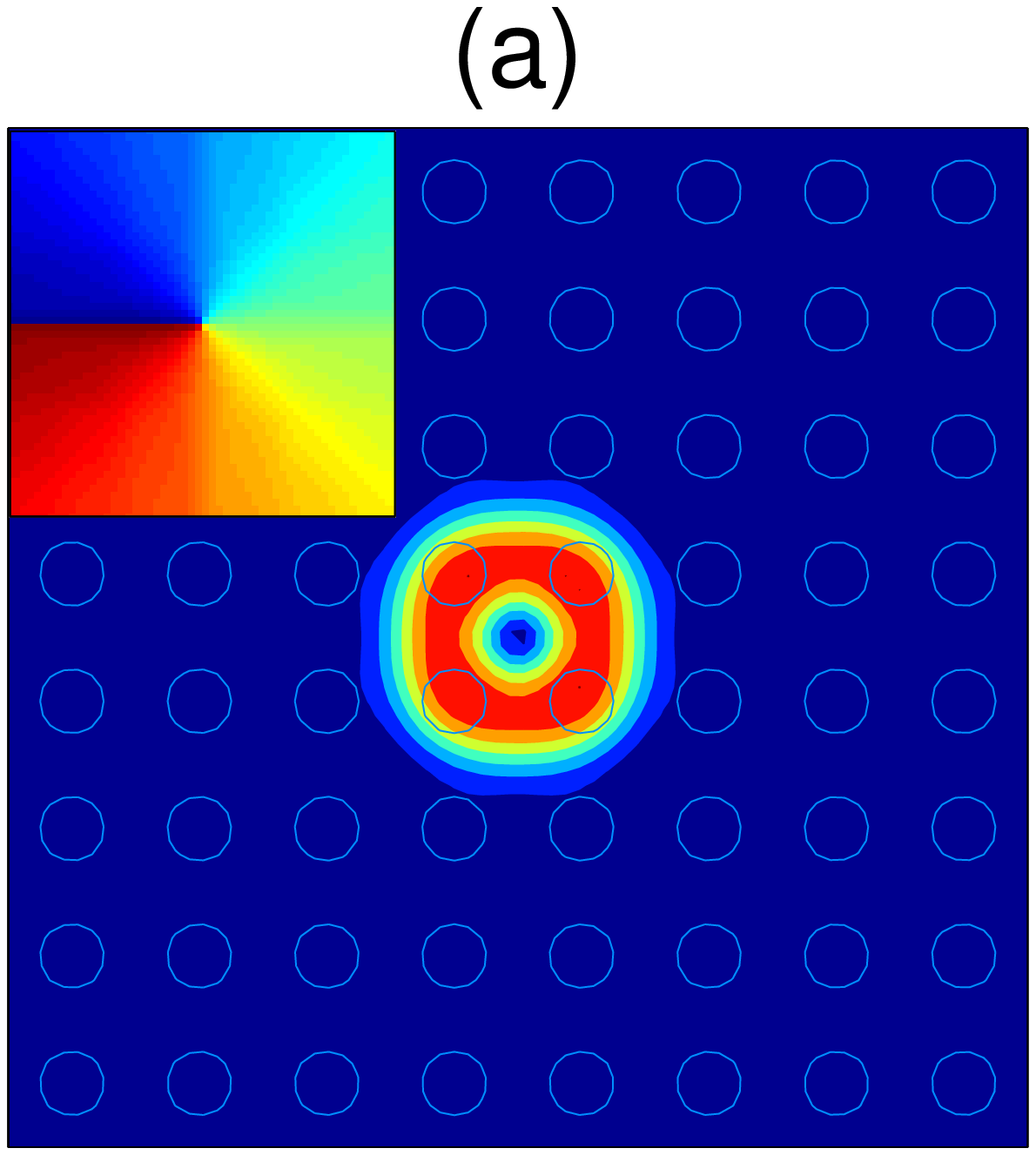}
\includegraphics[width=0.15\textwidth]{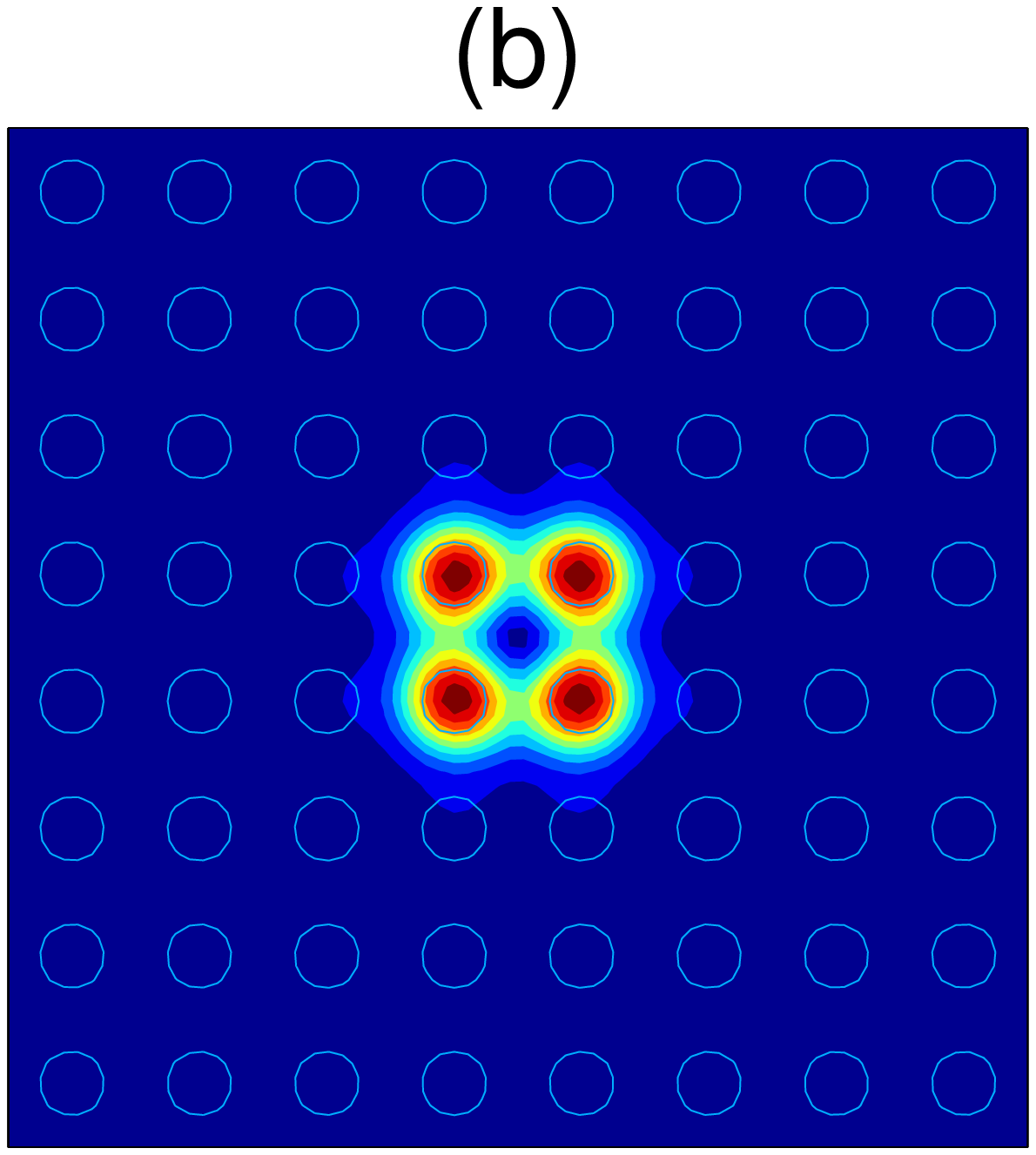}
\includegraphics[width=0.15\textwidth]{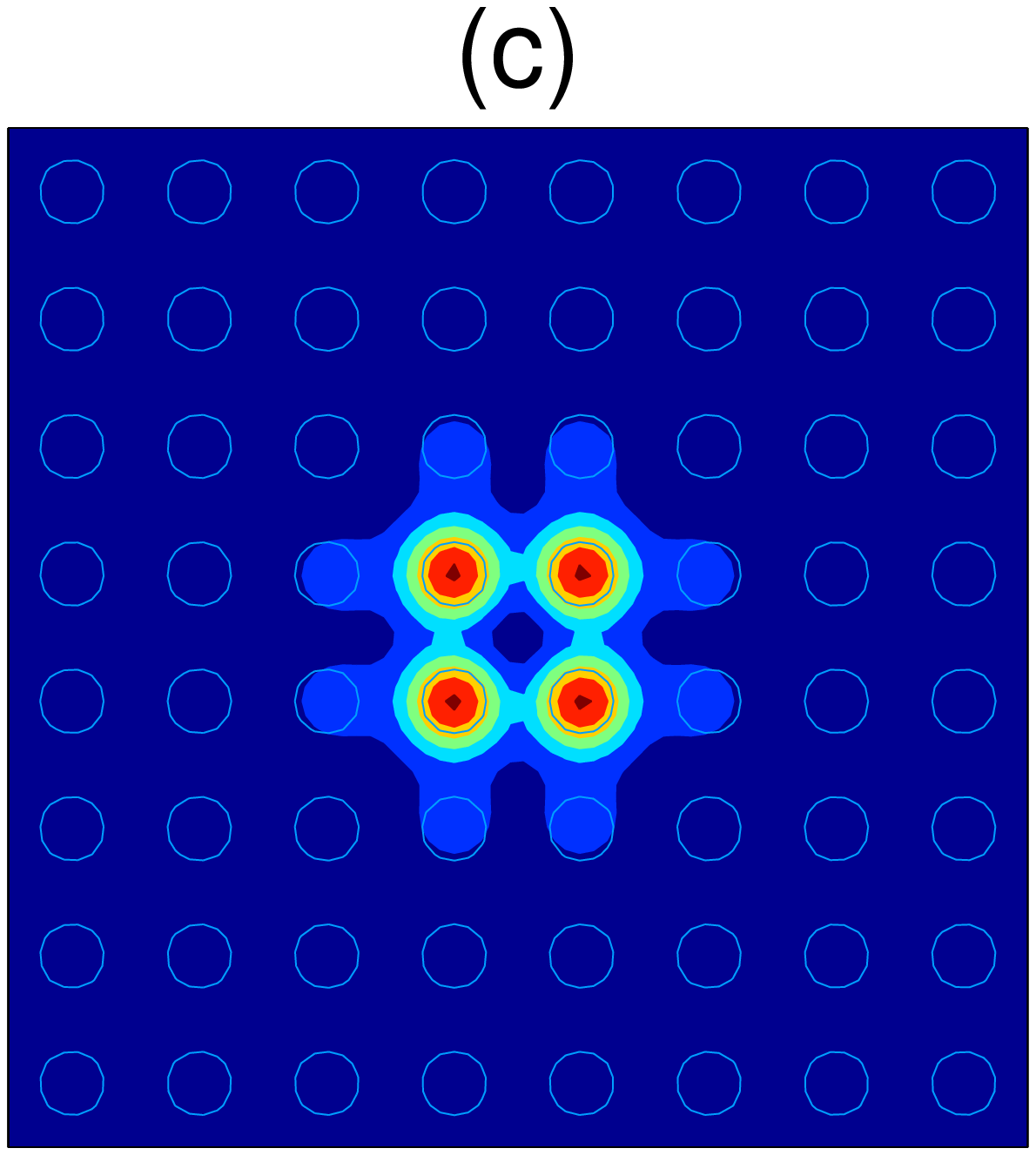}
\includegraphics[width=0.15\textwidth]{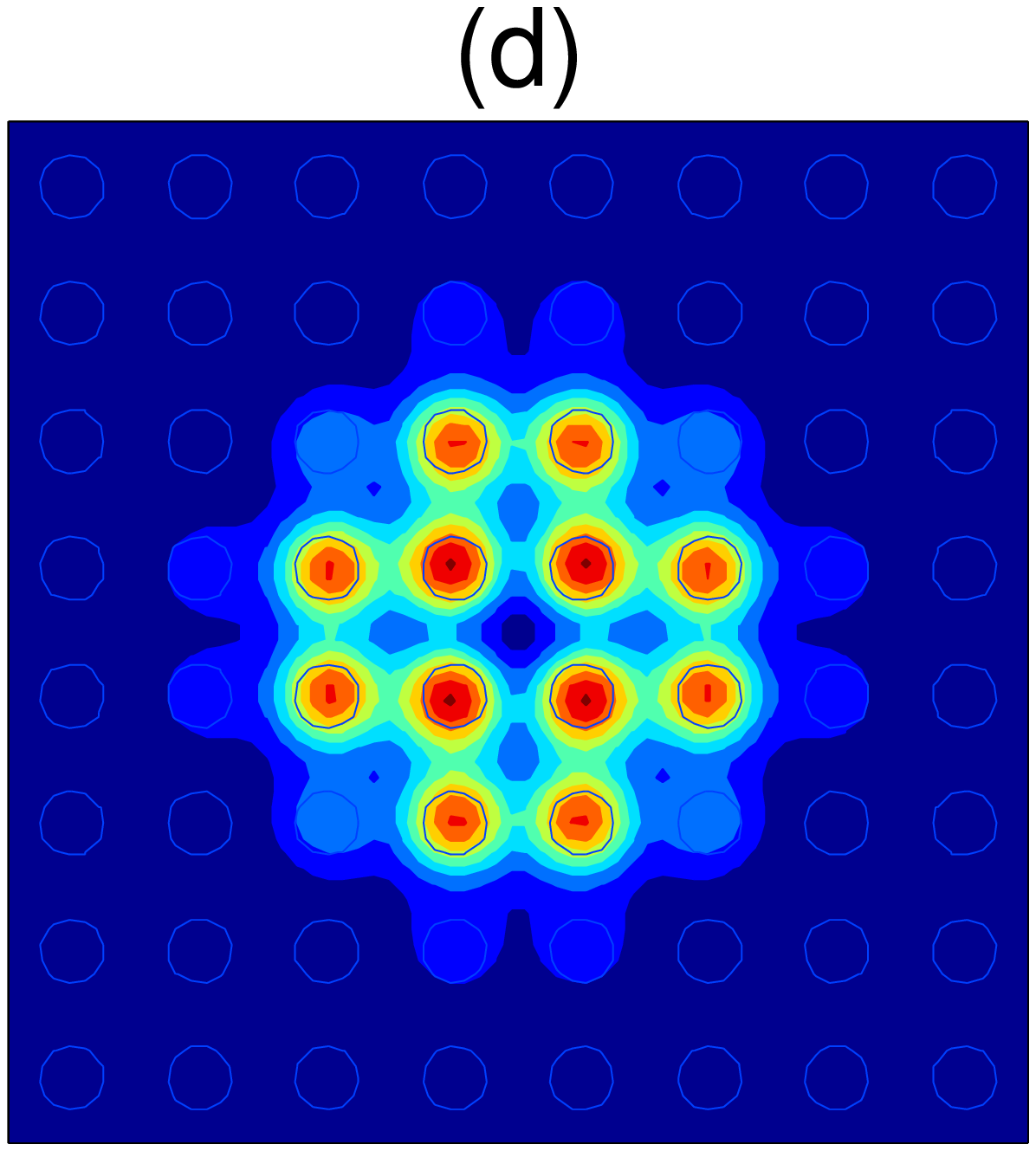}
\includegraphics[width=0.15\textwidth]{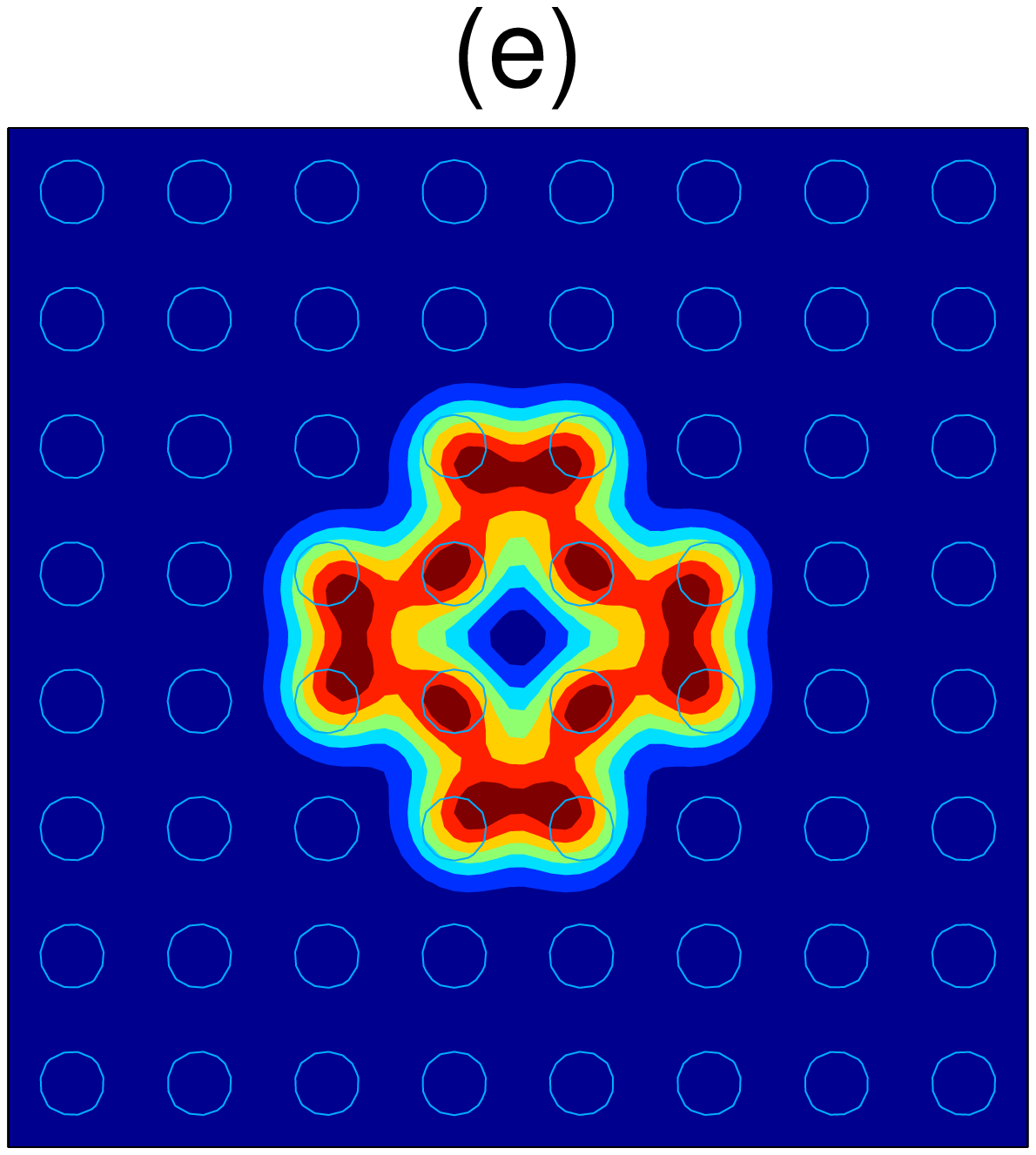}
\includegraphics[width=0.15\textwidth]{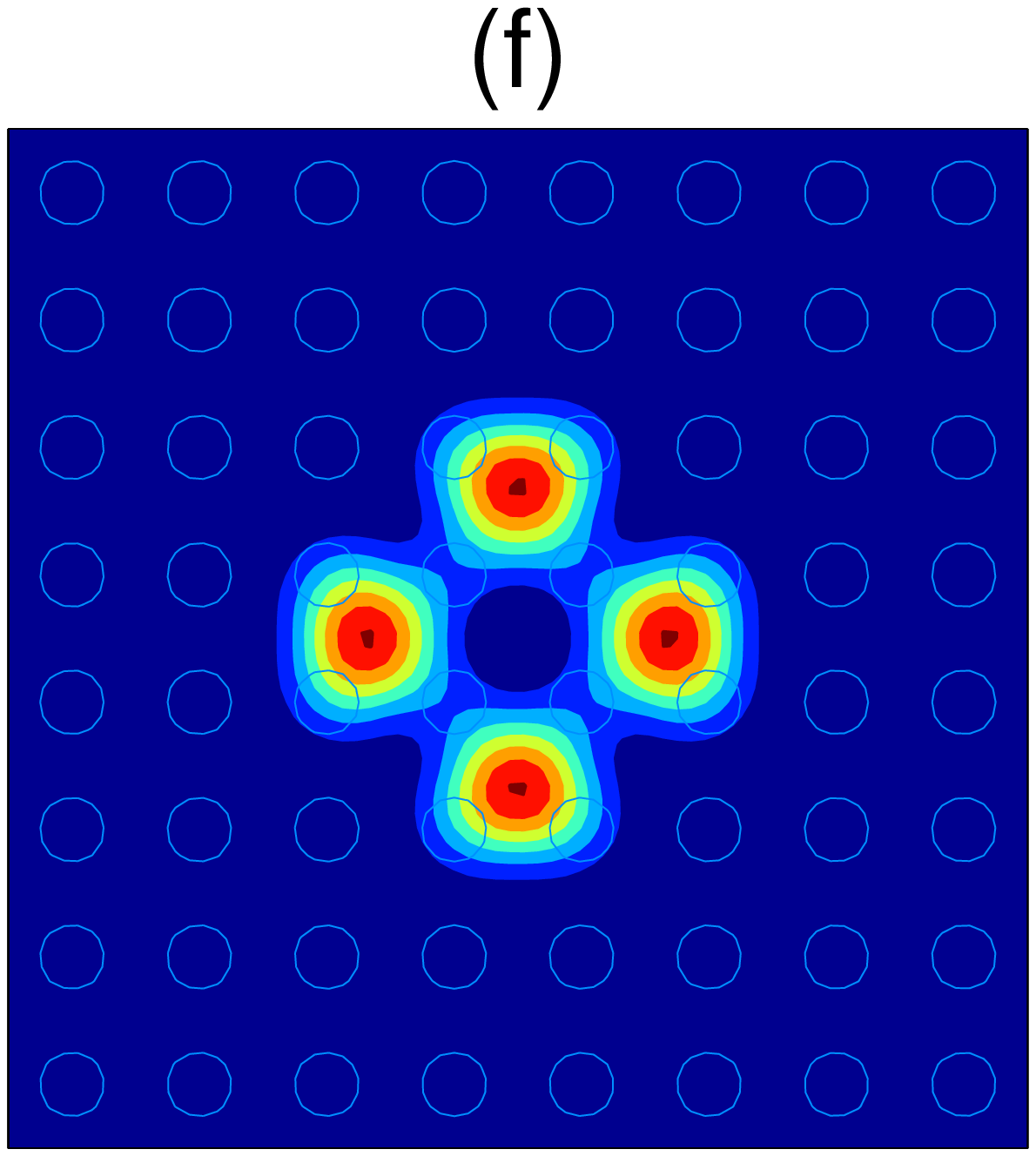}
\caption{(Color online) Off-site vortex profiles ($|u|$)
corresponding to the marked points in Fig. \ref{figure10} under
focusing saturable nonlinearity. The inset in (a) is the typical
phase structure of all these vortices. }\label{figure11}
\end{figure}

Now we consider off-site vortex solitons under focusing saturable
nonlinearity. Again, many families of them are found. One such
family is shown in Fig. \ref{figure10} (power curve) and Fig.
\ref{figure11} (vortex profiles). We see that this power curve as
well as its vortex reshaping process are simpler. Indeed, unlike the
above on-site saturable case, the power curve here has no twists. It
does have a turning point near the Bloch band though, just like all
other vortex families reported in this paper. At high powers, the
off-site vortex has a ring profile [Fig. \ref{figure11} (a)]. As its
power decreases (with increasing $\mu$ values), it reshapes itself
into a four-humped off-site vortex occupying four adjacent lattice
sites [Fig. \ref{figure11} (b, c)]. These vortices have been
theoretically analyzed and experimentally observed in
\cite{Yang_NJP,Chen_vortex,Segev_vortex}. Comparing these vortices
with those of off-site Kerr vortices in Fig. \ref{figure3}, we can
see that this saturable-vortex family is the counterpart of the
first off-site Kerr-vortex family. As the propagation constant $\mu$
moves from the lower branch to the upper one, this saturable vortex
continuously reshapes itself, and eventually turns into four humps
with $\pi/2$ phase delay between each other (as in the on-site
case). In all vortices of this off-site family, their phase fields
have simple $2\pi$-winding around the vortex center [see inset in
Fig. \ref{figure11} (a)], thus having charge one.

When the saturable nonlinearity is of defocusing type ($E_0<0$),
infinite families of vortex solitons exist in the first band gap. In
this case, we have found that saturable vortices show similar
behaviors (such as power curves and vortex shapes) as defocusing
Kerr vortices in Figs. \ref{figure5}-\ref{figure7}, thus such
results will not be displayed here.

\section{Summary and Discussion}
In this paper, we examined various families of charge-one vortex
solitons in a 2D periodic media. For both Kerr and saturable
nonlinearities, we found that there exist infinite families of
on-site and off-site vortex solitons in the semi-infinite gap (for
focusing nonlinearity) and the first gap (for defocusing
nonlinearity). We further showed that all these vortex families do
not bifurcate from edges of Bloch bands. Rather, before reaching
band edges, they turn back and move into band gaps again. Within
each vortex family, we revealed that topological structures of
vortex solitons undergo drastic changes (such as evolving from a
square configuration to a cross configuration) as the propagation
constant varies.

Regarding the linear stability of vortex solitons, it has been shown
before that certain on-site and off-site charge-one vortex solitons
in the semi-infinite gap (under focusing nonlinearity) are stable in
deep lattice potentials or with weak intersite couplings
\cite{Malomed,YangMuss,Yang_NJP,Peli_Panos,Kaiser}. Certain on-site
and off-site gap vortices under defocusing nonlinearity are stable
as well \cite{EAOstrovskaya2,Kaiser}. Such vortices generally
correspond to the lower branches of first-family vortices reported
in this paper [see Figs. 3(a) and \ref{figure7}(b) for instance]. It
will be interesting to investigate the linear stability of upper
branches of first-family vortices as well as branches of
higher-family vortices obtained in this paper. The stability results
in \cite{Peli_Panos} imply that some of the upper branches of
vortices are linearly unstable [such as Fig. 2(d)]. In such cases,
one can ask where on the vortex branches the instability starts to
appear. Note that all vortices reported in this paper have charge
one. It is known that vortices with more intensity humps can be
stabilized by increasing their charges
\cite{Peli_Panos,stable_charge2}. For instance, the vortex in Fig.
2(d) can be stabilized if its charge increases to three
\cite{Peli_Panos}. Thus it would be interesting to study vortex
families with higher charges as well as their stability properties.
These questions will be left for future studies.

\section{Acknowledgments}
This work was supported in part by the Air Force Office of
Scientific Research under Grant No. USAF 9550-05-1-0379.

\end{document}